\pdfoutput=1

\documentclass[11pt,twoside,a4paper,cmspaper,final,collab]{cms-tdr}
\def\svnVersion{379373}\def\cmsCernNoTag{CERN-EP/2016-097}\def\cmsCernDate{\today}\def\cmsMessage{Published in Physical Review D as \href{http://dx.doi.org/10.1103/PhysRevD.94.112009}{\doi{10.1103/PhysRevD.94.112009.}}}

\begin{document}\cmsNoteHeader{SUS-14-003}

\hyphenation{had-ron-i-za-tion}
\hyphenation{cal-or-i-me-ter}
\hyphenation{de-vices}
\RCS$Revision: 373727 $
\RCS$HeadURL: svn+ssh://svn.cern.ch/reps/tdr2/papers/SUS-14-003/trunk/SUS-14-003.tex $
\RCS$Id: SUS-14-003.tex 373727 2016-11-14 15:00:13Z cawest $

\newcommand{\ST}{\ensuremath{S_\mathrm{T}}\xspace}
\newcommand{\ETcone}{\ensuremath{E_\text{T,\,cone}}\xspace}
\newcommand{\Njet}{\ensuremath{N_{\text{jet}}}\xspace}
\newcommand{\Njets}{\ensuremath{N_{\text{jets}}}\xspace}
\providecommand{\tauh}{\ensuremath{\tau_\mathrm{h}}\xspace}
\providecommand{\CLs}{\ensuremath{\mathrm{CL}_\mathrm{s}}\xspace}
\ifthenelse{\boolean{cms@external}}{\providecommand{\CL}{C.L.\xspace}}{\providecommand{\CL}{CL\xspace}}
\ifthenelse{\boolean{cms@external}}{\providecommand{\NA}{\ensuremath{\cdots}}\xspace}{\providecommand{\NA}{---\xspace}}
\ifthenelse{\boolean{cms@external}}{\providecommand{\cmsTableResize[1]}{\relax{#1}}}{\providecommand{\cmsTableResize[1]}{\resizebox{\columnwidth}{!}{#1}}}
\newlength\cmsFigWidth
\ifthenelse{\boolean{cms@external}}{\setlength\cmsFigWidth{0.85\columnwidth}}{\setlength\cmsFigWidth{0.4\textwidth}}
\ifthenelse{\boolean{cms@external}}{\providecommand{\cmsLeft}{top\xspace}}{\providecommand{\cmsLeft}{left\xspace}}
\ifthenelse{\boolean{cms@external}}{\providecommand{\cmsRight}{bottom\xspace}}{\providecommand{\cmsRight}{right\xspace}}
\ifthenelse{\boolean{cms@external}}{\providecommand{\cmsLLeft}{Top\xspace}}{\providecommand{\cmsLLeft}{Left\xspace}}
\ifthenelse{\boolean{cms@external}}{\providecommand{\cmsRRight}{Bottom\xspace}}{\providecommand{\cmsRRight}{Right\xspace}}
\cmsNoteHeader{SUS-14-003}
\title{\texorpdfstring{Searches for $R$-parity-violating supersymmetry in $\Pp\Pp$ collisions at $\sqrt{s}=8$\TeV in final states with 0--4 leptons}{Searches for R-parity-violating supersymmetry in pp collisions at sqrt(s)=8 TeV in final states with 0--4 leptons}}

\date{\today}

\abstract{
Results are presented from searches for $R$-parity-violating supersymmetry in events produced in $\Pp\Pp$ collisions at $\sqrt{s}=8$\TeV at the LHC. Final states with 0, 1, 2, or multiple leptons are considered independently. The analysis is performed on data collected by the CMS experiment corresponding to an integrated luminosity of 19.5\fbinv. No excesses of events above the standard model expectations are observed, and 95\% confidence level limits are set on supersymmetric particle masses and production cross sections. The results are interpreted in models featuring $R$-parity-violating decays of the lightest supersymmetric particle, which in the studied scenarios can be either the gluino, a bottom squark, or a neutralino. In a gluino pair production model with baryon number violation, gluinos with a mass less than 0.98 and 1.03\TeV are excluded, by analyses in a fully hadronic and one-lepton final state, respectively. An analysis in a dilepton final state is used to exclude bottom squarks with masses less than 307\GeV in a model considering bottom squark pair production. Multilepton final states are considered in the context of either strong or electroweak production of superpartners and are used to set limits on the masses of the lightest supersymmetric particles. These limits range from 300 to 900\GeV in models with leptonic and up to approximately 700\GeV in models with semileptonic $R$-parity-violating couplings.
 }

\hypersetup{%
pdfauthor={CMS Collaboration},%
pdftitle={Searches for R-parity-violating supersymmetry in pp collisions at sqrt(s)=8 TeV in final states with 0-4 leptons},%
pdfsubject={CMS},%
pdfkeywords={CMS, physics, R-parity violation}}

\maketitle

\section{Introduction}
Supersymmetry (SUSY) is an attractive extension of the standard model (SM) because it can solve the hierarchy problem and can ensure gauge coupling unification~\cite{Nilles:1983ge,Haber:1984rc}. The majority of the searches for SUSY focus on $R$-parity-conserving (RPC) models.
The $R$-parity of a particle is defined by $R = (-1)^{3B+L+2s}$, where $B$ and $L$ are its baryon and lepton numbers, respectively, and $s$ is the particle spin~\cite{Farrar:1978xj}. In RPC SUSY, the lightest supersymmetric particle (LSP) is stable, which ensures proton stability and provides a dark matter candidate. All SM particles have $R = +1$; SUSY posits the existence of a superpartner with $R = -1$ corresponding to each SM particle.

The most general gauge-invariant and renormalizable superpotential violates $R$-parity, and so $R$-parity violation is expected unless it is forbidden by some symmetry.
Supersymmetric models with $R$-parity-violating (RPV) interactions can break baryon or lepton number conservation~\cite{Barbier:2004ez,PDG}.
The superpotential $W_\mathrm{RPV}$ includes a bilinear term proportional to the coupling $\mu^{\prime}_{i}$ and three trilinear terms parameterized by the couplings $\Lam_{ijk}$, $\Lamp_{ijk}$, and
$\Lampp_{ijk}$:
\ifthenelse{\boolean{cms@external}}{
\begin{multline}
\label{eqn:wrpv}
W_\mathrm{RPV} =
\frac{1}{2}\, \lambda_{ijk}L_iL_j\overline{E}_k+
\lambda^\prime_{ijk}L_iQ_j\overline{D}_k\\
+\,\frac{1}{2}\, \lambda^{\prime \prime}_{ijk}\overline{U}_i\overline{D}_j\overline{D}_k+
\mu^\prime_{i} H_{u} L_{i},
\end{multline}
}{
\begin{equation}
\label{eqn:wrpv}
W_\mathrm{RPV} =
\frac{1}{2}\, \lambda_{ijk}L_iL_j\overline{E}_k+
\lambda^\prime_{ijk}L_iQ_j\overline{D}_k\,+\,
\frac{1}{2}\, \lambda^{\prime \prime}_{ijk}\overline{U}_i\overline{D}_j\overline{D}_k+
\mu^\prime_{i} H_{u} L_{i},
\end{equation}
}
where $i,j,$ and $k$ are generation indices; $L$,  $Q$ and $H_u$ are the lepton, quark, and up-type Higgs $SU(2)_L$ doublet superfields, respectively; and $\overline{E}$, $\overline{D}$, and $\overline{U}$ are the charged lepton, down-type quark, and up-type quark $SU(2)_L$ singlet superfields, respectively.  The third term violates baryon number conservation, while the other terms violate lepton number conservation.
The final term, involving the lepton and up-type Higgs doublets, is also allowed in the superpotential but the effects of this term are not considered in this analysis.

Experimental bounds on leptonic, semileptonic, and hadronic RPV couplings are complementary due to the strong constraint on the product of RPV couplings from nucleon stability measurements.
For example, for squark masses of 1\TeV, stringent experimental limits on proton decay result in the constraint $|\lambda^{\prime}_{ijk}\lambda^{\ast}_{i^{\prime}j^{\prime}k^{\prime}}|<\mathcal{O}(10^{-9})$ for all generation indices~\cite{Smirnov:1996bg}.
Much stronger (by a factor of up to $\approx 10^{18}~$\cite{Barbier:2004ez}) constraints are possible for couplings involving light generations, and similar constraints exist for products of other RPV couplings.

A subset of RPV scenarios focus on the RPV extension of the minimal supersymmetric model (MSSM) when the assumption of minimal flavor violation (MFV) is imposed~\cite{Nikolidakis:2007fc}. Under this assumption, the only sources of $R$-parity violation are the SM Yukawa couplings,
 and the RPV couplings are therefore related to the components of the Cabibbo--Kobayashi--Maskawa matrix and the fermion masses.
In some of these models, $\lambda^{\prime\prime}_{332}$ is the largest RPV coupling~\cite{MFV} and will be a focus of the searches involving hadronic $R$-parity violation.

The missing transverse momentum vector $\ptvecmiss$ is defined as the negative of the vector sum of momenta in the transverse direction. Its magnitude $\ETmiss$ is often used in searches for RPC SUSY as the LSP is stable and leaves the detector undetected, leading to large values of $\MET$.
In the RPV models considered, the LSP decays promptly to SM particles and therefore no large $\MET$ is expected.
Instead, we employ a variety of methods to search for the different types of RPV decays. 

We search for hadronic RPV SUSY, which arises when any of the $\lambda^{\prime \prime}_{ijk}$ are nonzero, in events with zero or one lepton using the jet and $\PQb$-tagged jet multiplicities of the event, and in dilepton events by means of a kinematic fit to reconstruct the bottom squark mass.
We focus on couplings that involve top quarks, as motivated by MFV; the leptons in the final state are the result of leptonic $\PW$ decay.
The results of the fully hadronic and one-lepton analyses are interpreted in a model in which the gluino decays via $\PSg\rightarrow\cPaqt\sTop$, followed by decay of the top squark via a nonzero $\lambda''_{323}$ coupling: $\sTop\rightarrow\cPaqb\cPaqs$ (charge conjugate reactions are implied throughout this paper). Here the top squark is considered to be much heavier than the gluino, resulting in an effective three-body decay of the gluino.
The analysis of the dilepton final state considers pair production of bottom squarks, which decay to a top quark and either a down or a strange quark.

To search for leptonic and semileptonic RPV SUSY, which arise when $ \lambda_{ijk}$ and $ \lambda^{\prime}_{ijk}$ respectively, are nonzero, we examine events with three or more leptons, binned in the multiplicity of reconstructed objects.
Both strong and electroweak production of superpartners are considered.
An analysis of a four-lepton final state targets production of squarks and gluinos in which the lightest neutralino decays to final states with electrons or muons.
We also study final states with at least three leptons, which are sensitive to electroweak production of winos and Higgsinos.

In all analyses considered, the LSP is assumed to decay promptly, meaning the decay vertex is indistinguishable from the primary interaction. This generally implies $\lambda > 10^{-6}$~\cite{Farrar:1978xj}.
All RPV couplings are assumed to be zero, except for the specific coupling under study.

In this paper, we present the results of these searches with interpretations in a variety of different RPV models.  The data set was recorded with the CMS detector at the CERN LHC in proton-proton collisions at a center-of-mass energy of 8\TeV and corresponds to an integrated luminosity of 19.3--19.5\fbinv.

Searches for multijet resonances, a prominent signal when hadronic RPV SUSY is present, have been performed by CDF~\cite{cdfmultijets}, ATLAS~\cite{gluinoATLAS}, and CMS~\cite{gluino2011,cmsmultijets,Chatrchyan:2013gia}. The ATLAS Collaboration has also performed a search for RPV SUSY in high-multiplicity events~\cite{Aad:2015lea}. Searches for RPV interactions in multilepton final states have been carried out at LEP~\cite{Heister:2002jc,Abdallah:2003xc,Achard:2001ek}, the Tevatron~\cite{Abazov:2006ii,Abazov:2006nw,Abulencia:2007mp}, HERA~\cite{Collaboration:2010ez,zeus:lqdstop}, and the LHC~\cite{RA7PLB,Chatrchyan:2012mea,ATLAS:2012kr,Aad:2012ypy,Chatrchyan:2013xsw,Aad:2014iza}.

The paper is organized as follows.
Section~\ref{sec:detector} presents an overview of the CMS detector, and a description of simulated signal and background samples is given in Section~\ref{sec:simulation}.
The analyses described in this paper use a common limit-setting procedure.
This procedure, as well as the treatment of signal samples, is described in Section~\ref{sec:signals}.
The event selections that are common to all analyses in this paper are described in Section~\ref{sec:selections}.
The searches reported in this paper cover a wide range of signatures induced by the various RPV couplings.
Sections~\ref{sec:hadronic} and~\ref{sec:onelepton} detail searches for hadronic $R$-parity violation in zero- and one-lepton final states, respectively.
Section~\ref{sec:dilepton} describes a search for bottom squarks that decay to a top quark and either a down or a strange quark.
Finally, searches for $R$-parity violation induced by leptonic and semileptonic RPV couplings in multilepton final states are described in Sections~\ref{sec:multilepton} and~\ref{sec:multilepton_broad}, respectively.
The results of this paper are summarized in Section~\ref{sec:conclusions}.

\section{CMS detector and reconstruction}
\label{sec:detector}
A detailed description of the CMS detector, together with a definition of the coordinate system used, can be found in Ref.~\cite{Chatrchyan:2008zzk}.
The central feature of the CMS apparatus is a superconducting solenoid with an internal diameter of 6\unit{m}, which generates a 3.8\unit{T} uniform magnetic field along the axis of the LHC beams. A silicon pixel and strip tracker, a crystal electromagnetic calorimeter (ECAL), and a brass and scintillator hadron calorimeter (HCAL) are located within the magnet. Muons are identified and measured in gas-ionization detectors embedded in the outer steel magnetic flux-return yoke of the solenoid.
The silicon tracker, the muon system, and the barrel and endcap calorimeters cover the pseudorapidity ranges $\abs{\eta} < 2.5$, $\abs{\eta} < 2.4$, and $\abs{\eta} < 3.0$, respectively.
The first level of the CMS trigger system, composed of custom hardware processors, uses information from the calorimeters and muon detectors to select the events most relevant for analysis in a fixed time interval of less than 4\mus. The high-level trigger processor farm further decreases the event rate from around 100\unit{kHz} to less than 1\unit{kHz}, before data storage.

The particle-flow event algorithm~\cite{PFT-10-004,PFT-08-001,PFT-10-002} reconstructs and identifies individual particles with an optimized combination of information from the various elements of the CMS detector. The energy of photons is directly obtained from the ECAL measurement. The energy of electrons is determined from a combination of the electron momentum at the primary interaction vertex as determined by the tracker, the energy of the corresponding ECAL cluster, and the energy sum of all bremsstrahlung photons spatially compatible with originating from the electron track. The energy of muons is obtained from the curvature of the corresponding track. The energy of charged hadrons is determined from a combination of their momentum measured in the tracker and the matching ECAL and HCAL energy deposits. Finally, the energy of neutral hadrons is obtained from the corresponding corrected ECAL and HCAL energy.

\section{Simulation}
\label{sec:simulation}
The Monte Carlo (MC) simulation is used to estimate some of the SM backgrounds and to understand the efficiency of the signal models, including geometrical acceptance. The SM background samples are generated using \MADGRAPH 5.1.3.30~\cite{Maltoni:2002qb}, with parton showering and fragmentation modeled using \PYTHIA (version 6.420)~\cite{Sjostrand:2007gs}, and passed through a \GEANTfour-based~\cite{Agostinelli:2002hh} representation of the CMS detector.
QCD multijet samples are generated with up to four partons in the matrix element and \ttbar+jets events are generated with up to three extra partons in the matrix element.
The parton shower is matched to the matrix element with the MLM prescription~\cite{Mangano:2006rw}.
Signal samples~\cite{Evans:2012bf} are generated with \MADGRAPH and \PYTHIA. Most of these samples are then passed through the CMS fast-simulation package~\cite{Abdullin:2011zz}; the others are simulated with the same full simulation used for background processes.
The CTEQ6L1~\cite{Pumplin:2002vw} set of parton distribution functions (PDFs) is used throughout.
Background yields, when taken from simulation, are normalized to next-to-leading-order (NLO) or next-to-next-to-leading-order (NNLO) cross sections, when available.
There is an additional uncertainty of 2.6\% in these yields due to the imperfect knowledge of the integrated luminosity~\cite{CMS-PAS-LUM-13-001}.
The modeling of multiple proton-proton primary interactions in a single bunch crossing, referred to as pileup, is corrected so that the pileup profile matches that of the data. In the data set used in this paper, the mean number of interactions per bunch crossing is 21.

\section{Common procedures for signal samples and limits}
\label{sec:signals}

The analyses described in this paper use many shared procedures for the modeling of signal components and the setting of cross section limits, which we describe in this section.

The analyses are interpreted in simplified models of SUSY~\cite{ArkaniHamed:2007fw,Alwall:2008ag,Alves:2011wf}.
In all of the interpretations, supersymmetric particles not explicitly considered are assumed to have very large masses so that their effect is negligible.
However, the masses of intermediate states are assumed to be small enough that all supersymmetric particles decay promptly.

The separate analyses probe different sectors of RPV parameter space, and the models used to interpret the results vary accordingly.
The hadronic and one-lepton analyses are sensitive mainly to RPV SUSY in the hadronic sector and therefore assume that $\lambda^{''}_{332}\neq 0$ and all other RPV couplings are zero; the experimental signature of a nonzero $\lambda^{''}_{331}$ coupling is identical as there is no discrimination between \PQs~and \cPqd~quarks.
Similarly, the two-lepton analysis assumes $\lambda^{''}_{332}\neq 0$ or $\lambda^{''}_{331}\neq 0$, with no other nonzero RPV couplings.
As the multilepton searches analyze different lepton flavors separately, they are interpreted in terms of several models with nonzero lepton-flavor-violating couplings, $\lambda_{ijk}\neq 0$ or $\lambda^{'}_{ijk}\neq 0$, for several values of $i$, $j$, and $k$.

The uncertainty in the knowledge of the PDFs is obtained by applying the envelope prescription of the PDF4LHC working group~\cite{Alekhin:2011sk,Botje:2011sn} with three different PDF sets (CTEQ6.6~\cite{Nadolsky:2008zw}, MSTW2008nlo68cl~\cite{Martin:2009iq} and NNPDF2.0-100~\cite{Ball:2010de}).
Scale factors are applied to the fast-simulation samples, as a function of the transverse momentum \pt and $\abs{\eta}$, to reproduce the \PQb~jet identification and misidentification efficiencies obtained from a full simulation of the CMS detector.
The uncertainty in the modeling of initial- and final-state radiation (ISR and FSR, respectively) is obtained from the discrepancies between data and simulation observed in Z+jets, diboson+jets, and $\ttbar$+jets events as a function of the \pt of the system recoiling against the ISR jets~\cite{Chatrchyan:2013xna}.

Limits are calculated using the \CLs~\cite{Junk:1999kv,Read:2002hq} method. The LHC-style test statistic is used, within the formalism developed by the CMS and ATLAS collaborations in the context of the LHC Higgs Combination Group~\cite{CMS-NOTE-2011-005}.
For each cross section $\sigma$ being tested, the likelihood is profiled with respect
to the nuisance parameters; that is, the nuisance parameters are treated as fit parameters subject to external constraints on their magnitude and distribution.
We find the one-sided $p$-value of the
observed data in the signal-plus-background hypothesis, denoted
$p_\sigma$.  This is the fraction of pseudoexperiments with test statistic
$\lambda_p(\sigma)$ less than the value measured in data.
We also generate pseudoexperiments with the signal cross section
set to zero to construct the distribution of
$\lambda_p(\sigma)$ under the background-only hypothesis.
From this distribution we obtain the $p$-value of data in the
background-only hypothesis, denoted $p_0$.  Then
\CLs is defined as $p_\sigma/(1-p_0)$.  If
$\CLs<0.05$, that value of $\sigma$ is deemed to be excluded at a 95\% confidence level (\CL).
The largest cross section not excluded corresponds to the
\CLs upper limit.

Cross sections for SUSY signal processes, calculated at NLO with next-to-leading-log (NLL) resummation, are taken from the LHC SUSY Cross Sections Working Group~\cite{Beenakker:1999ch,Kulesza:2008jb,Kulesza:2009kq,Beenakker:2009ha,Beenakker:2011fu}.
To account for theoretical uncertainties conservatively, mass exclusions are quoted using a signal production cross section that is reduced from the nominal value by the amount of the theoretical uncertainty.

\section{Object selection}
\label{sec:selections}
Electrons and muons are reconstructed using the tracker, calorimeter, and muon systems. Details of the reconstruction and identification for electrons and muons can be found in Ref.~\cite{Khachatryan:2015hwa} and Ref.~\cite{Chatrchyan:2012xi}, respectively.
In the leptonic analyses, we require that at least one electron or muon in each event has $\pt > 20$\GeV. Additional electrons and muons must have $\pt > 10$\GeV and all leptons must be within $\abs{\eta} < 2.4$.

The majority of hadronic decays of $\tau$ leptons ($\tauh$) yield either a single charged particle (one-prong) or three charged particles (three-prong), with or without additional electromagnetic energy from neutral pion decays. We use one- and three-prong $\tauh$ candidates with $\pt > 20$\GeV, reconstructed with the ``hadron plus strips'' method~\cite{HPStaus}, which has an efficiency of approximately 70\%.
Leptons produced in $\tau$ decays are included with other electrons and muons.

To ensure that electrons, muons, and $\tauh$ candidates are isolated, we use the variable $\ETcone$, defined as the transverse energy in a cone of radius $\DR \equiv \sqrt{\smash[b]{(\Delta \eta)^2 + (\Delta \phi)^2}}=0.3$ around the candidate, excluding the candidate itself.
We remove energy from additional simultaneous proton-proton collisions by subtracting an average energy density computed on a per-event basis~\cite{PFT-10-002,Khachatryan:2015hwa}. For electrons and muons, we divide $\ETcone$ by the lepton $\pt$ to obtain the relative isolation $I_\text{rel} = \ETcone/\pt $, which is required to be less than 0.15. We require $\ETcone < 2$\GeV for $\tauh$ candidates.

The difference in the reconstruction efficiencies of muons and electrons between data and simulations is estimated with a standard technique that uses dilepton decays of \Z~bosons. Scale factors (SFs) are applied to simulation to match the data efficiencies and are \pt- and $\eta$-dependent.
The combined muon identification and isolation efficiency uncertainty is 11\% at muon $\pt$ of 10\GeV and 0.2\% at 100\GeV. The corresponding uncertainties for electrons are 14\% and 0.6\%.

We reconstruct jets from particle flow (PF) candidates using the anti-\kt algorithm~\cite{Cacciari:2008gp} with a distance parameter of 0.5.
Jets are required to have $\abs{\eta} < 2.5$ and $\pt > 30$\GeV and have $\Delta R > 0.3$ with respect to any isolated electron, muon, or $\tauh$ candidate.
The jet energy scale (resolution) is corrected using $\pt$- and $\eta$-dependent data-to-simulation scale (resolution) ratios~\cite{jets}. Jet four-momenta are varied using the uncertainty on these correction factors to account for the uncertainty in the jet energy scale measurement. We account for any additional discrepancy in $\MET$ between simulation and data~\cite{PFT-10-002} arising from PF candidates that are not clustered into jets and find that this discrepancy results in a negligible systematic uncertainty.

To determine if a jet originated from a bottom quark, we use the combined secondary-vertex (CSV) algorithm, which calculates a likelihood discriminant from the track impact parameter and secondary-vertex information~\cite{Chatrchyan:2012jua}.
A loose, medium, and tight discrimination selects \PQb~jets with average efficiencies of 85\%, 70\%, and 50\%, \cPqc~jets with average misidentification probabilities of 40\%, 20\%, and 5\%, and light-parton jets (\cPqu, \cPqd, \PQs, \Pg) with average misidentification probabilities of 10\%, 1.5\%, and 0.1\%, respectively.
Scale factors, depending on \pt and $\abs{\eta}$, are measured in data control samples of $\ttbar$ and
$\Pgm$+jets events and are used to correct the tagging efficiencies obtained from simulation.
A weight is applied to the response of the \PQb-tagging algorithm for each jet that is matched to a bottom quark.
A similar procedure is applied to model the mistag probability for jets originating from light partons and $\cPqc$~quarks.
The $\PQb$-, $\cPqc$-, and light-parton-tagging efficiencies are varied separately within their statistical uncertainties, and data-to-simulation SFs are applied and varied within the measured uncertainties~\cite{Chatrchyan:2012jua,btagging13}. The $\PQb$ and $\cPqc$ quark SFs are treated as correlated, and the light-parton SFs are treated as uncorrelated with the heavy-flavor SF.

\section{Fully hadronic final state}
\label{sec:hadronic}

Many signatures for physics beyond the standard model (BSM) result in long decay chains that produce high-multiplicity final states.
Most searches for SUSY involve either leptonic final states or missing transverse momentum, but fully hadronic final states that do not result in missing transverse momentum have been explored less thoroughly.
This section presents a search in a high-multiplicity, fully hadronic final state with no missing transverse momentum requirement.
The multiplicity of $\PQb$-tagged jets is used as a discriminating variable.

\begin{figure}[!t]
\centering
\includegraphics[width=0.45\textwidth]{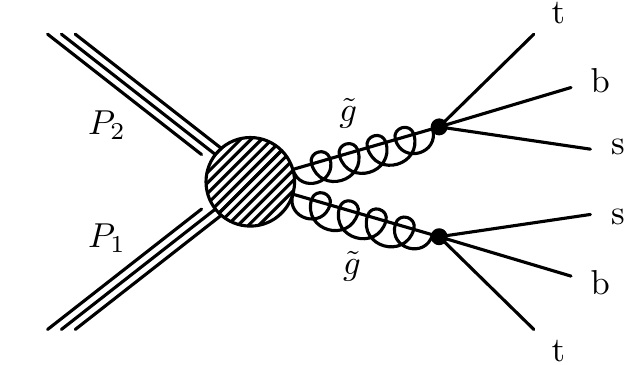}
\caption{Diagram for pair production of gluinos that decay to $\PQt\PQb\PQs$.
\label{fig:gluinodiagram}}
\end{figure}

Results are interpreted in a model in which pair-produced gluinos each decay via $\PSg\to \PQt\PQb\PQs$, which is allowed when $\lambda''_{332}\neq0$, so that a top antisquark couples directly to $\PQb$ and $\PQs$ quarks.
The top squark is assumed to be much heavier than the gluino, resulting in the three-body decay of the gluino shown in Fig.~\ref{fig:gluinodiagram}.
All supersymmetric particles other than the top squark and the gluino are assumed to be decoupled.
The top-squark mass and $\lambda''_{332}$ are assumed to take values such that the gluino decays promptly.
Because the coupling $\lambda''_{332}$ involves heavy quarks, it is relatively unconstrained by measurements of nucleon stability or neutrino masses~\cite{Barbier:2004ez}.

\subsection{Event selection}
\label{sec:EventSelection}

Events are selected by the trigger via a requirement on the sum of the \pt of the jets in the event that varied between 650 and 750\GeV over the course of data taking.

Substantial background suppression is achieved through the application of
multiplicity requirements on the jets reconstructed in the event, together
with \pt threshold requirements.
We require at least four jets with $\pt > 50\GeV$, where at least one jet must additionally satisfy $\pt > 150\GeV$.
All jets with $\pt > 50\GeV$ are used to calculate the offline $\HT$ of the event, which is required to be greater than 1.0\TeV.
With this selection, the trigger efficiency, measured with prescaled triggers that have lower $\HT$ requirements, is consistent with 100\%; separate studies with leptonic triggers have shown that there is no source of inefficiency common to all triggers with $\HT$ requirements.
The tight CSV requirement is used to identify jets arising from $\PQb$ quarks.
At least two such $\PQb$-tagged jets are required.

Events are required to have no isolated muons or electrons with $\pt>10$\GeV.
This requirement renders backgrounds due to feed-down from leptonic final states essentially negligible and ensures that this analysis is disjoint from the leptonic variant described in Section~\ref{sec:onelepton}.

The data are divided into bins of jet multiplicity \Njet and the scalar sum of the transverse momenta of the jets, $\HT$.  The $\HT$ bins are $1.0 < \HT < 1.75$\TeV and $\HT > 1.75$\TeV and the \Njet bins are 4, 5, 6, 7, and $\geq$8.  In each of these ten bins we fit the multiplicity of $\PQb$-tagged jets, $N_{\PQb}$.

\subsection{Standard model background}
\label{sec:Overview}
The dominant background in this analysis is from QCD multijet events (hereafter labeled as QCD), with contributions from $\ttbar$ becoming important at large values of $N_{\PQb}$.  Background sources other than multijet events are estimated directly from simulation.  These backgrounds include \ttbar production, hadronic decays of $\PW$ and $\Z$ bosons, single top quark production, $\Z\Z$ production, rare processes that include a \ttbar pair ($\ttbar\PW$, $\ttbar\Z$, $\ttbar\PH$, and $\ttbar\ttbar$), and leptonically decaying $\PW$ bosons in which the lepton is not reconstructed correctly or the lepton is a hadronically decaying $\tau$ lepton.

As the dominant background in this analysis arises from QCD multijet events, the modeling of this component is crucial.
We proceed by deriving corrections from data to the simulated QCD background to predict the distribution of the number of $\PQb$-tagged jets.
There are three main concerns: the modeling of the \Njet and $\HT$ distributions, the flavor composition of the QCD events, and the \PQb~quark production mechanisms.
The theoretical uncertainty on the $N_{\PQb}$ distribution arising from mismodeling of the \Njet and $\HT$ distributions is avoided by binning the sample in these two variables.
The modeling of the flavor composition is corrected to match the data (Section~\ref{sec:Flavor}).
The modeling of the \PQb~quark production mechanism is also validated with the data (Section~\ref{sec:GluonSplitting}).
With this procedure, we obtain an estimate of the QCD background from data in the variables $\Njet$, $\HT$, and $N_{\PQb}$.
Measurements of \ttbar+~jets events, which are the second-largest background to this analysis, have demonstrated that the jet multiplicity distribution is modeled within the uncertainties~\cite{Khachatryan:2015mva}.

\subsubsection{Flavor composition correction}
\label{sec:Flavor}

Although this analysis will determine the overall normalization of the QCD multijet background from data, an uncertainty arises from the poorly known flavor composition of this background.
To ensure that the simulated QCD events have the appropriate flavor composition, events are reweighted to match the flavor composition measured in data.
The coefficients used in the reweighting procedure are derived from a fit to the distribution of the CSV discriminant.
This fit is performed in a control region comprising events with only four or five reconstructed jets, to exclude a potential bias from signal contamination, and with the slightly tightened CSV discriminator requirement CSV $>$ 0.9 (compared with the nominal 0.898).
Additionally, to avoid bias due to the large weights arising from the low equivalent luminosity of the simulated QCD samples with $\HT < 1.0$\TeV, the $\HT$ requirement is increased slightly to $\HT>1.1$\TeV; it has been verified that the flavor composition corrections are statistically compatible for requirements of 1.0 or 1.1\TeV.

The fit of the distribution of the CSV discriminant is performed, including the statistical uncertainty in the MC prediction as nuisance parameters in the fit via the Barlow--Beeston method~\cite{ref:BarlowBeeston}.
The overall QCD contribution is normalized to the data yield, less the expected non-QCD yield (obtained from simulation).
Reconstructed jets are matched to the corresponding simulated jets, and templates for the CSV discriminant are formed for each flavor.
The relative normalization of templates corresponding to jets matched at the generator level to bottom and charm partons are allowed to vary in the fit.
The small contributions of non-QCD events (mainly $\ttbar$) and light-parton jets are fixed in the fit, with the uncertainty in the light fraction considered as a systematic uncertainty.

\begin{figure}[!t]
\centering
\includegraphics[width=0.45\textwidth]{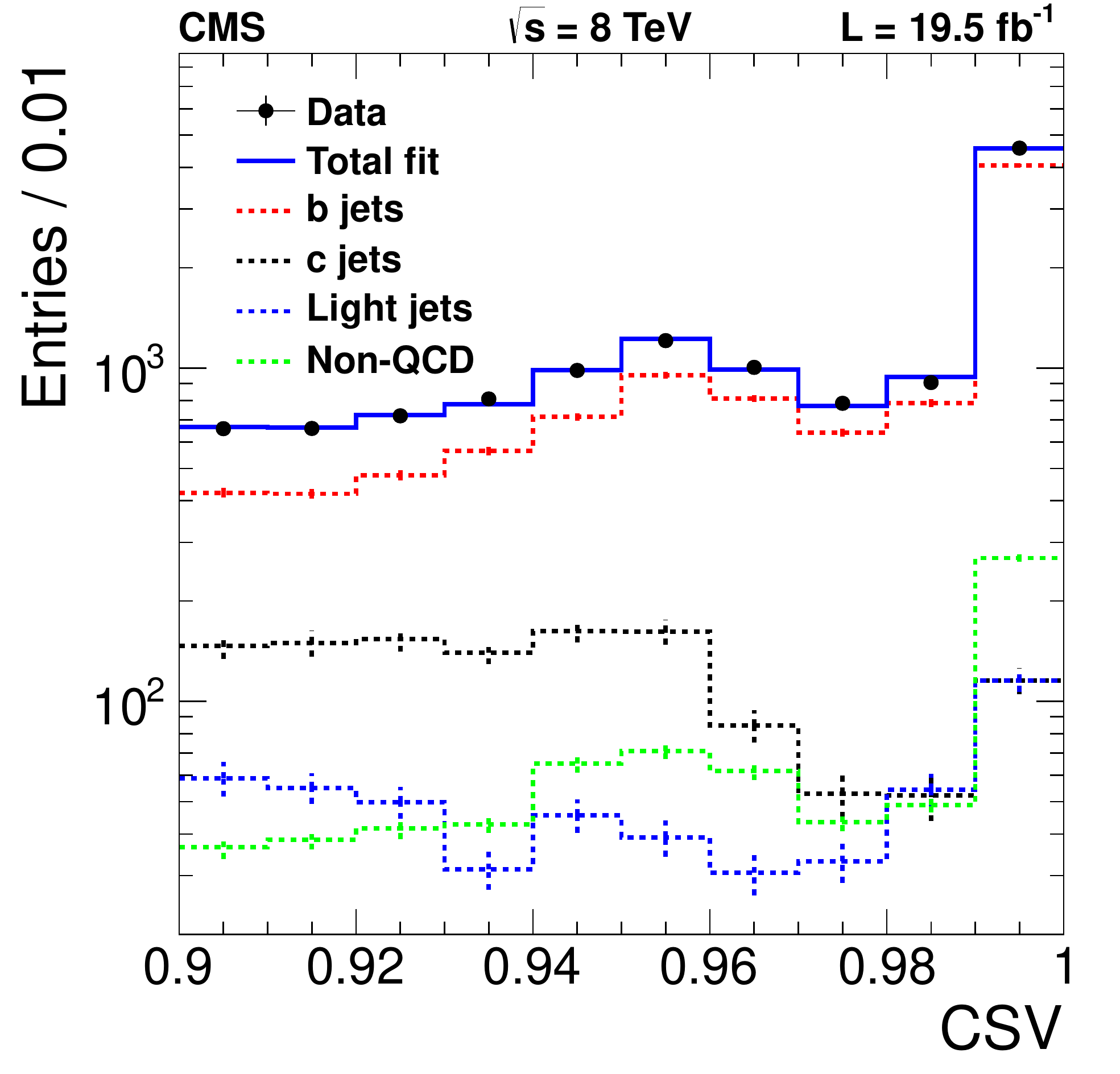}
\caption{The CSV distribution in data for $4\le \Njet\le 5$, $\HT>1.1$\TeV, and $\mathrm{CSV}>0.9$.
The solid line is the result of a fit to the data with MC templates.
Error bars reflect statistical uncertainties in the data (smaller than the marker) and MC samples.
\label{fig:flavorFit}}
\end{figure}

The fractions of bottom, charm, and light-parton jets prior to the fit are $f_{\PQb}$, $f_{\PQc}$, and $f_\text{light}$, respectively.
The fit provides new fractions $f'_{i}$ defined as
\begin{equation}
f'_{i}=\frac{n_{i}}{n_{\PQb}+n_{\PQc}+n_\text{light}+n_{\text{non-QCD}}}
\end{equation}
where $n_{\PQb}$ and $n_{\PQc}$ are the fitted yields of bottom and charm jets, respectively; $n_\text{light}$ and $n_{\text{non-QCD}}$ are the fixed yields of light-parton and non-QCD jets, respectively.
The index $i$ corresponds to $\PQb$, $\cPqc$, light-parton, and non-QCD events.
The values of $w_{i}^\text{\PQb jet}=f'_{i}/f_{i}$ are listed in Table~\ref{tab:flavorFit} (before reweighting) and the fit of the CSV distribution is shown in Fig.~\ref{fig:flavorFit}.
The fit quality is good, with $\chi^2/$d.o.f. = 7.0/8 (where d.o.f. is the number of degrees of freedom in the fit), providing confidence in the modeling of the CSV distribution.

For each simulated QCD multijet event, a weight is assigned based on the flavor fractions:
\begin{equation}
w_\text{event} = \prod_\text{\PQb jet} w^\text{\PQb jet}_{i},
\end{equation}
\noindent where $w_{i}^\text{\PQb jet}$ is a per-jet weight that is assigned to each $\PQb$-tagged jet and depends on the flavor $i$ of the parton matched to the reconstructed jet.
This form of the per-event reweighting is motivated by treating the corrections as independent corrections to the per-jet efficiency; an alternate reweighting procedure that reweights only $\bbbar$ pairs gives similar results.
The reweighting procedure has a small effect on the $N_{\PQb}$ distributions, with no $N_{\PQb}$ bin changing by more than 1.2 standard deviations due to the reweighting.

Although the fit models the data well, the good agreement between the model and the data could occur if mismodeled distributions accidentally have a linear combination that is consistent with the data.
To eliminate this possibility, fits are performed with variations of the fit range.
Even with an extreme variation in which the most sensitive region of the fit ($\mathrm{CSV}>0.98$) is removed, the fit results are still consistent with the nominal fit.
There is no evidence for any systematic effect.

As an additional cross-check, the fit is iterated: after the simulated events have been reweighted, the reweighted templates are fit to the data again.
The resulting heavy-flavor weights are consistent with unity, as shown in Table~\ref{tab:flavorFit}.

\begin{table}[htbp]
\centering
\topcaption{\label{tab:flavorFit}The \PQb~jet weights $w^\text{\PQb jet}_{i}$ derived from the fit of the $\Njet = 4$--5 control region before and after reweighting the QCD MC sample. The last column shows the result of the validation by iteration of the fit.}
\begin{scotch}{lccccc}
Flavor & Before reweighting & After reweighting \\
\hline
\PQb &  $0.94 \pm 0.03$ & $1.02 \pm 0.03 $\\
\PQc &  $2.00 \pm 0.43$ & $0.84 \pm 0.18$\\
Light & Fixed to 1.0 & Fixed to 1.0 \\
\end{scotch}
\end{table}

\begin{figure*}[!t]
\centering
\includegraphics[width=0.30\textwidth]{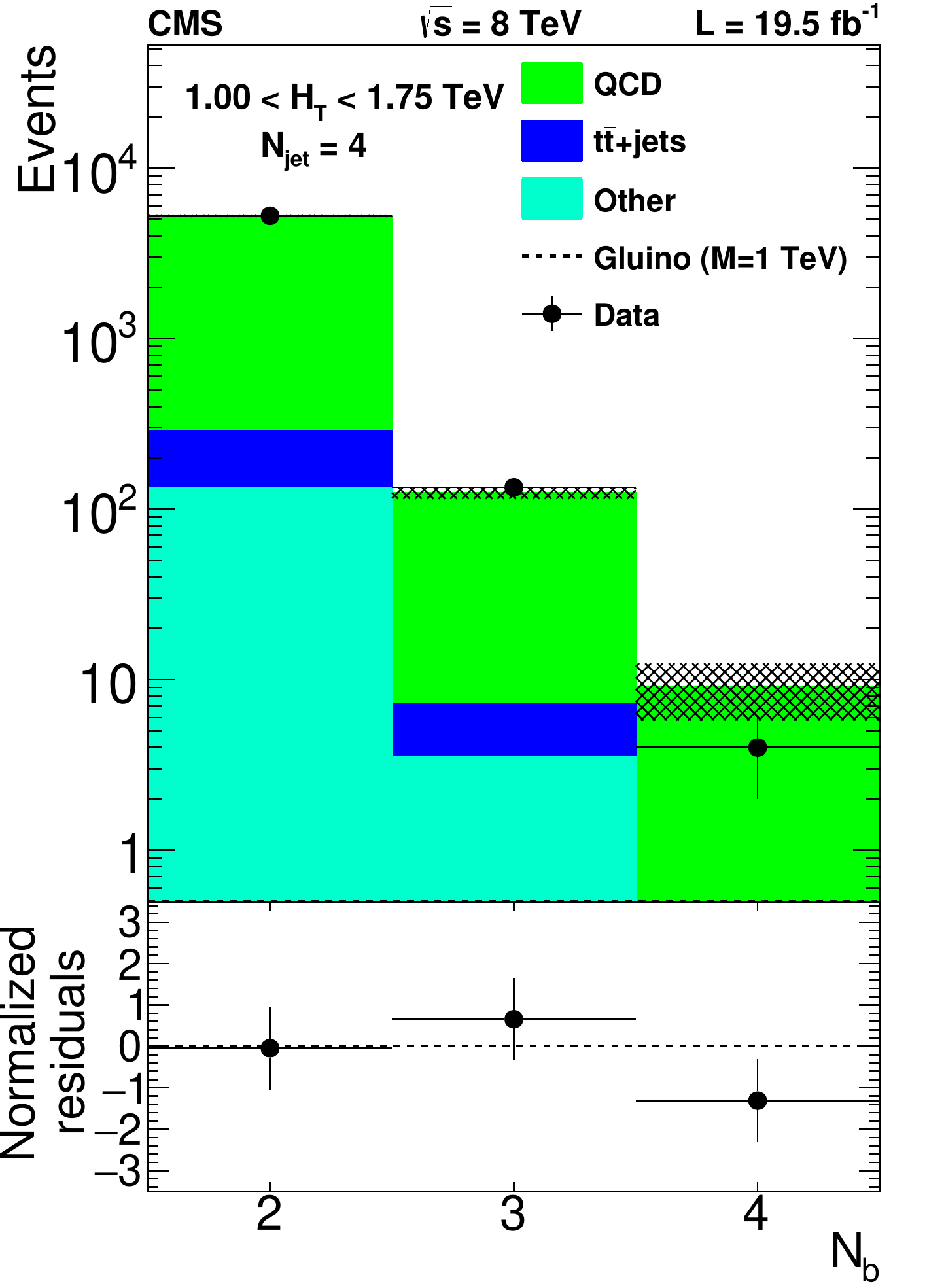}
\includegraphics[width=0.30\textwidth]{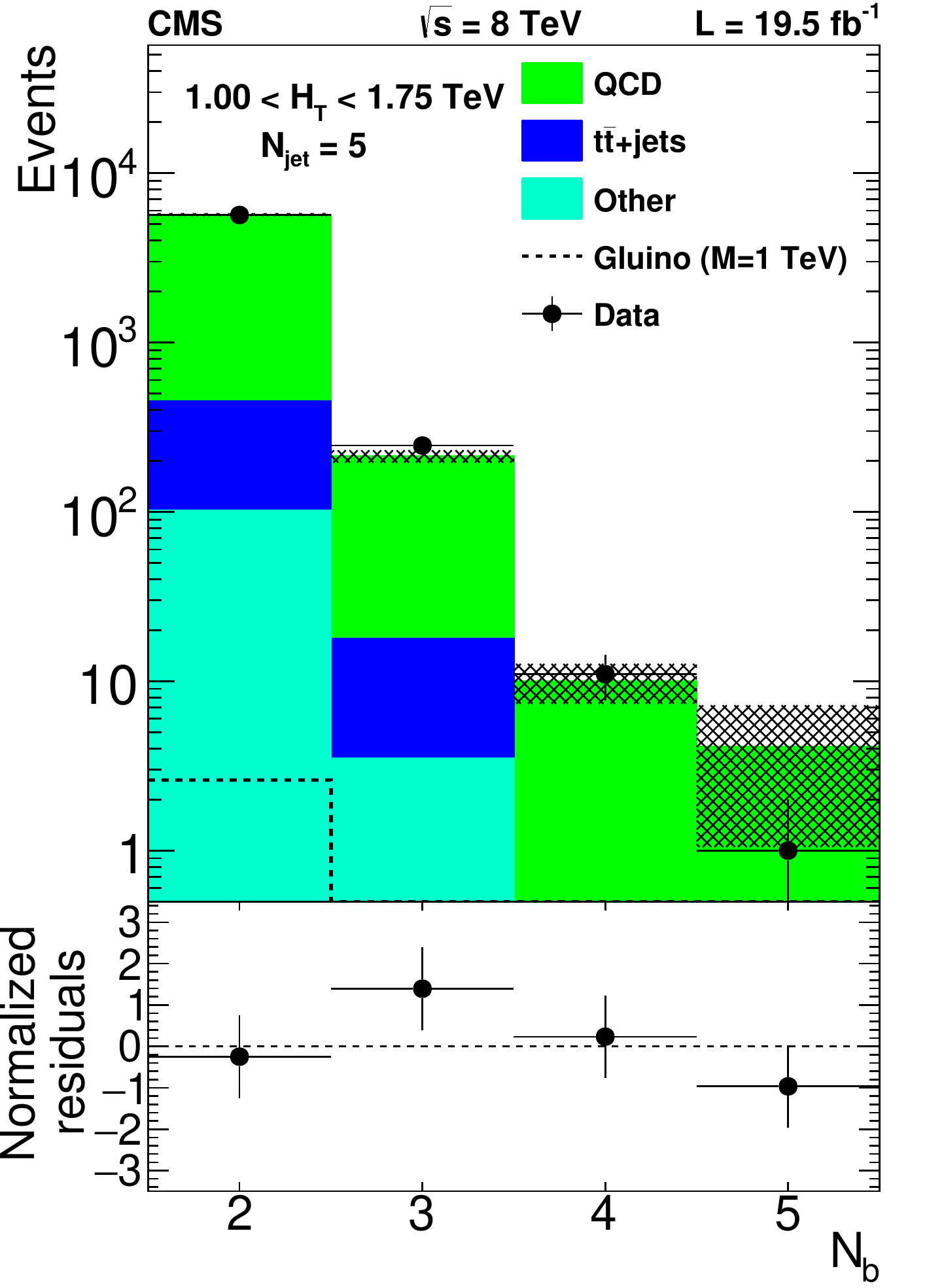}
\includegraphics[width=0.30\textwidth]{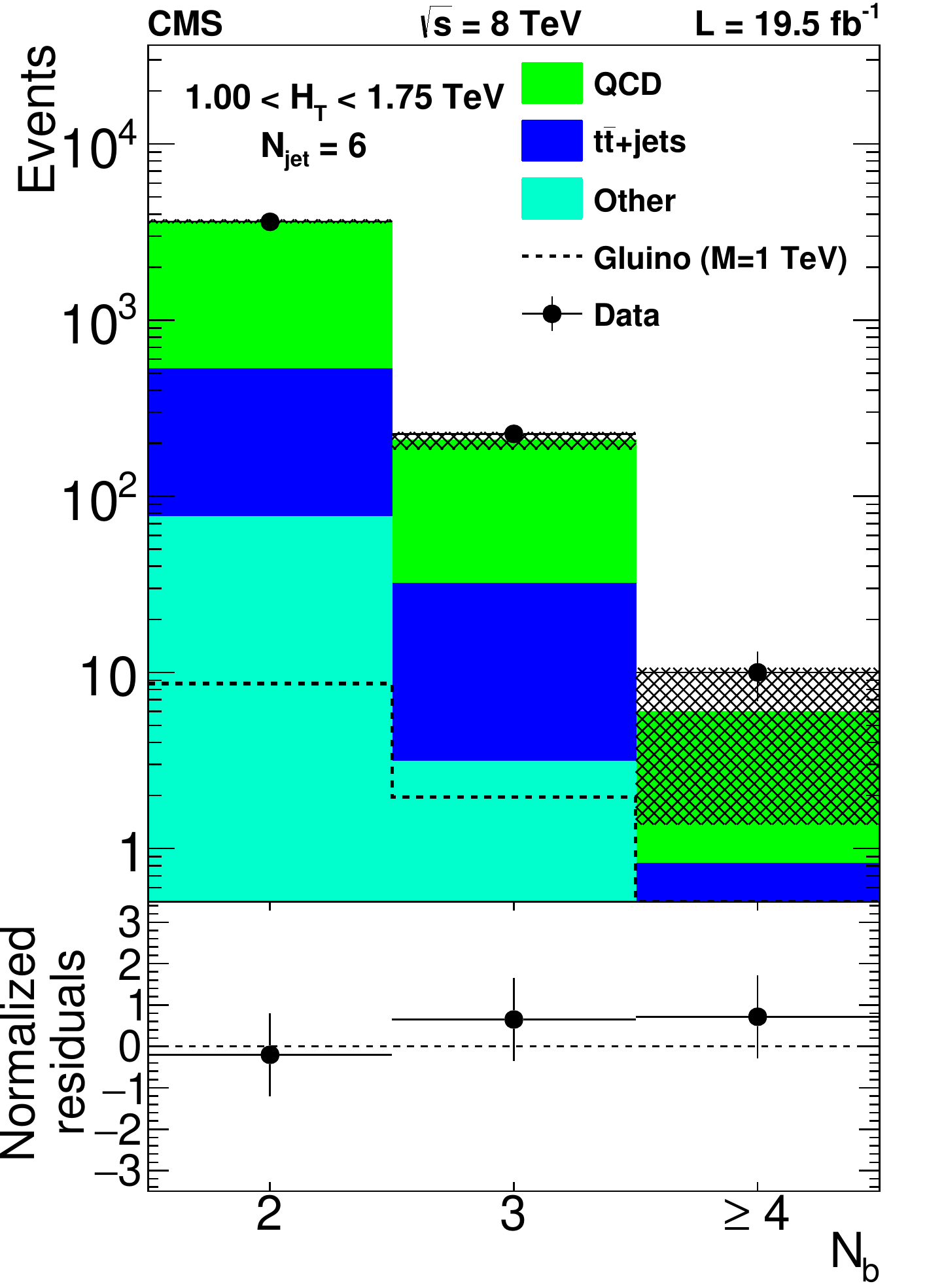}
\includegraphics[width=0.30\textwidth]{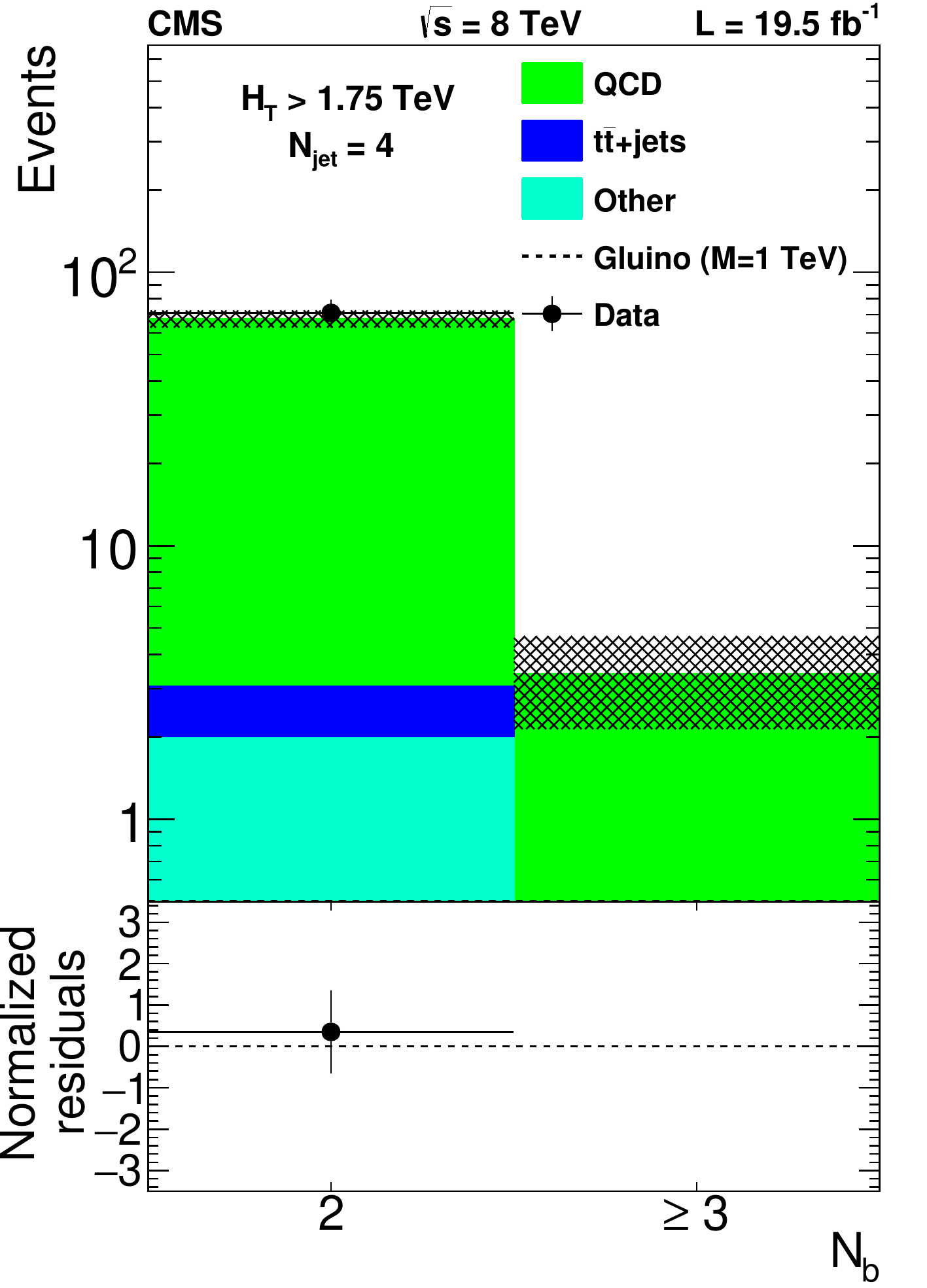}
\includegraphics[width=0.30\textwidth]{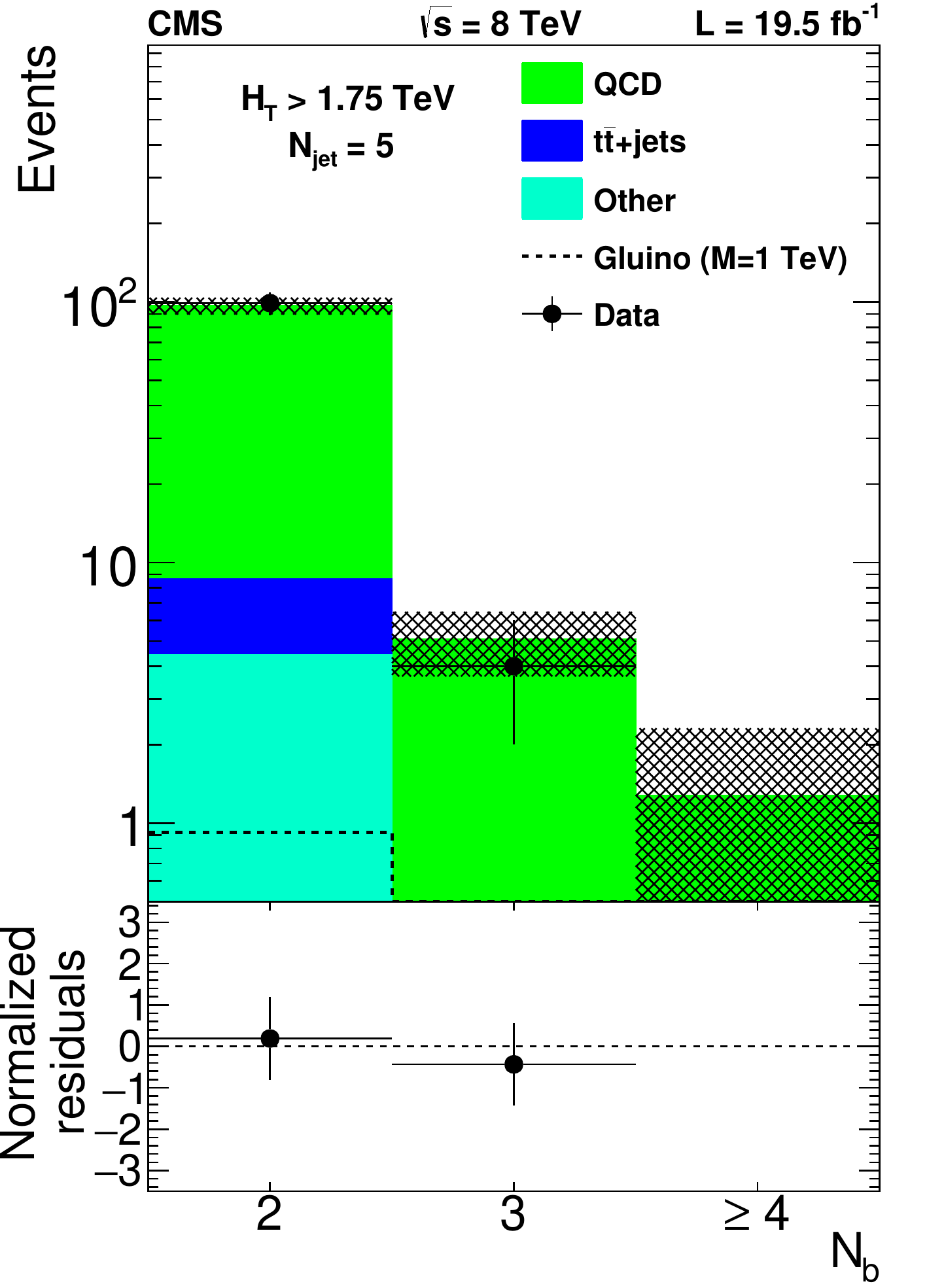}
\includegraphics[width=0.30\textwidth]{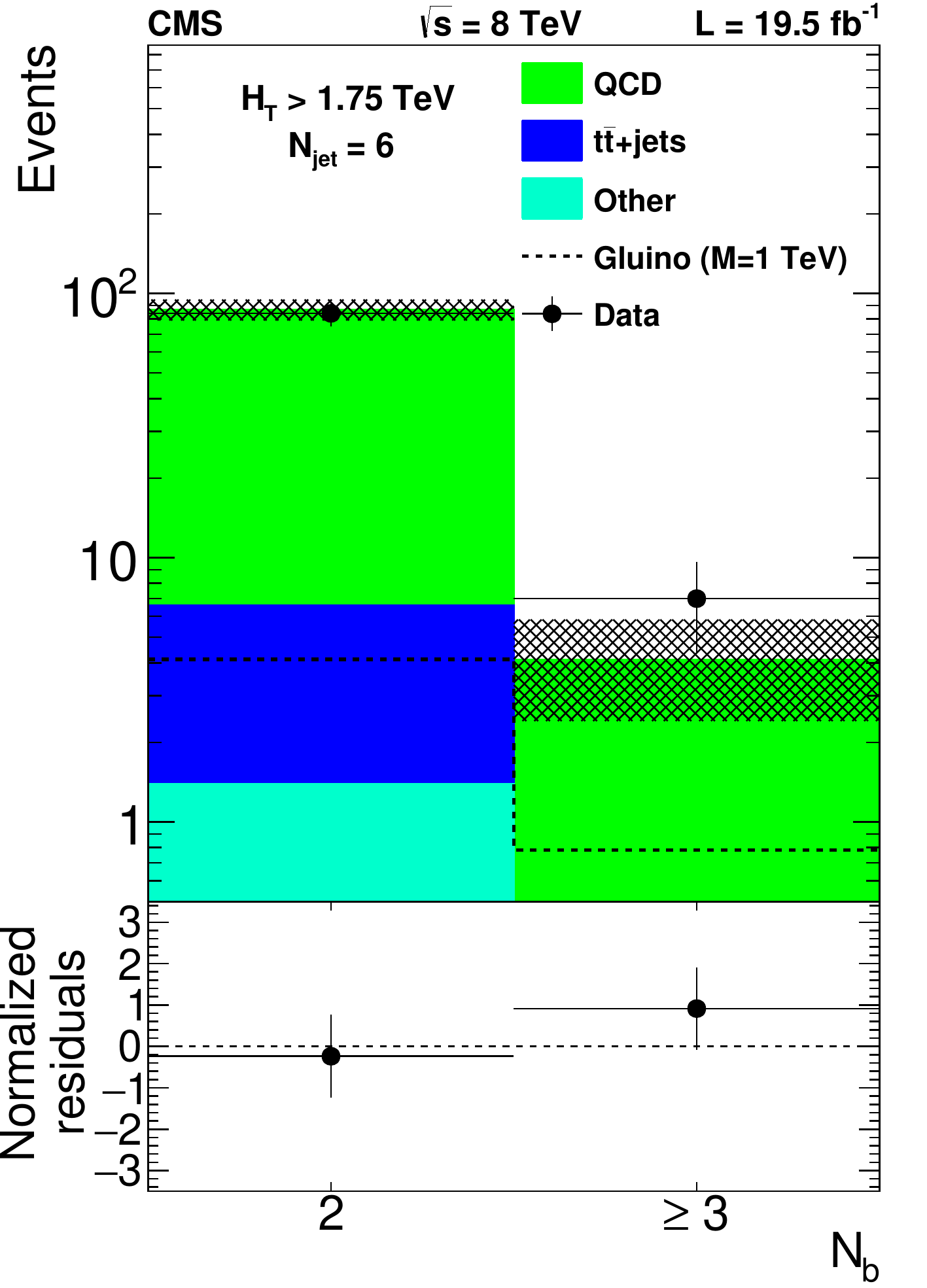}
\caption{Distribution of $N_{\PQb}$ for data~(dots with error bars) and corrected predictions.
The upper (lower) row shows data in which events are required to have $1.00<\HT<1.75$\TeV ($\HT>1.75$\TeV).
The jet multiplicity requirements are $\Njet=4$ (left), $\Njet=5$ (middle), and $\Njet=6$ (right).
The hatched bands shows the statistical uncertainty of the simulated data.
The bottom panels of each plot show the difference between the data and corrected prediction divided by the sum in quadrature of the statistical uncertainties associated with each.
\label{fig:controlRegion}}
\end{figure*}

It is important to demonstrate that the weights derived in this fit of the $\Njet = $~4--5 control region are applicable to the $\Njet\geq 6$ signal regions.
This has been verified in two ways.
First, the weights have been applied to the $\Njet = 6$ region, which has a negligible signal contribution.
The corrected predictions show good agreement with the data, as seen in Fig.~\ref{fig:controlRegion}.
Second, as the expected signal yield is extremely small compared to the background in the region defined by $\HT>1.1$\TeV, $N_{\PQb}\geq 2$, and $\Njet\geq 6$, the CSV distribution can be fit directly
in this region.
The reweighting parameters resulting from this fit are all within one standard deviation of those from the low-\Njet control region.
The fit of the $\Njet\geq 6$ region is not used in the reweighting procedure because of the larger statistical uncertainty, as well as because of the potential bias arising from signal contamination.

\subsubsection{Gluon splitting systematic uncertainty}
\label{sec:GluonSplitting}

In QCD multijet events, jets containing $\PQb$ quarks are produced in three different ways: pair production ($\Pq\Paq \to\bbbar$), flavor excitation ($\PQb\Paq\to\PQb\Paq$ and charge conjugate), and gluon splitting ($\Pg\to\bbbar$).
The first two processes are important primarily in the initial hard scatter, with the second suppressed owing to the small intrinsic $\PQb$-quark content of the proton.
Pair production is known to be well modeled by the \MADGRAPH generator, but the rate of gluon splitting is known to be off by up to a factor of two in the region of phase space dominated by the parton shower~\cite{ref:GluonSplitting}, necessitating an additional systematic uncertainty derived from data.

The systematic uncertainty is obtained by constructing an alternative template to be used in the signal extraction fit described in Section~\ref{sec:ControlSampleFit}. The alternative template for the QCD component differs from the nominal template by an additional $\Pg\to \bbbar$ component, which may have a negative normalization if the simulation overpredicts gluon splitting. The $N_{\PQb}$ distribution shape of this component is derived from $\Pg\to \bbbar$ simulated events.  Its normalization is obtained by comparing the $\Delta R_{\bbbar}$ distributions for data and simulation, where $\Delta R_{\bbbar}$ is the angular distance computed between any two $\PQb$-tagged jets in the event.  The simulated distribution, shown in Fig. 4, is normalized to data in the high-$\Delta R_{\bbbar}$ region, $\Delta R_{\bbbar} > 2.4$.  The difference between data and simulation in $\Delta R_{\bbbar}<1.6$ is assumed to arise entirely from $\Pg\to \bbbar$ events, and this difference provides the normalization for the (negative) correction that we add to the QCD template to obtain an estimate of the systematic uncertainty of the $\Pg \to \bbbar$ component.
The assumption that the difference arises entirely from gluon splitting to \bbbar provides a conservative estimate of the systematic uncertainty because, for example, events with gluon splitting to \cPqc\cPaqc~would generate a smaller difference in the distribution of the multiplicity of $\PQb$-tagged jets because of the smaller \PQb-tagging efficiency for charm jets.
The difference between data and simulation is determined separately for each $(\HT, \Njet)$ bin.
The constraint on the modeling of gluon splitting at low $\Delta R_{\bbbar}$ is used as an uncertainty rather than a correction because in the large-\Njet regions statistical fluctuations at low $\Delta R_{\bbbar}$ are larger than the size of the correction.

\begin{figure}[!t]
\centering
\includegraphics[width=0.45\textwidth]{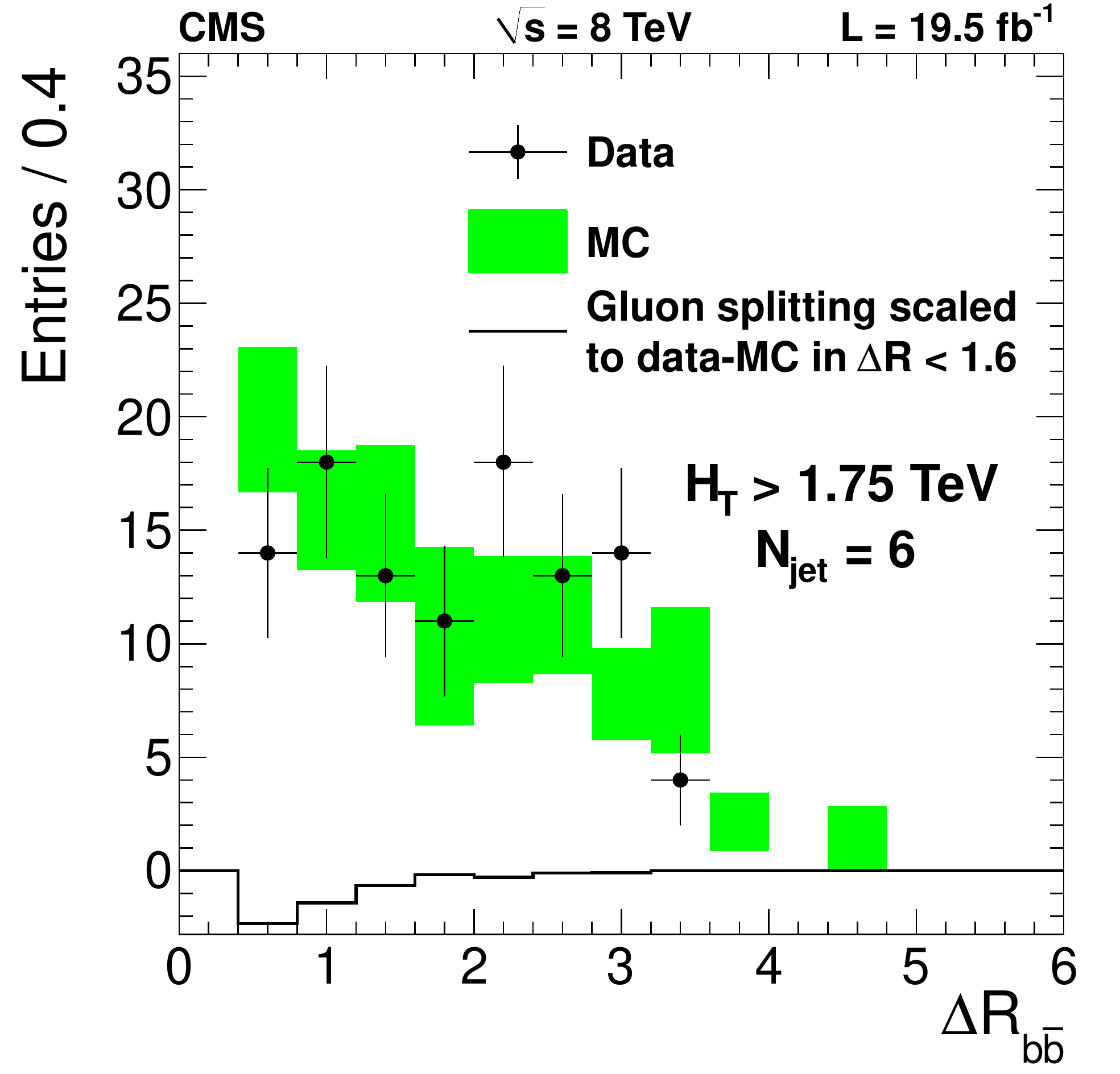}
\caption{Distribution of the angular distance between any two $\PQb$-tagged jets for data~(dots with error bars), uncorrected MC prediction (band), and generator-matched gluon splitting events scaled to the difference between data and simulation in $\Delta R_{\bbbar}<1.6$ (histogram) for events with $\HT>1.75$\TeV and $\Njet=6$.
The error bars and bands include the statistical uncertainty in the data and MC simulation, respectively.
There are no entries with $\Delta R_{\bbbar}<0.4$ as a consequence of the jet definition.
\label{fig:gluonSplittingHighHT}}
\end{figure}

\subsection{Systematic uncertainties}

The systematic uncertainties due to the modeling of the QCD background constitute an important part of the total uncertainty.
The uncertainty on the QCD flavor composition and gluon splitting are evaluated as described in Section~\ref{sec:Flavor} and Section~\ref{sec:GluonSplitting}, respectively.

As $\ttbar$ is a subdominant background, the effect of its uncertainties are generally small.
The tune of the underlying event, as well as variations of the renormalization, factorization, and matching scales are considered.
Also, the inclusive $\ttbar$ production cross section is varied according to its NNLO+(next-to-NLL) uncertainty~\cite{ref:ttbarXSec}.
The top quark \pt spectrum, which is not modeled well by simulation~\cite{Khachatryan:2015oqa}, is reweighted to agree with the data. The background contribution arising from $\ttbar\bbbar$~production is doubled to match the data~\cite{Khachatryan:2015mva}, with a 100\% uncertainty.

The cross sections of the remaining backgrounds are varied by 50\%.
The uncertainties in the jet \pt scale, the jet resolution, and the \PQb-tagging SFs for heavy-flavor and light-parton jets are evaluated as discussed in Section~\ref{sec:selections}.

The QCD MC simulation is affected by large statistical uncertainties, which are taken into account by variations in which single bins of each $N_{\PQb}$ histogram are varied according to their statistical uncertainty.
This is the largest systematic uncertainty in most $N_{\PQb}$ bins, and is about 40\% in the most sensitive signal region ($\HT>1.75$\TeV and $\Njet\geq 8$).
The systematic uncertainties are individually calculated for each ($\HT$, \Njet, $N_{\PQb}$) bin.  In general, the statistical uncertainty from data, the statistical uncertainty of the QCD MC simulation, and the sum of all other systematic uncertainties are of similar magnitude in each bin, ranging from 1\% to more than 50\% across the bins.

No uncertainty is assigned for trigger efficiency, which is consistent with 100\% with per mille uncertainties.

The signal samples are generated with a fast simulation.
The efficiency for tagging \PQb~jets and the mistag rate for charm and light-flavor quarks is corrected to match the efficiency predicted by full simulation.
Nuisance parameters parameterizing the uncertainty in these corrections for bottom jets, charm jets, and light-parton jets are considered separately and assumed to be mutually uncorrelated.

Most signal systematic uncertainties are modeled as modification of the templates of the $N_{\PQb}$ distribution.
The only exceptions are that of the luminosity uncertainty and the PDF uncertainty, which are modeled assuming a log-normal distribution for the corresponding nuisance parameter for each $(\Njet, \HT)$ bin.

\subsection{Control sample fit}
\label{sec:ControlSampleFit}

Signal-depleted control regions at low \Njet ($\Njet= 4$, 5, and 6) are studied before examining the signal region.
For low jet multiplicities, $\ttbar$ backgrounds are less important, giving a largely pure sample of QCD events.

A binned maximum likelihood fit of the $N_{\PQb}$ distributions is performed in which systematic uncertainties are profiled.
Systematic uncertainties are included as shape uncertainties by interpolating between $N_{\PQb}$ histograms corresponding to $\pm 1$~standard deviation variations.
As the $\HT$ and \Njet dependence of the QCD contribution may not be modeled well, a separate normalization of the QCD contributions is allowed in each bin of $\HT$ and \Njet.
The likelihood used in the fit of the yields $N_{ijk}$ in the $N_{\PQb}$ distributions of the signal and control regions is
\ifthenelse{\boolean{cms@external}}{
\begin{multline}
\label{eq:likelihood}
L=\prod_{\substack{i\in {\HT~\text{bins}}\\j \in {\Njet~\text{bins}}\\k \in {N_{\PQb}~\text{bins}}\\n\in \text{syst}}} P(N_{ijk}|\theta_n)\text{Poisson}\Bigl(N_{ijk}|\mu_\text{signal}\nu_{ijk, \text{signal}}\\
   \quad+\mu_{ij,\mathrm{QCD}}\nu_{ijk, \mathrm{QCD}}+\nu_{ijk, \text{other}}\Bigr).
\end{multline}
}{
\begin{equation}
\label{eq:likelihood}
L=\prod_{\substack{i\in {\HT~\text{bins}}\\j \in {\Njet~\text{bins}}\\k \in {N_{\PQb}~\text{bins}}\\n\in \text{syst}}} P(N_{ijk}|\theta_n)\text{Poisson}\Bigl(N_{ijk}|\mu_\text{signal}\nu_{ijk, \text{signal}}
   \quad+\mu_{ij,\mathrm{QCD}}\nu_{ijk, \mathrm{QCD}}+\nu_{ijk, \text{other}}\Bigr).
\end{equation}
}
Here $\mu_\text{signal}$ and $\mu_{ij,\mathrm{QCD}}$ are normalization constants.
The parameters $\mu_\text{signal}$ and $\mu_{ij,\mathrm{QCD}}$ do not have a dependence on $k$ because the $N_{\PQb}$ input distribution is fixed for a given $\HT$ and $\Njet$ bin.
The yields of signal, QCD background, and non-QCD background are relative to the nominal values specified by $\nu_{ijk,\text{signal}}$, $\nu_{ijk,\mathrm{QCD}}$, and $\nu_{ijk,\text{other}}$, respectively.
In other words, there is a floating QCD normalization in each $(\HT,\Njet)$ bin and fixed non-QCD background yields.
The systematic uncertainties are included with nuisance parameters $\theta_n$ that can affect the interpolation between the $\pm$1~standard deviation templates; the parameters $\nu_{ijk,\text{signal}}$, $\nu_{ijk,\mathrm{QCD}}$, and $\nu_{ijk,\text{other}}$ are dependent on these nuisance parameters.
These parameters are the same for all $(\HT,\Njet)$ bins except for those associated with MC statistics, which have separate parameters for each $(\HT,\Njet,N_{\PQb})$ bin.

In the control sample fit, the product over \Njet bins is restricted to $\Njet= 4,$ 5, 6, and the signal yields are fixed to zero.
The data and MC simulation inputs to this fit are shown in Fig.~\ref{fig:controlRegion}.

All of the nuisance parameters are consistent within one standard deviation of their prefit uncertainties, except for a 1.4 standard deviation difference in the light jet fraction nuisance parameter, which is however a subdominant uncertainty in the high-\Njet signal region.

\subsection{Results for fully hadronic final state}
\label{sec:Results}

\begin{figure*}[t!]
\centering
\includegraphics[width=0.45\textwidth]{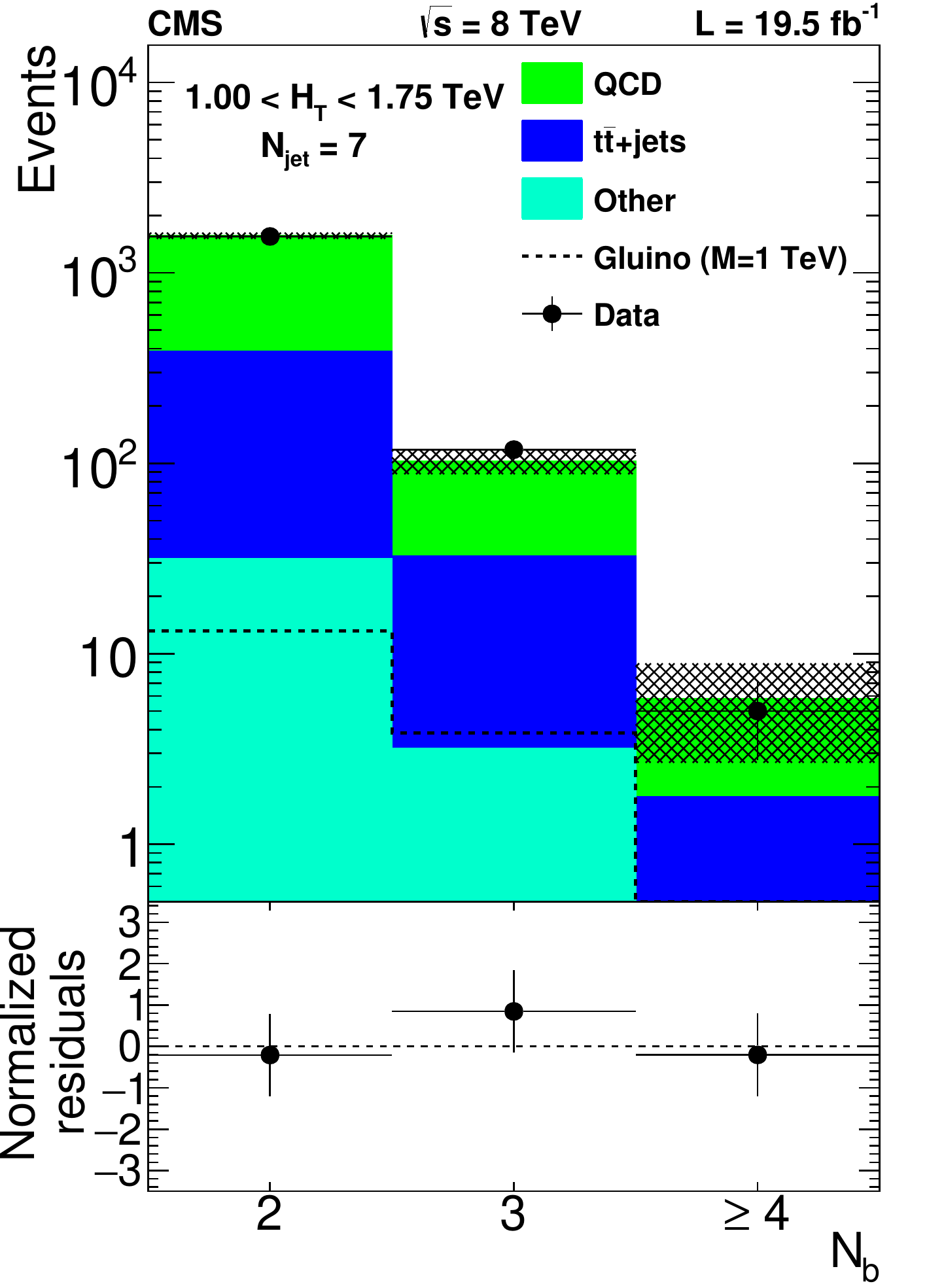}
\includegraphics[width=0.45\textwidth]{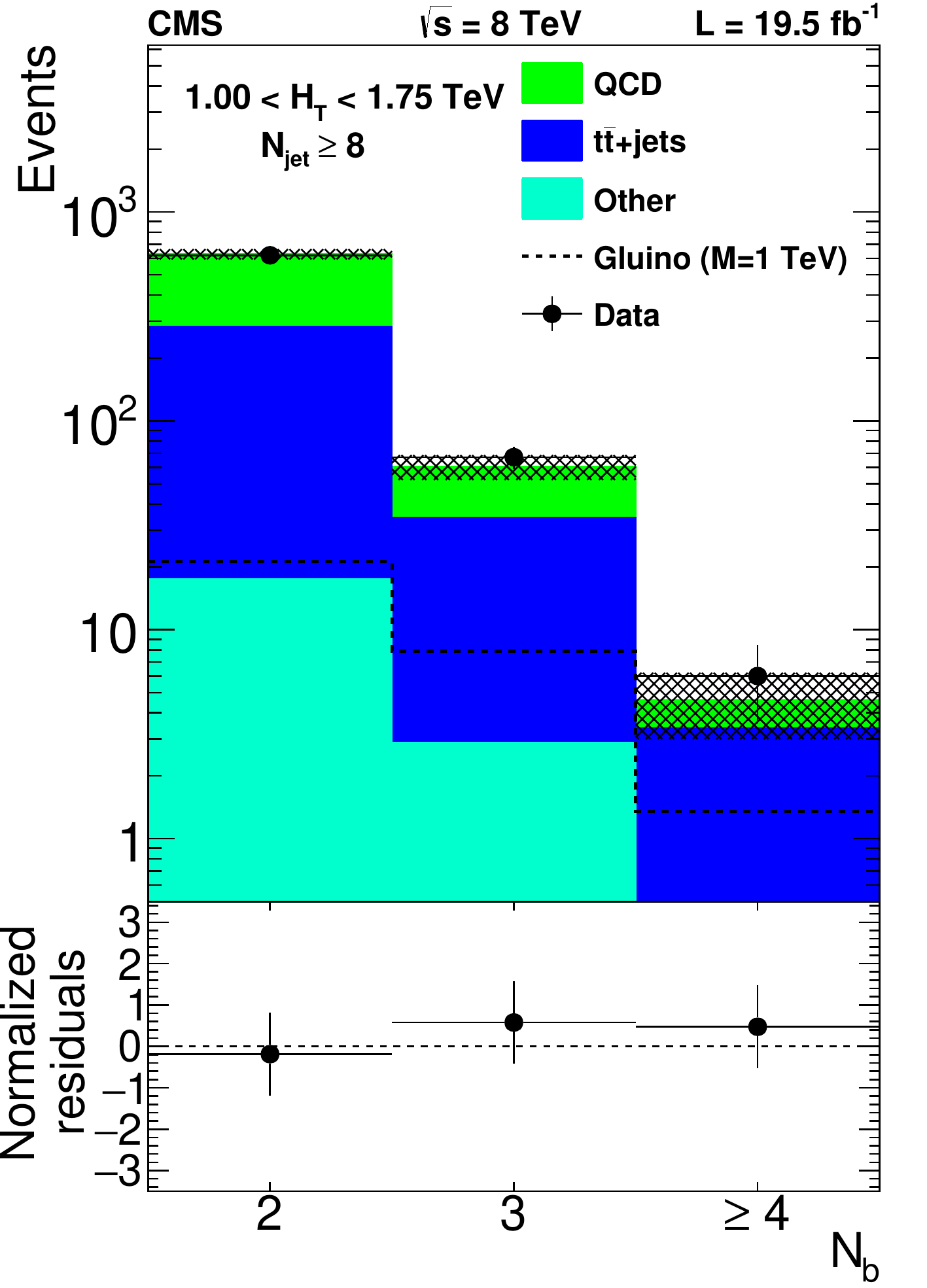}
\includegraphics[width=0.45\textwidth]{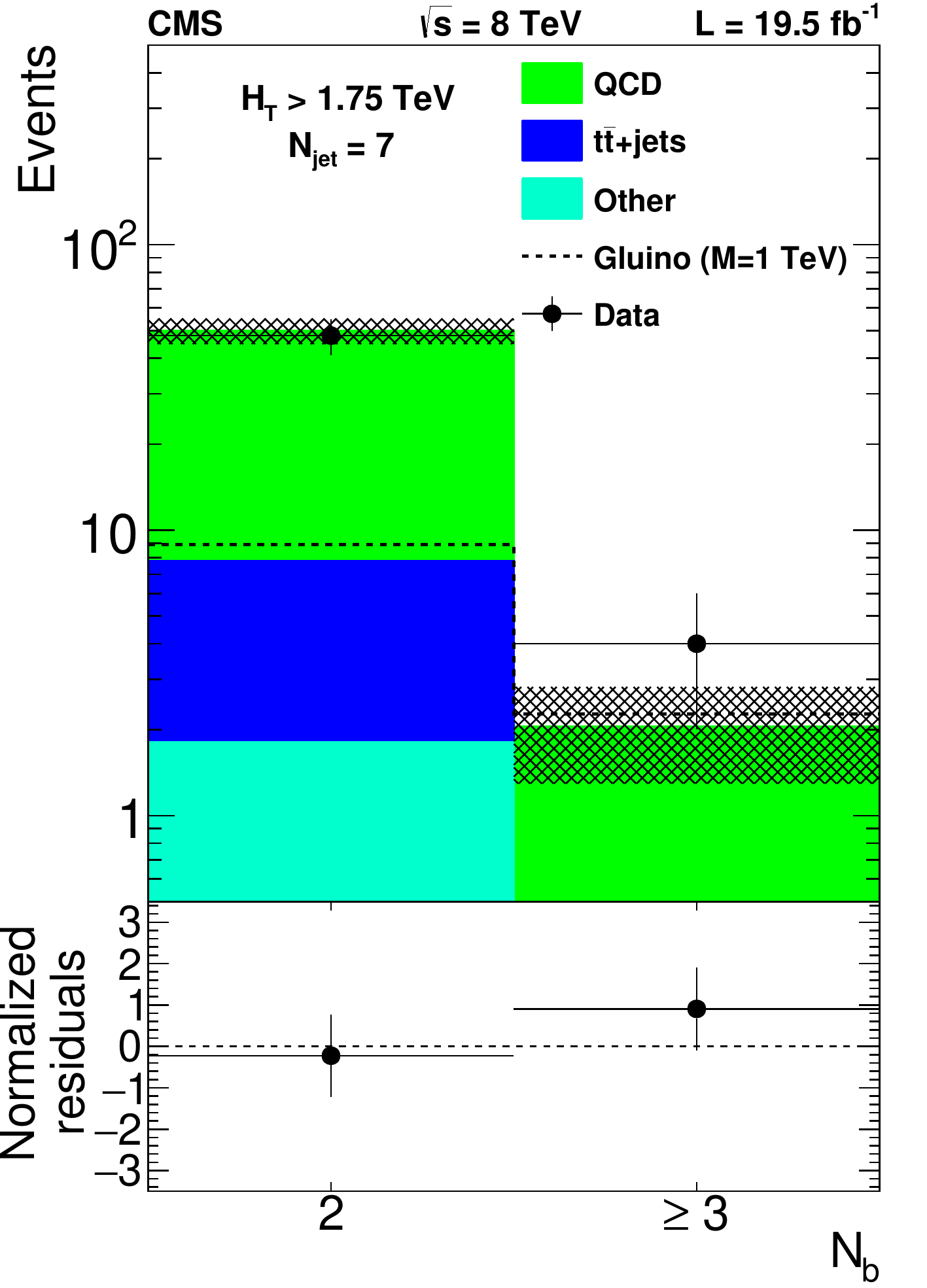}
\includegraphics[width=0.45\textwidth]{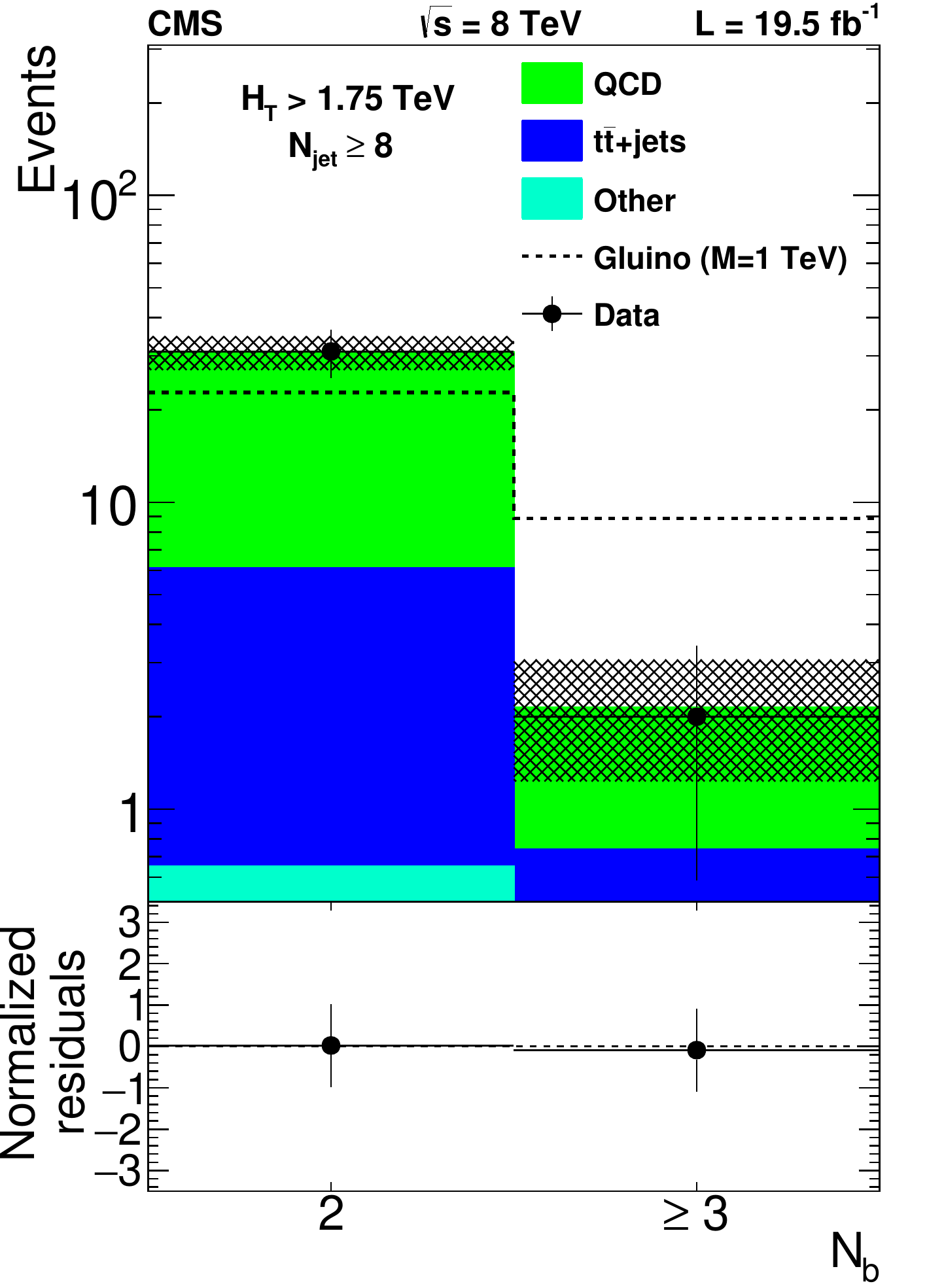}
\caption{Data~(dots with error bars) and the corrected prediction of the $N_{\PQb}$ distribution in the high-\Njet signal region.
The hatched band shows the MC statistical uncertainty.
The upper (lower) row shows events with $1.00<\HT<1.75$\TeV ($\HT>1.75$\TeV).
The jet multiplicity requirements are $\Njet=7$ (left) and $\Njet\geq 8$ (right).
The bottom panels of each plot show the difference between the data and corrected prediction divided by the sum in quadrature of the statistical uncertainties associated with each.
\label{fig:signalRegion}}
\end{figure*}

The likelihood used in the fit of the signal region is that given by Eq.~(\ref{eq:likelihood}), with $\Njet=6,$ 7, $\geq$8, and $\mu_\text{signal}$ left free.
Figure~\ref{fig:signalRegion} shows a comparison of the data with the corrected simulation, where the QCD component has been scaled to the data yield minus the non-QCD background yields obtained from simulation. The yields corresponding to the $\Njet \geq 8$ and $\HT > 1.75$\TeV region in Fig.~\ref{fig:signalRegion} are shown in Table~\ref{tab:yields_hadronic}.

\begin{table}[!htb]
\centering
\topcaption{\label{tab:yields_hadronic}Summary of prefit expected background, expected signal for $m_{\PSg} = 1$\TeV, and observed yields for  $\Njet \geq 8$ and $\HT>1.75$\TeV. Uncertainties are statistical only.}\begin{scotch}{l r r}
Background & \multicolumn{1}{c}{$N_{\PQb}=2$} & \multicolumn{1}{c}{$N_{\PQb}\geq 3$}\\
\hline
QCD multijet & 24.7 $\pm$ 3.8  &1.4 $\pm$ 0.9   \\
$\ttbar$+jets & 5.5 $\pm$ 0.6  &0.7 $\pm$ 0.2   \\
Other & 0.6 $\pm$ 0.4 & \multicolumn{1}{c}{$<$ 0.1} \\
\hline
Total background & 30.9 $\pm$ 3.9  &2.2 $\pm$ 0.9   \\
\hline
Data & \multicolumn{1}{c}{31}  & \multicolumn{1}{c}{2}    \\
\hline
Signal ($m_{\PSg}=1.0$\TeV) & 22.8 $\pm$ 0.3  &8.9 $\pm$ 0.2   \\
\end{scotch}
\end{table}

\begin{figure}[t!]
\centering
\includegraphics[width=0.45\textwidth]{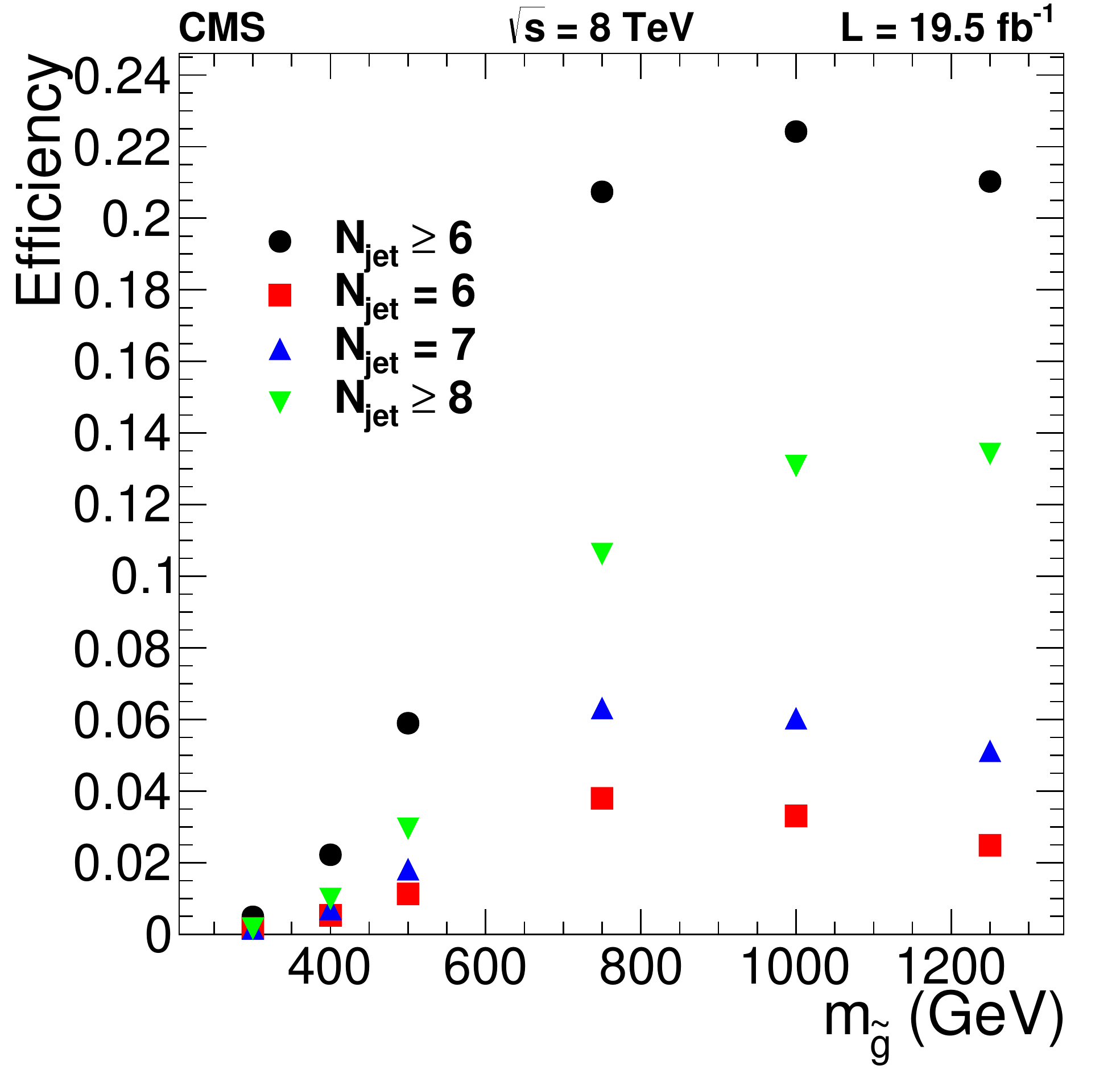}
\caption{
Signal efficiencies as a function of $m_{\PSg}$ for $N_{\PQb}\geq2$, $\HT>1$\TeV, and $\Njet\geq6$, together with the breakdown among \Njet bins.
\label{fig:efficiency}}
\end{figure}

\begin{figure}[t!]
\centering
\includegraphics[width=0.45\textwidth]{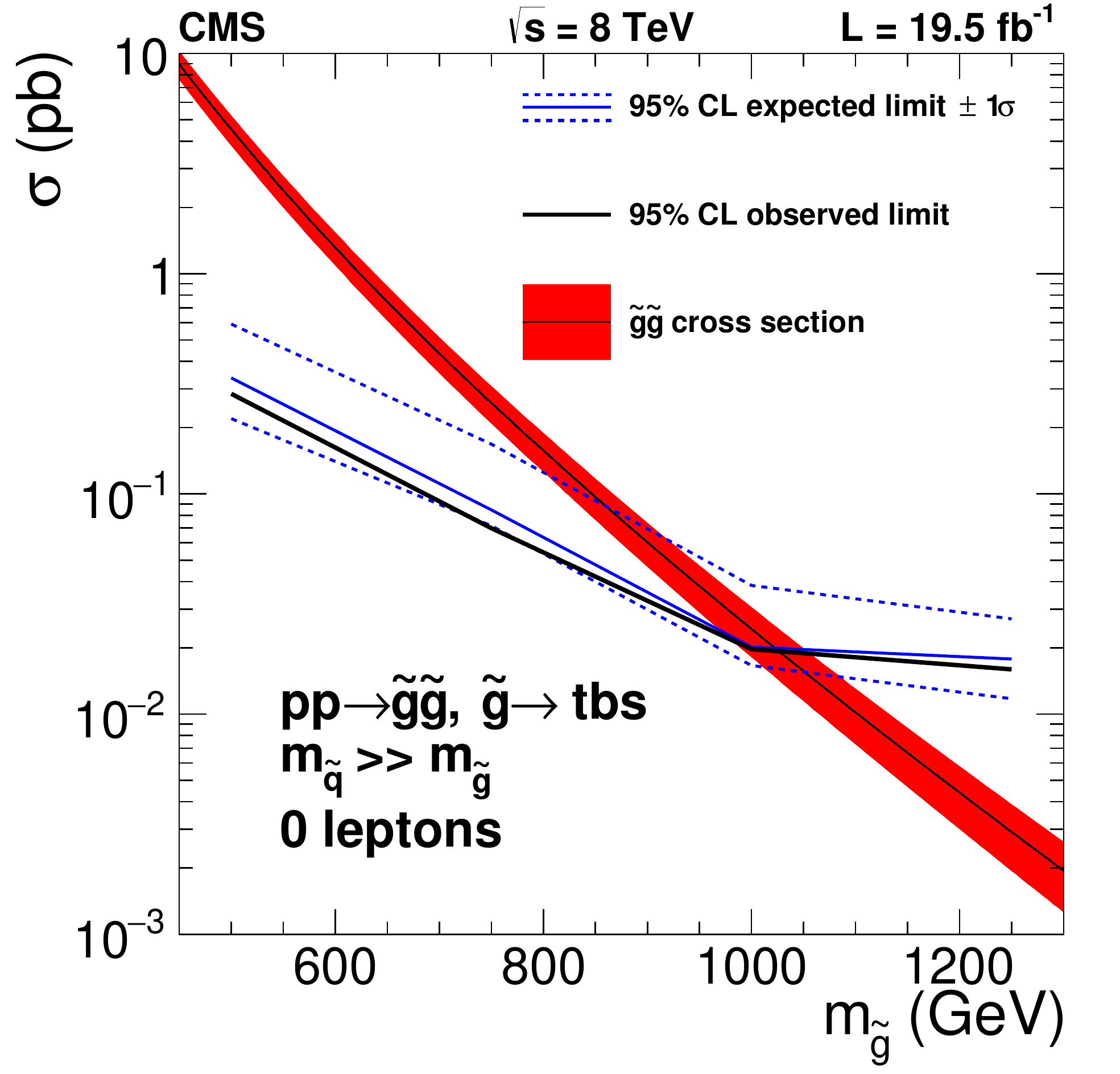}
\caption{
The $95\%$ \CL limit on the gluino pair production cross section as a function of $m_{\PSg}$ in the analysis of all-hadronic final states. The signal
    considered is $\Pp\Pp \to \PSg\PSg$, followed by the decay
    $\PSg\to\PQt\PQb\PQs$. The red band shows the theoretical cross section and its uncertainty. The blue dashed lines show the uncertainty on the expected limit.
\label{fig:limit}}
\end{figure}

At each gluino mass, the best fit returns zero signal events.
The efficiency after applying all the selection criteria is shown in Fig.~\ref{fig:efficiency} and reaches a plateau of around 20\% for $m_{\PSg}>0.7$\TeV.
Figure~\ref{fig:limit} shows the expected and observed limits compared to the gluino pair production cross section.

To summarize the fully hadronic search, the data in the signal regions are well described by the background predictions.
The results are interpreted in terms of a specific model of RPV SUSY in which gluinos are pair produced and each gluino decays promptly via $\PSg\to\PQt\PQb\PQs$.
Cross section limits are calculated and result in a 95\%\CL lower limit on the gluino mass of $0.98$\TeV within this simplified model.

\section{Single-lepton final state}
\label{sec:onelepton}

This section describes a search for $\PSg \to \PQt\PQb\PQs$ events in which the top quark undergoes semileptonic decay. The strategy is similar to that of the previous section but with a requirement of at least one isolated charged lepton, which rejects most of the QCD background.

We select final states with one lepton and multiple jets. Then we use the number of $\PQb$-tagged jets as a discriminating variable to separate the signal from the background.

\subsection{Event selection}\label{sec:sel}
The analysis considers events selected by a trigger requiring an electron with $\pt>27$\GeV and $\abs{\eta}<2.5$ or a trigger requiring a muon with $\pt>24$\GeV and $\abs{\eta}<2.1$.
Both triggers include loose isolation requirements.
The offline selection raises the $\pt$ threshold to $\pt> 35$\GeV for both muons and electrons while the $\abs{\eta}$ selection remains the same.

We measure the trigger efficiency with respect to the offline selection in bins of $\eta$ and $\pt$ of the lepton, and find it to vary from approximately $92\%$ for $\abs{\eta}<0.8$ to $65\%$ for $1.2<\abs{\eta}\leq 2.4$ in the case of muons. Electron efficiencies are within a few percent of those quoted for muons.
For both lepton flavors, the variation of the efficiency over $\pt$ in a given $\eta$ range is 1--2\% for $\pt > 35$\GeV.

The baseline selection requires at least one lepton, and six jets with $\pt > 30$\GeV.
The medium working point of the CSV \PQb~tagging discriminator is used and at least one jet must pass this selection.
Events with two identified leptons are allowed in the sample, but to avoid double counting we veto events in the electron sample if they also contain an identified muon.

The electron and muon samples are distinguished to allow cross-checks; however most systematic effects are correlated between the two samples and this is considered when fitting the $\PQb$-tagged jet multiplicity distribution to extract the signal.

\subsection{Standard model background}\label{sec:bkg}

The selected events are divided into three signal regions, according to their
jet multiplicity: 6, 7, and $\geq$8~jets. For each signal region, the
signal and background yields are obtained by comparing the observed
multiplicity distribution of \PQb-tagged jets with their respective background shapes.

The main source of background to this search is the
production of pairs of top quarks in association with jets. Additional contributions from single top quark, vector bosons, and QCD multijet production are relevant for low \PQb~jet multiplicities, becoming negligible
for events with at least three \PQb~jets. Events with at least three \PQb~jets also have a small
contamination, typically below 1\%, from $\ttbar V$ events (where $V=\PW$ or $\Z$). The background from
SM $\ttbar\ttbar$ production and $\ttbar\PH$ production is negligible,  due to the small cross section for these processes.

We study the \PQb~jet multiplicity of the background sources using
MC simulation.
Events are corrected for the different response of the \PQb~tagging algorithm in
simulation and data as described in Section~\ref{sec:selections}.
We verified that the \PQb~tagging efficiency and the mistag rate vary negligibly in semileptonic $\ttbar$ events going from four to nine jets, always remaining within their uncertainty. Furthermore, we compared data and simulation in control regions with one or two leptons and four or five jets and found good agreement within the uncertainties in the \PQb-tagging SFs.
We account for residual small discrepancies by allowing for an MC mismodeling of SM events with four $\PQb$ quarks, as discussed in more detail in Section~\ref{sec:sys}.

The corrected \PQb~jet multiplicity provides the prediction for the SM background, which is compared with data in order
to check for the presence of a signal. The final signal extraction fit obtains the background normalization from data using an extended likelihood. Therefore only the shape of the multiplicity distribution of \PQb~tagged jets is taken from simulation. This considerably reduces the systematic uncertainty in the background, since this shape is very weakly dependent on the jet energy scale and on the choices of matching and renormalization scales.

\subsection{Systematic uncertainties}

\subsubsection{Background}\label{sec:sys}

The background shape is affected by the jet energy scale uncertainty; the uncertainty in the \PQb~tagging efficiency SFs; the variation of renormalization, factorization, and matching scales; and the MC statistical uncertainty. Furthermore, we include a systematic effect parameterizing the mismodeling of the fraction of events with four bottom quarks.

We evaluate the effect of jet energy, matching, renormalization, and factorization scales by repeating the selection procedure on MC simulated samples with the scales shifted up or down. The jet energy scale is varied as described in Section~\ref{sec:selections}.
The matching, renormalization, and factorization scales are fixed to factors of 2.0 and 0.5 with respect to the nominal scale, for positive and negative variations, respectively.
The renormalization and factorization scales are varied simultaneously. The uncertainty from the \PQb~tagging SF is computed by comparing the $\PQb$-tagged jet multiplicity distributions obtained by correcting the tagging efficiencies with SFs shifted by $\pm$1 standard deviation. The uncertainties in the \PQb~jet and \cPqc~jet SFs are taken to be correlated.

\begin{figure*}[htb]
  \centering%
  \includegraphics[width=0.48\textwidth]{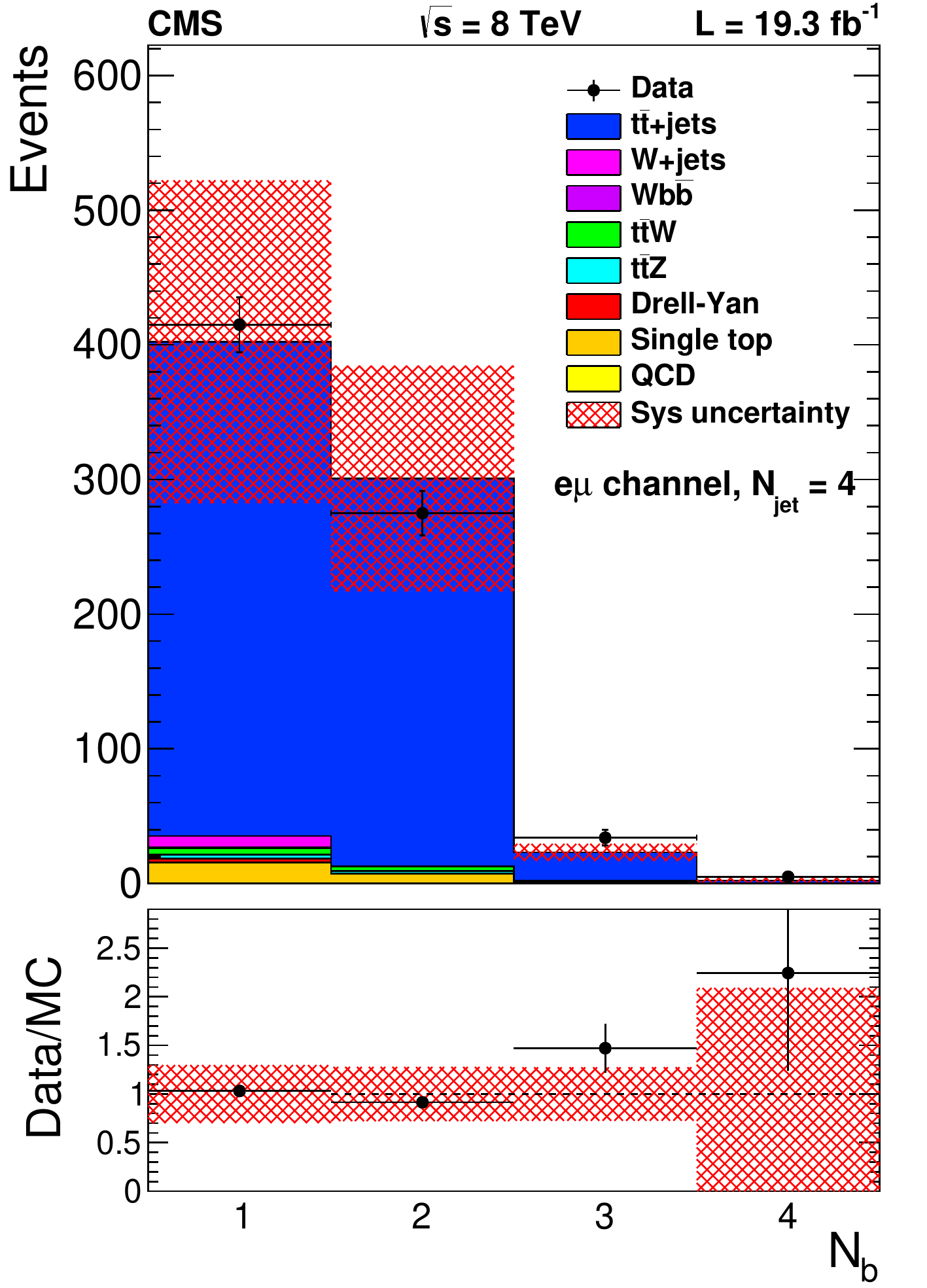}
  \includegraphics[width=0.48\textwidth]{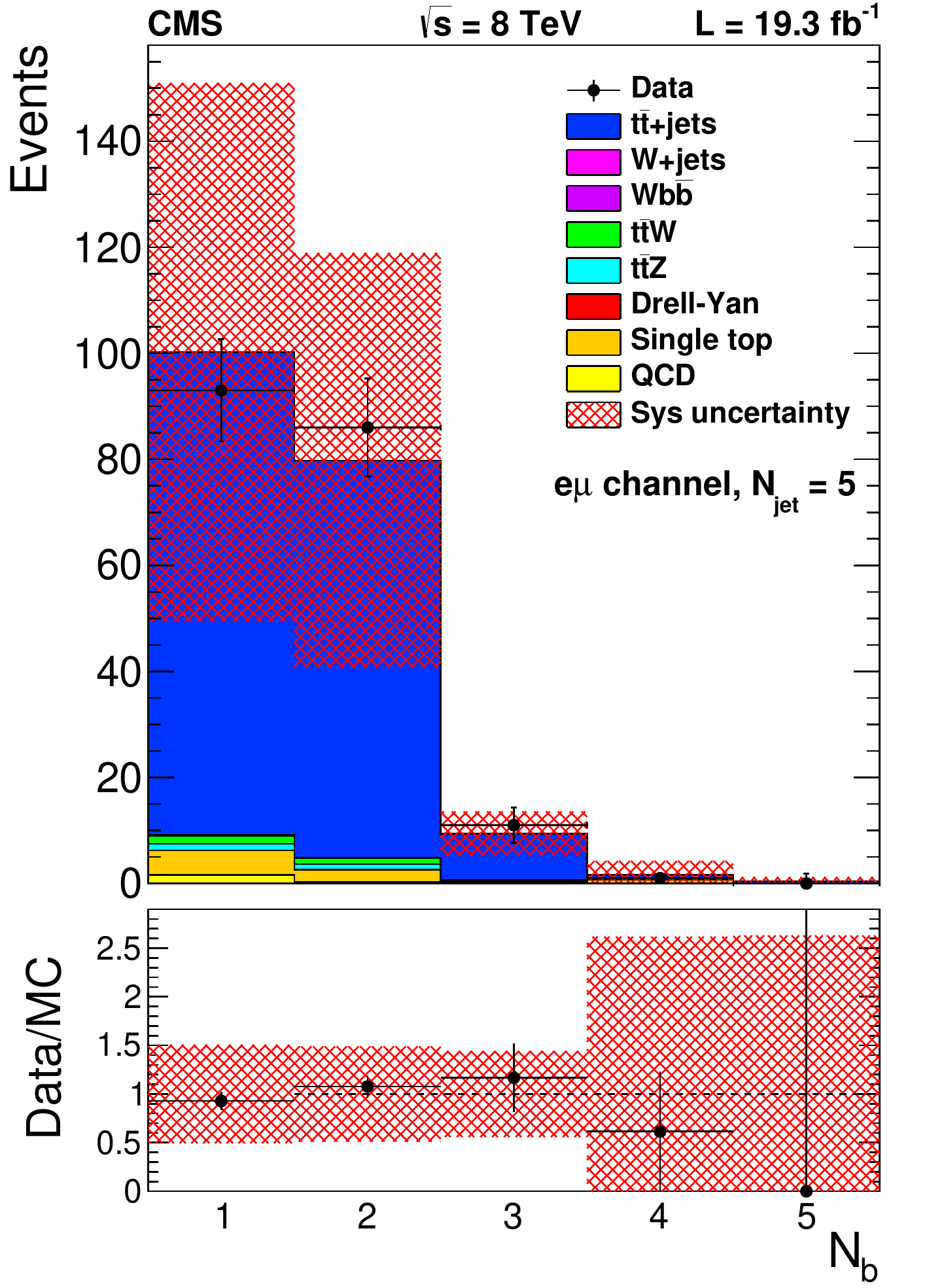}
  \caption{Distribution of the number of $\PQb$-tagged jets for events with one
    electron, one muon, and $\Njets= 4$ (left), or $\Njets=5$ (right) in data, compared to the background prediction from simulation corrected for the \PQb~tagging response. The hatched region represents the total uncertainty on the background yield.}\label{fig:gluonsplitcontrol}
\end{figure*}

The parameterization of the mismodeling of events with four $\PQb$ quarks is not as straightforward as the computation of the other uncertainties. While the data-corrected MC distribution is expected to
account for events with multiple $\PQb$-tagged jets originating from mistags, the
contribution from gluon splitting to $\bbbar$, and of SM four $\PQb$ quark events in general, is sensitive
to the details of the MC modeling~\cite{Khachatryan:2015mva}, as discussed in Section~\ref{sec:GluonSplitting}. We constrain the uncertainty in this contribution by
studying the agreement between data and simulation in events with one
identified electron, one identified muon, and associated jets. We
consider separately events with four or five jets. Furthermore, we use single-lepton control regions by selecting events with one electron or one muon and four or five jets. These control regions  provide a high-purity sample of $\ttbar$+jets events, for
which the signal contamination is expected to be negligible.
Figure~\ref{fig:gluonsplitcontrol} shows that the largest difference
between the prediction used in the analysis and the observed
yield in the dileptonic control sample is an excess of less than one standard deviation in the
three and four \PQb~jet bins for the four-jet sample.
Total uncertainties are shown in this figure, including uncertainties that affect only the normalization of the background prediction and not the shape.
The single-lepton control regions have similarly small discrepancies for $N_{\PQb} \geq 3$. These are the bins of the \PQb~tag multiplicity distribution that are most sensitive to the signal so we parameterize this effect and include it in the analysis as a systematic uncertainty.

Using the data yields in the \PQb~tag multiplicity bins of the control samples defined above, we construct a system of three equations and
three unknowns for each control region.
In a sample with $N$ events and $J$ jets, we write the number of events with $n$ $\PQb$-tagged jets $N(n, J)$ as a function of the \PQb-tagging efficiency  $\epsilon_{\PQb}$ and the mistag rate $\epsilon_\text{mis}$. Assuming that the sample consists mainly of \bbbar events, i.e. that \PQb-tag multiplicities $>2$ originate from mistagged jets, one finds:
\begin{equation}
\begin{split}
\label{eq:Ntwob}
N(n, J)=&N \sum_{i=0}^2\theta(n-i)\theta(J-n-2+i)\begin{pmatrix}2\\i\end{pmatrix}\begin{pmatrix} J-2 \\    n-i \end{pmatrix}\\
&\times(1-\epsilon_{\PQb})^{2-i}\epsilon_{\PQb}^i (1-\epsilon_\text{mis})^{J-2-n+i}\epsilon_\text{mis}^{n-i},
\end{split}\end{equation}
where $\theta(m)$ is the Heaviside step function, where $\theta(m)=1$ for $m\geq 0$ and $\theta(m)=0$ for $m<0$ and the standard representation for the binomial coefficient is used.
The index $i$ runs over the number of true $\PQb$ quarks that are within the acceptance and tagged.
Once a subset $\Delta N$ of the events is allowed to contain four
real \PQb~quarks, Eq.~\ref{eq:Ntwob} is modified as follows:
\begin{widetext}
\begin{equation}\begin{split}
N^\prime(n, J)=&(N-\Delta N)\frac{N(n,J)}{N}\\
&+\Delta N \sum_{i=0}^4 \theta(n-i)\theta(J-n-4+i)\begin{pmatrix}4\\ i \end{pmatrix}\begin{pmatrix} J-4\\ n-i \end{pmatrix}\\
&\times(1-\epsilon_{\PQb})^{4-i}\epsilon_{\PQb}^i (1-\epsilon_\text{mis})^{J-4-n+i}\epsilon_\text{mis}^{n-i}.
\end{split}\end{equation}
Taking as input the yield observed for three values of $n$ at a given $J$, one can solve a system of three equations and three unknowns
and derive values for $\epsilon_{\PQb}$, $\epsilon_\text{mis}$, and $\Delta N$. Rather than solving for $\Delta N$, we introduce
\begin{equation}
\Delta f_{4\PQb} = \left.\frac{\Delta N}{N}\right|_\text{data} - \left.\frac{\Delta N}{N}\right|_\mathrm{MC}
\end{equation}
and solve for each control region the set of equations
\begin{equation}\begin{aligned}
&\frac{\epsilon_\text{mis}^2\epsilon_{\PQb}^2 (1 - \Delta f_{4\PQb}) + \epsilon_\text{\PQb}^4 \Delta f_{4\PQb}}{\left[\epsilon_{\PQb}^2(1-\epsilon_\text{mis})^2+\epsilon_\text{mis}^2(1-\epsilon_{\PQb})^2+4\epsilon_\text{mis}(1-\epsilon_{\PQb})\epsilon_{\PQb}(1-\epsilon_\text{mis})\right](1-\Delta f_{4\PQb})+6\epsilon_{\PQb}^2(1-\epsilon_{\PQb})^2 \Delta f_{4\PQb}}&= \frac{N_4}{N_2} \\[10pt]
&\frac{\left[2 \epsilon_{\PQb}^2\epsilon_\text{mis}(1-\epsilon_\text{mis})+2\epsilon_{\PQb}(1-\epsilon_{\PQb})\epsilon_\text{mis}^2\right] (1 - \Delta f_{4\PQb}) + 4 \epsilon_{\PQb}^3 (1 - \epsilon_{\PQb}) \Delta f_{4\PQb}}{\left[\epsilon_{\PQb}^2(1-\epsilon_\text{mis})^2+\epsilon_\text{mis}^2(1-\epsilon_{\PQb})^2+4\epsilon_\text{mis}(1-\epsilon_{\PQb})\epsilon_{\PQb}(1-\epsilon_\text{mis})\right](1-\Delta f_{4\PQb})+6\epsilon_{\PQb}^2(1-\epsilon_{\PQb})^2 \Delta f_{4\PQb}} &= \frac{N_3}{N_2}\\[10pt]
&\frac{\left[\epsilon_{\PQb}^2(1-\epsilon_\text{mis})^2+\epsilon_\text{mis}^2(1-\epsilon_{\PQb})^2+4\epsilon_\text{mis}(1-\epsilon_{\PQb})\epsilon_{\PQb}(1-\epsilon_\text{mis})\right](1-\Delta f_{4\PQb})+6\epsilon_{\PQb}^2(1-\epsilon_{\PQb})^2 \Delta f_{4\PQb}}{\left[2(1-\epsilon_{\PQb})\epsilon_{\PQb}(1-\epsilon_\text{mis})^2+2(1-\epsilon_{\PQb})^2\epsilon_\text{mis}(1-\epsilon_\text{mis})\right](1-\Delta f_{4\PQb})+4 \epsilon_{\PQb}(1-\epsilon_{\PQb})^3 \Delta f_{4\PQb}}&=\frac{N_2}{N_1}.
\end{aligned}
\end{equation}
\end{widetext}
Here $N_{i}$ indicates the yield in the \PQb~tag multiplicity bin with $N_{\PQb}=i$, and the unknowns are $\epsilon_{\PQb}$, $\epsilon_\text{mis}$, and $\Delta f_{4\PQb}$. We solve the system of equations numerically, neglecting terms quadratic in the mistag rate. Since $\Delta f_{4\PQb}$ is common to all control regions, we use the resulting values of the average tagging efficiency and the average mistag rate determined in each control region to construct a global $\chi^2$:
\ifthenelse{\boolean{cms@external}}{
\begin{multline}
\chi^2\left(\Delta f_{4\PQb}\right) =\\
 \sum_{\substack{i=1,\ldots,N_{\PQb}\\j\in \mathrm{CR}}} \frac{\left(N_\text{obs}^{ij}-N_\mathrm{MC}^{ij}-N_{4\PQb}(\epsilon_{\PQb}^j, \epsilon_\text{mis}^j, \Delta f_{4\PQb})\right)^2}{\sigma_{ij}^2},
\end{multline}
}{
\begin{equation}
\chi^2\left(\Delta f_{4\PQb}\right) = \sum_{\substack{i=1,\ldots,N_{\PQb}\\j\in \mathrm{CR}}} \frac{\left(N_\text{obs}^{ij}-N_\mathrm{MC}^{ij}-N_{4\PQb}(\epsilon_{\PQb}^j, \epsilon_\text{mis}^j, \Delta f_{4\PQb})\right)^2}{\sigma_{ij}^2},
\end{equation}
}
where the sum over $j$ spans the different control regions, the index $i$ gives the bin of the multiplicity of $\PQb$-tagged jets, and $\sigma_{ij}$ is the sum in quadrature of the statistical uncertainty in data and total uncertainty in simulation. Minimizing the $\chi^2$ results in an improved determination of $\Delta f_{4\PQb}$ from the data in all control regions.

We associate a systematic uncertainty with the shape of the background, by determining $\Delta f_{4\PQb}$ with the information from both the dilepton and single-lepton control regions, obtaining $\Delta f_{4\PQb}=-0.011\pm0.049$. The choice of combining the two control regions is justified by the fact that fitting them separately we obtain compatible results. We use $\pm$1~standard deviation variations of $\Delta f_{4\PQb}$ to construct two new background shapes in the 6 jets, 7 jets, and $\geq$8 jets signal regions. The difference between these shapes is used to evaluate a systematic uncertainty that is used in the signal extraction fit.  The determination of the systematic uncertainty depends on the values of the efficiencies used in the fit, on their uncertainty and on the choice of control regions. For this reason we compute the limit on the signal cross section for several different choices of control regions, tagging efficiencies, and mistag rates. The observed variations are below $10^{-3}$ of the cross section value obtained with the value of $\Delta f_{4\PQb}$ from the combined fit.

In the bins with fewer than three $\PQb$-tagged jets the dominant sources of uncertainty come from the jet energy, renormalization, and factorization scales. For higher multiplicities of $\PQb$-tagged jets, the uncertainty in the tagging SFs and the mismodeling of events with four \PQb~quarks become the main sources of uncertainty.

\subsubsection{Signal}

The uncertainties in the signal efficiency and signal shape include the jet energy scale uncertainty, the \PQb~tagging SF uncertainty,
the uncertainty in the PDFs, the uncertainty in the measured integrated luminosity, the uncertainty in trigger and identification efficiencies for leptons, and the uncertainty in the MC modeling of ISR and FSR.

The jet energy scale and \PQb~tagging SF uncertainties are computed in the same way as for the background. However, since the signal samples are processed through a fast rather than full detector simulation, we apply additional compensating SFs for the \PQb-tagging efficiencies.
The uncertainties in the reconstruction and trigger efficiencies of muons and electrons are estimated from $\Z \to \ell \ell$ events in bins of $\eta$ and $\pt$ of the lepton; both are found to be always below 1\%. The nominal efficiency from simulation is also corrected to reflect the lepton
efficiency measured in data.
The total efficiency for $\Njet \geq 6$ is 15.8\% for $m_{\PSg} = 1.0$\TeV.
\subsection{Results for the single-lepton final state}\label{sec:res}

\begin{figure*}[!ht]
  \centering%
  \includegraphics[width=0.30\textwidth]{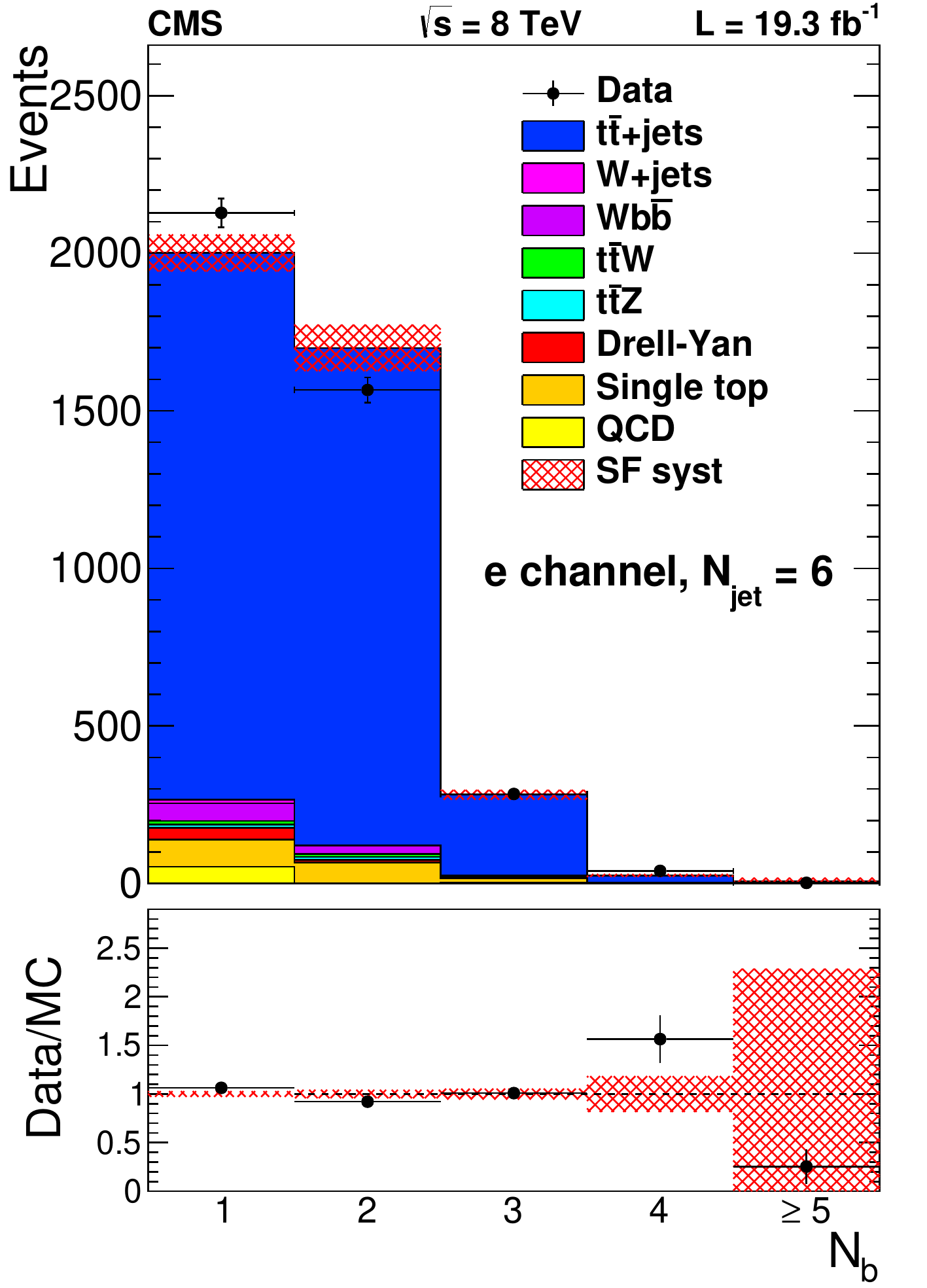}
  \includegraphics[width=0.30\textwidth]{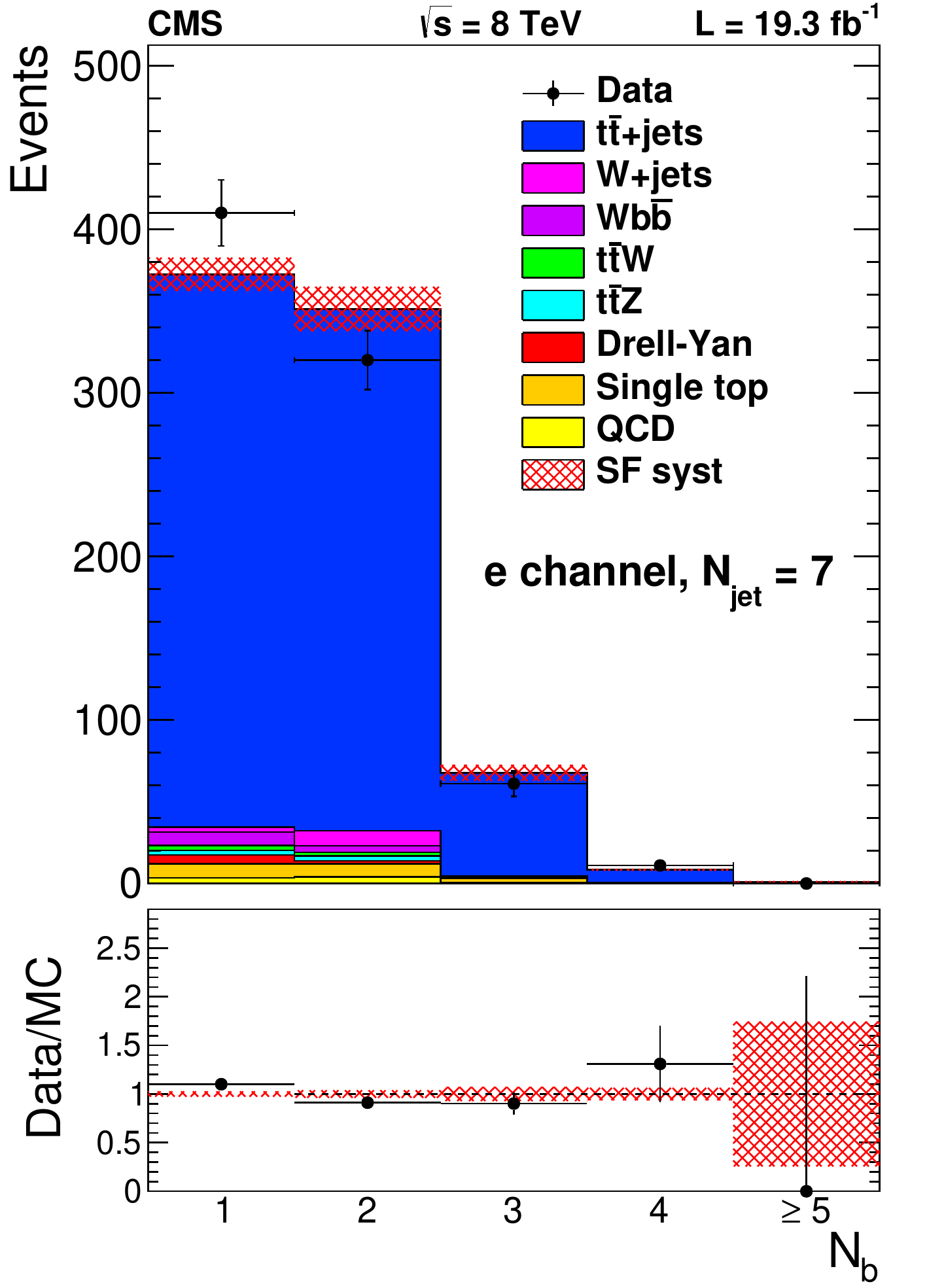}
  \includegraphics[width=0.30\textwidth]{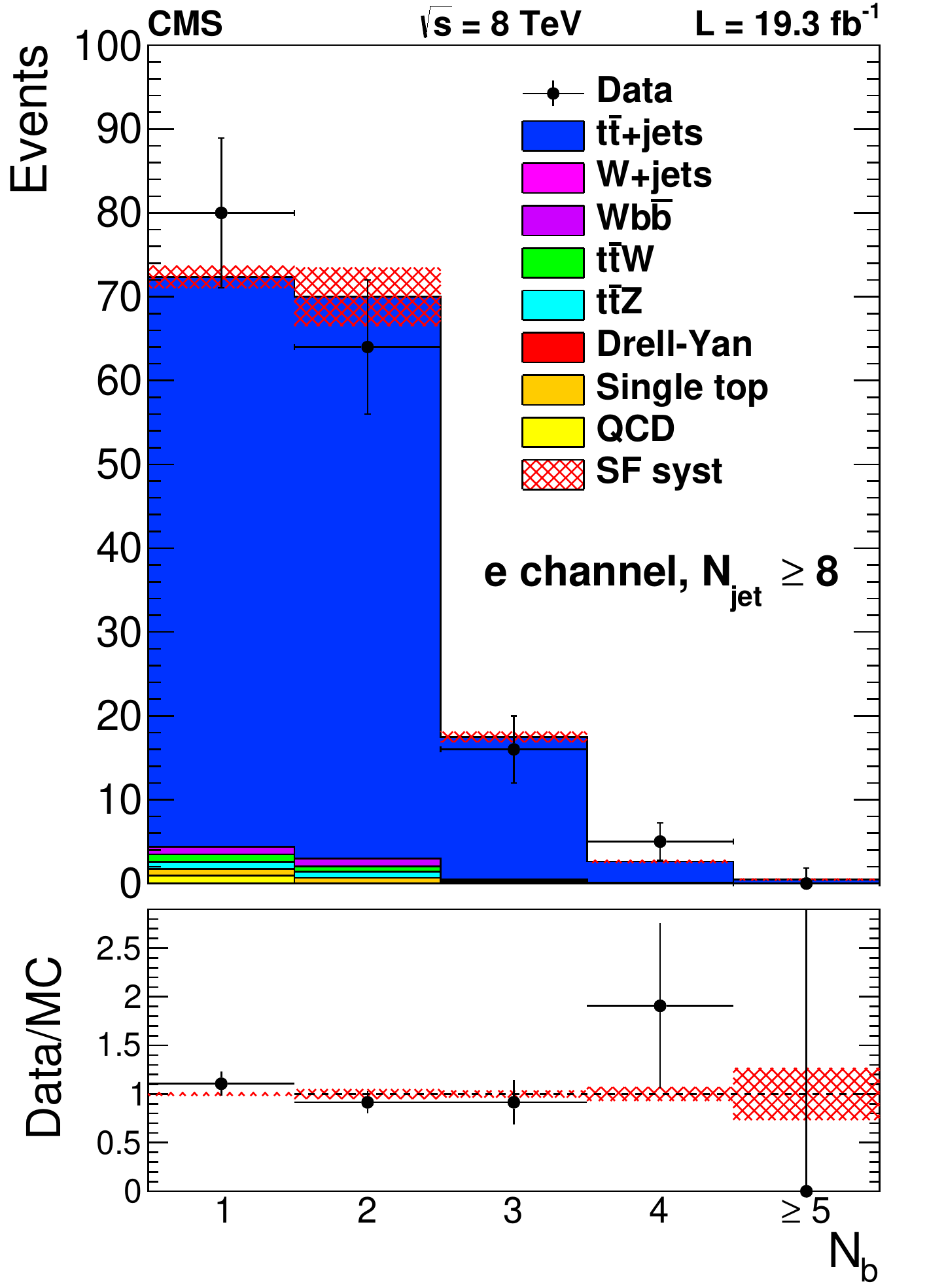}
  \includegraphics[width=0.30\textwidth]{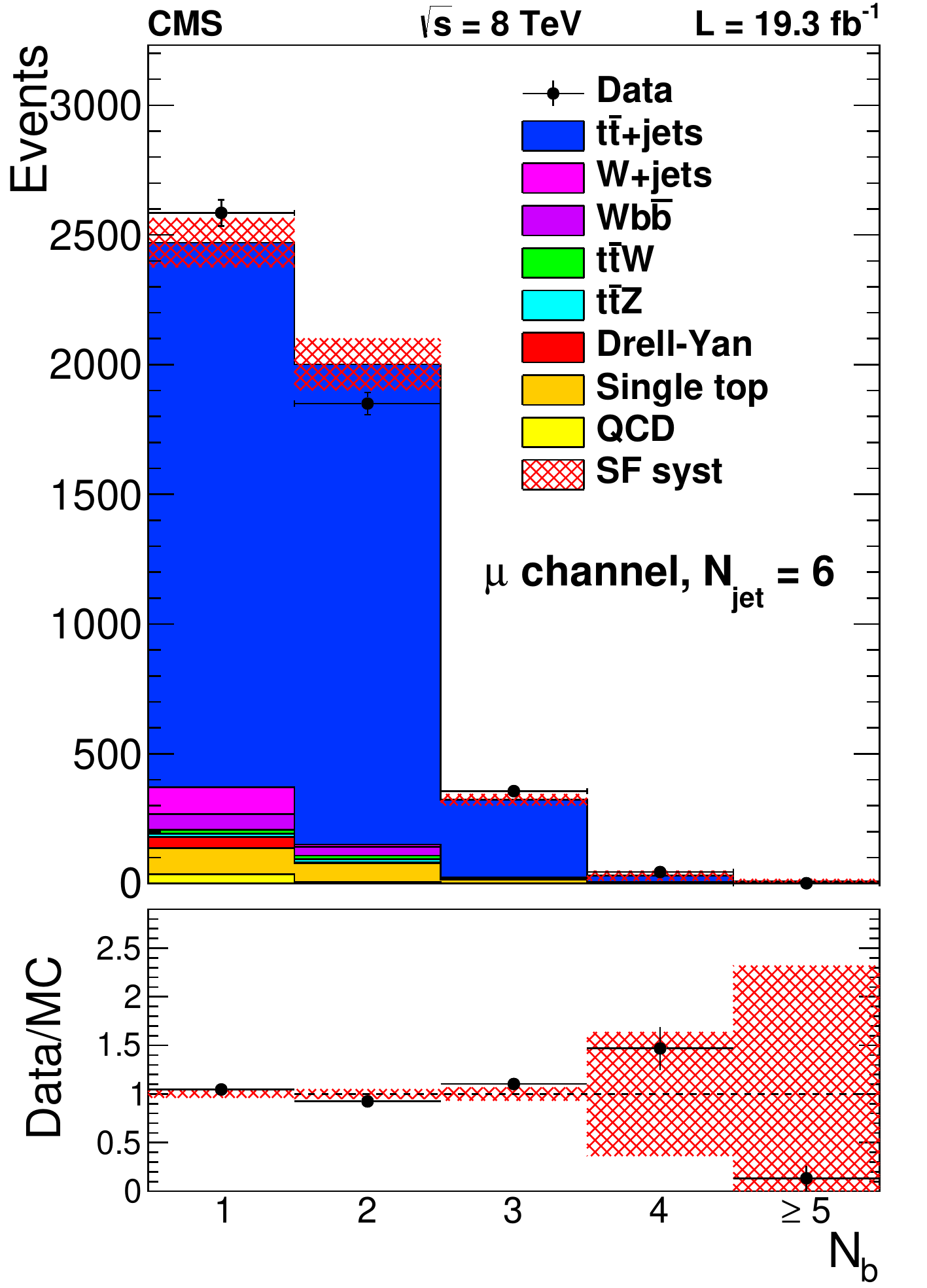}
  \includegraphics[width=0.30\textwidth]{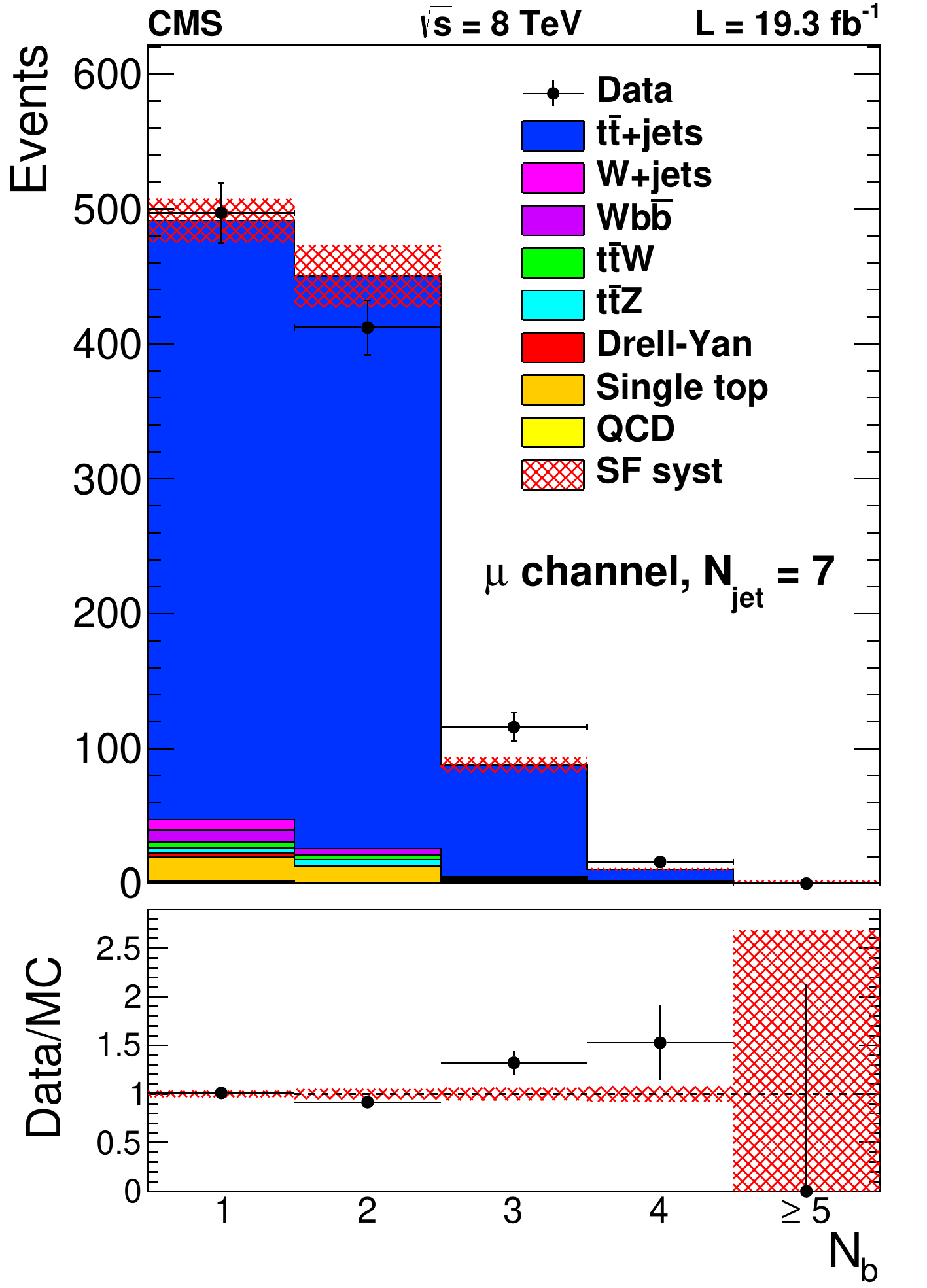}
  \includegraphics[width=0.30\textwidth]{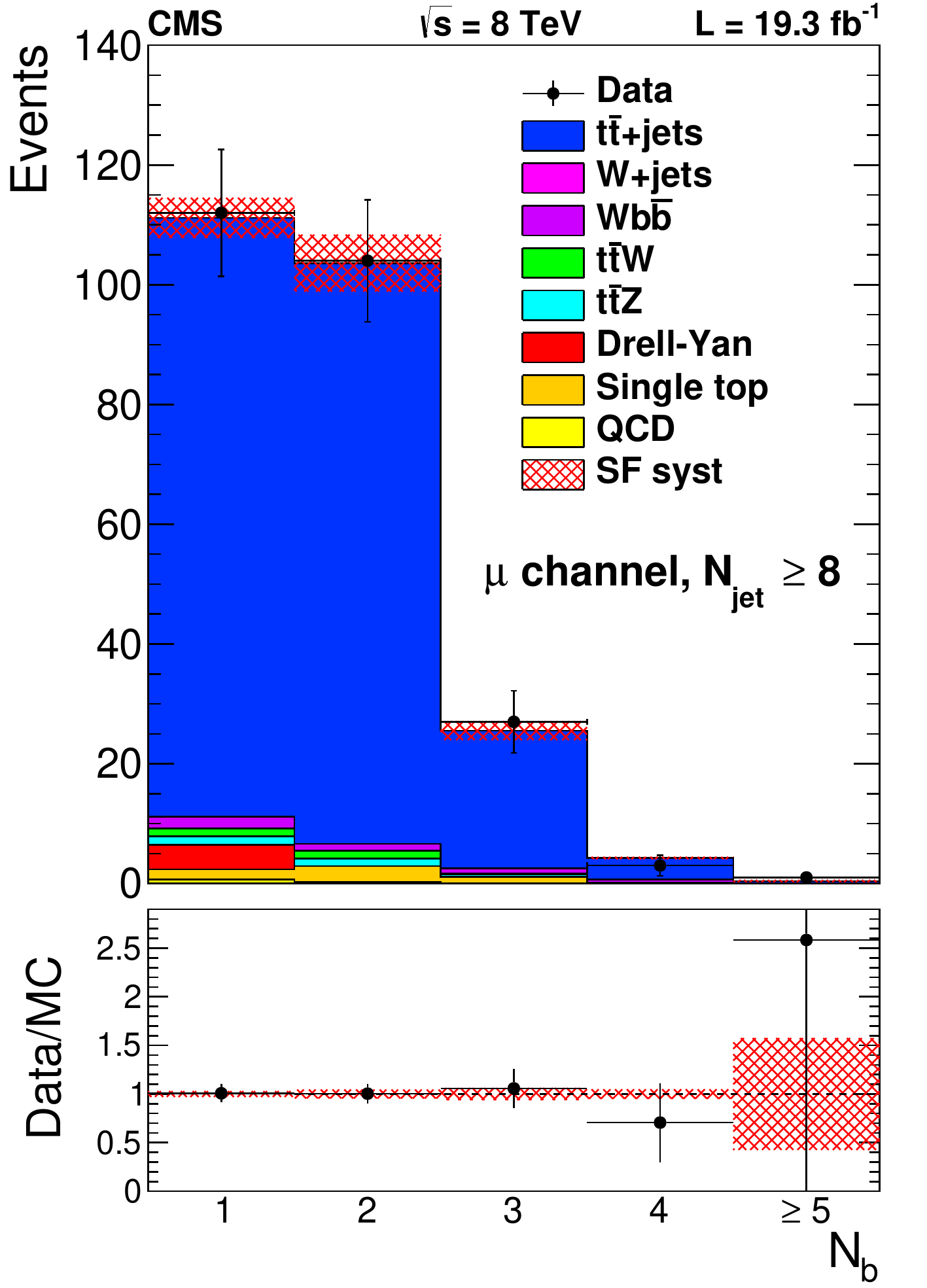}
  \caption{Distribution of the number of $\PQb$-tagged jets for events with one
    electron (upper) or muon (lower) and 6 (left), 7 (middle) or $\geq$8 (right) jets, compared to the background
    prediction from simulation corrected for the \PQb~tagging response in data. The hatched region represents the uncertainty originating from the uncertainty in the \PQb~tagging correction factors. Most other uncertainties affect only the normalization and will cancel in the fit.\label{fig:NbData6}}
\end{figure*}

Figure~\ref{fig:NbData6} shows the \PQb~tag multiplicity for the 6 jets,
7 jets, and $\geq$8 jets signal regions, for events with at least one well-identified electron and for events with at least one well-identified muon. The hatched region represents the uncertainty propagated from the \PQb-tagging correction factors.
We summarize the expected background, the expected signal for $m_{\PSg}=1$\TeV, and the observed yield in each signal region in Tables~\ref{tab:resultELE} and~\ref{tab:resultMU}.
The postfit uncertainties in the background, shown in Tables~\ref{tab:resultELE} and~\ref{tab:resultMU}, are considerably reduced with respect to the uncertainties in the prediction. The fit, described below, extracts the background normalization from data, reducing the total uncertainty to an uncertainty in the shape of the multiplicity distribution of $\PQb$-tagged jets. Therefore uncertainties coming from jet energy, matching, renormalization, and factorization scales that affect mostly the total yield become almost negligible. The central values for the background prediction do not change significantly in the fit. The electron and muon samples, although presented separately to facilitate reinterpretations, are fitted simultaneously; for the same reason, this fit does not assume a signal model and instead sets the signal yield to zero.

Tables~\ref{tab:resultELE} and~\ref{tab:resultMU} also give the signal prediction for $m_{\PSg}=1$\TeV. Combining all the signal uncertainties gives a total uncertainty of $\sim$10--20\% on the individual bins of $\PQb$-tagged jet multiplicity. At high gluino masses ($m_{\PSg} \geq 1$\TeV) the uncertainty is dominated by the PDF uncertainty, while for lower masses the jet energy scale uncertainty constitutes the most important source of uncertainty, followed in magnitude by the uncertainty in the modeling of ISR and FSR.

No sizable deviation from the expected SM yields is observed. We interpret the absence of an excess as an
upper bound on the cross section for SUSY models predicting final
states with one lepton and multiple $\PQb$-tagged jets.
The cross section limit is obtained with a maximum likelihood fit to the shape of the $\PQb$-tagged jet multiplicity distribution and is converted to a bound on the mass of the gluino.

\begin{table*}[!ht]
\centering
\topcaption{Summary of expected background, expected signal for $m_{\PSg}=1$\TeV, observed yields, and total background after the background-only fit for the electron samples considered in the analysis. The uncertainties given include all statistical and systematic uncertainties.\label{tab:resultELE}}
\begin{scotch}{cccccc}
e + $6$ jets & One \PQb-tag & Two \PQb-tags & Three \PQb-tags & Four \PQb-tags & Five \PQb-tags \\
\hline
Background prediction &  2003$\pm$827 & 1701$\pm$762 & 281$\pm$130 & 27$\pm$17 & 8.0$\pm$6.8 \\
Signal ($m_{\PSg}=1$\TeV)&  1.9$\pm$0.3 & 2.9$\pm$0.5 & 1.9$\pm$0.3 & 0.41$\pm$0.10 & 0.03$\pm$0.01 \\
Data & 2128 & 1566 & 284 & 40 & 2 \\
Background postfit &  1967$\pm$54 & 1636$\pm$53 & 296.1$\pm$9.5 & 33.6$\pm$3.0 & 1.9$\pm$1.2 \\
\hline
\Pe + $7$ jets & One \PQb-tag & Two \PQb-tags & Three \PQb-tags & Four \PQb-tags & Five \PQb-tags \\
\hline
Background prediction &  373$\pm$200 & 352$\pm$199 & 67$\pm$39 & 8.7$\pm$6.3 & 1.1$\pm$1.1 \\
Signal ($m_{\PSg}=1$\TeV)& 2.0$\pm$0.3 & 3.4$\pm$0.5 & 2.7$\pm$0.4 & 0.86$\pm$0.15 & 0.07$\pm$0.02 \\
Data &  410 & 320 & 61 & 11 & 0 \\
Background postfit &  368$\pm$13 & 347$\pm$12 & 70.6$\pm$2.8 & 10.38$\pm$0.65 & 0.70$\pm$0.12 \\
\hline
\Pe + $8$ jets & One \PQb-tag & Two \PQb-tags & Three \PQb-tags & Four \PQb-tags & Five \PQb-tags \\
\hline
Background prediction &  73$\pm$51 & 70$\pm$49 & 18$\pm$15 & 2.7$\pm$2.1 & 0.47$\pm$0.38 \\
Signal ($m_{\PSg}=1$\TeV)&2.4$\pm$0.4 & 4.9$\pm$0.8 & 4.7$\pm$0.7 & 2.0$\pm$0.3 & 0.23$\pm$0.04 \\
Data &  80 & 64 & 16 & 5 & 0 \\
Background postfit &  74.9$\pm$3.1 & 71.0$\pm$3.1 & 18.94$\pm$0.93 & 3.40$\pm$0.20 & 0.44$\pm$0.03 \\
\end{scotch}
\end{table*}

\begin{table*}[!ht]
\centering
\topcaption{Summary of the expected background, expected signal for $m_{\PSg}=1$\TeV, observed yields, and total background after the background-only fit for the muon samples. The uncertainties given include all statistical and systematic uncertainties.\label{tab:resultMU}}
\begin{scotch}{cccccc}
$\mu$ + $6$ jets & One \PQb-tag & Two \PQb-tags & Three \PQb-tags & Four \PQb-tags & Five \PQb-tags \\
\hline
Background prediction &  2474$\pm$977 & 2002$\pm$801 & 322$\pm$152 & 30$\pm$29 & 7.7$\pm$6.5 \\
Signal ($m_{\PSg}=1$\TeV)&  3.0$\pm$0.5 & 4.6$\pm$0.7 & 2.8$\pm$0.4 & 0.6$\pm$0.1 & 0.04$\pm$0.03 \\
Data &  2585 & 1850 & 356 & 44 & 1 \\
Background postfit &  2425$\pm$60 & 1985$\pm$49 & 340$\pm$11 & 43.0$\pm$3.5 & 3.1$\pm$1.1 \\
\hline
$\mu$ + $7$ jets & One \PQb-tag & Two \PQb-tags & Three \PQb-tags & Four \PQb-tags & Five \PQb-tags \\
\hline
Background prediction &  493$\pm$203 & 448$\pm$180 & 88$\pm$39 & 10.7$\pm$7.2 & 0.9$\pm$2.8 \\
Signal ($m_{\PSg}=1$\TeV)&  3.0$\pm$0.5 & 5.0$\pm$0.7 & 3.9$\pm$0.6 & 1.1$\pm$0.2 & 0.09$\pm$0.04 \\
Data &  497 & 412 & 116 & 16 & 0 \\
Background postfit &  506$\pm$15 & 462$\pm$13 & 95.9$\pm$3.0 & 14.81$\pm$0.99 & 1.03$\pm$0.18 \\
\hline
$\mu$ + $8$ jets & One \PQb-tag & Two \PQb-tags & Three \PQb-tags & Four \PQb-tags & Five \PQb-tags \\
\hline
Background prediction &  112$\pm$47 & 104$\pm$46 & 26$\pm$12 & 4.3$\pm$2.1 & 0.39$\pm$0.75 \\
Signal ($m_{\PSg}=1$\TeV) &  3.7$\pm$0.6 & 7.0$\pm$1.0 & 6.4$\pm$0.9 & 2.5$\pm$0.4 & 0.33$\pm$0.06 \\
Data &  112 & 104 & 27 & 3 & 1 \\
Background postfit &  119.7$\pm$4.3 & 110.7$\pm$3.6 & 29.0$\pm$1.0 & 5.63$\pm$0.34 & 0.54$\pm$0.07 \\
\end{scotch}
\end{table*}

In setting a limit on new physics, we consider the $\PSg \to \PQt\PQb\PQs$ simplified model, which assumes $\lambda^{\prime\prime}_{332} \neq 0$. The main ingredient
to the likelihood for a given multiplicity of $\PQb$-tagged jets and lepton
flavor in a given signal region (6 jets, 7 jets, $\geq$8 jets) is a Poisson
function for $n$ observed events, given an expected yield of $\epsilon
\mathcal{L} \sigma+B$:
\begin{equation}
P(n|\epsilon \mathcal{L} \sigma+B) = \frac{\re^{-(\epsilon \mathcal{L} \sigma+B)}}{n!} (\epsilon
\mathcal{L}\sigma +B)^n~.
\end{equation}
Here $B$ is the expected background yield, $\epsilon$ is the signal
efficiency, $\mathcal{L}$ is the integrated luminosity of the data set, and $\sigma$ is the
cross section on which we want to set the limit. The extended likelihood
is written as:
\ifthenelse{\boolean{cms@external}}{
\begin{multline}
L = \frac{\re^{-(\epsilon \mathcal{L} \sigma+B)}}{n!} (\epsilon
\mathcal{L}\sigma +B)^n
P_\text{LN}(\epsilon | \bar{\epsilon}, \delta \epsilon)\\
\times
P_\text{LN}(\mathcal{L} | \bar{\mathcal{L}}, \delta \mathcal{L})
P_\text{LN}( B | \bar{B}, \delta B). \label{eq:likOnebin}
\end{multline}
}{
\begin{equation}
L = \frac{\re^{-(\epsilon \mathcal{L} \sigma+B)}}{n!} (\epsilon
\mathcal{L}\sigma +B)^n
P_\text{LN}(\epsilon | \bar{\epsilon}, \delta \epsilon)
P_\text{LN}(\mathcal{L} | \bar{\mathcal{L}}, \delta \mathcal{L})
P_\text{LN}( B | \bar{B}, \delta B). \label{eq:likOnebin}
\end{equation}
}
We model the systematic uncertainty associated with the signal and the
background prediction as log-normal functions $P_\mathrm{LN}(x | \bar{x},
\delta x)$ for the measured value $x$, given an expected value $\bar
x$ and an uncertainty $\delta x$. The full likelihood is obtained as
the product of a set of likelihoods as the one in Eq.~(\ref{eq:likOnebin}). The product runs over each multiplicity of $\PQb$-tagged jets (one to five), lepton flavor
(\Pe, $\mu$), and each of the three signal regions (6 jets, 7 jets, $\geq$8
jets). The nuisance parameters are taken to be fully correlated across
the three signal regions with different jet multiplicities and the two lepton flavors: i.e. a common log-normal function for each nuisance multiplies the product of Poisson functions.

\begin{figure}[!ht]
  \centering
  \includegraphics[width=0.45\textwidth]{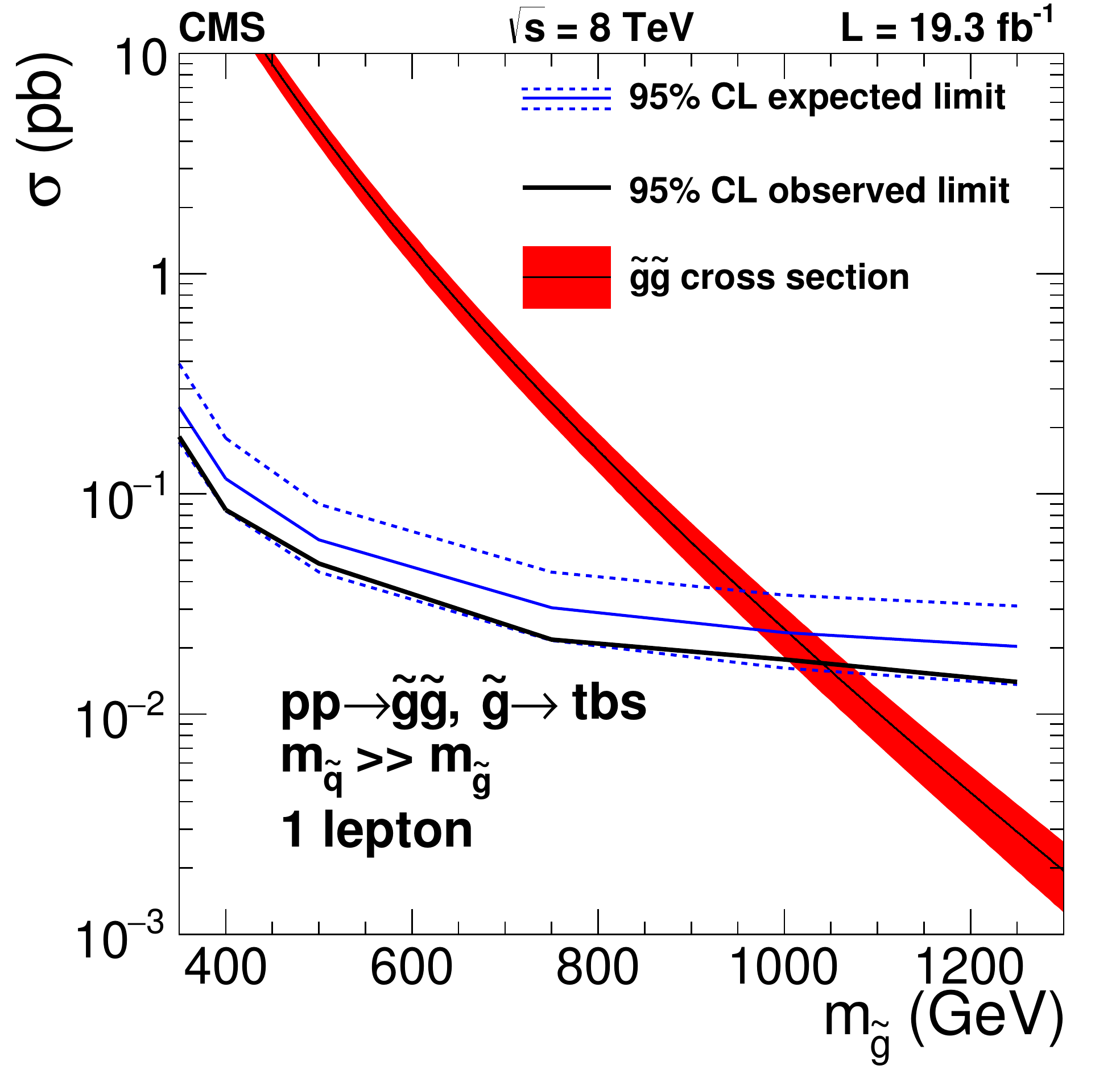}
  \caption{The $95\%$ \CL limit on the gluino pair production cross section as a function of $m_\text{\PSg}$ in the one-lepton analysis. The signal
    considered is $\Pp\Pp \to \PSg\PSg$, followed by the decay
    $\PSg\to\PQt\PQb\PQs$. The band shows the theoretical cross section and its uncertainty. The dashed lines show the uncertainty on the expected limit.}
\label{fig:limit_onelepton}
\end{figure}

For a given value of $\sigma$ under test the likelihood is profiled
with respect to the nuisance parameters ($\mathcal{L}$, $\epsilon$, and
$B$).
The result of this procedure is shown in Fig.~\ref{fig:limit_onelepton} and results in a 95\% \CL lower limit on the gluino mass
of 1.03\TeV when the gluino is assumed to decay exclusively to $\PQt\PQb\PQs$.
This is currently the strongest bound for this gluino decay mode.

\section{Dilepton final state}
\label{sec:dilepton}

We search for the RPV decays of the bottom squark ($\PSQb$) in an MSSM model featuring minimal flavor violation~\cite{MFV}. When the bottom squark is the LSP in this type of model, it can decay to a top quark and a down-type quark. We have chosen a model sensitive to the $\lambda''_{332}$ and $\lambda''_{331}$ hadronic RPV couplings, so the bottom squark decay of interest is to a top and either a strange or down quark. In contrast to Sections \ref{sec:hadronic} and \ref{sec:onelepton} which feature a gluino pair production model, this section focuses on a model with bottom squark pair production.

\begin{figure}[hbtp]
\centering
\includegraphics[width=0.45\textwidth]{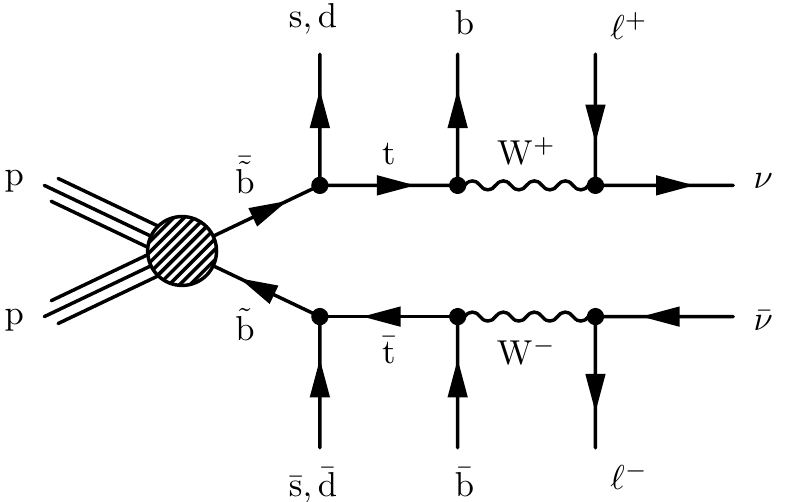}
\caption{Diagram for pair production of bottom squarks and their RPV decay.}
\label{fig:feynDiagram}
\end{figure}

We restrict this search to dilepton final states, where each top quark decays into a $\PW$ boson, which in turn decays leptonically. A diagram of this process is shown in Fig.~\ref{fig:feynDiagram}. To discriminate signal from background events, we analyze the reconstructed bottom squark mass and the transverse momenta of jets identified as coming from light quarks or gluons.

The trigger requires two leptons, one of which has  $\pt > 17$\GeV and the other $\pt > 8$\GeV. In the subsequent analysis at least two leptons passing identification and isolation criteria are required in each event. We require at least two selected jets to pass the loose \PQb~tagging selection, and in addition at least one of these must pass the medium \PQb~tagging selection. Additionally, we require that at least two jets fail the loose \PQb~tagging selection. This allows for unambiguous categorization of light-quark jets from the bottom squark decay and the \PQb~jets from the top decay.

\subsection{Signal and background discrimination}
\label{sec:model}

The dominant background to the signal originates from SM top quarks pair produced in association with jets from ISR or FSR. Other SM processes account for a small ($\approx$5\%) contribution: single top quark production, diboson production, Drell--Yan production, and top quark pair production in association with vector bosons.  Signal events contain a resonance that produces top quarks in association with a light-quark jet. The light-quark jet from this decay has a relatively high $\pt$. We use both of these properties to discriminate between signal and background by the construction of a three-dimensional probability distribution over the reconstructed resonance mass and the two \pt values of the light-flavor jets.

We associate the two highest \pt non-$\PQb$-tagged jets with light quarks from the bottom squark decays, the two highest \pt $\PQb$-tagged jets with bottom quarks from top quark decays, and the two highest \pt leptons with leptons from $\PW$ boson decays.
In total, 6478 events pass the selection requirements: 1723 in the $\Pe\Pe$ channel, 1365 in the $\mu\mu$ channel, and 3390 in the $\Pe\mu$ channel.
Eleven events contain more than two leptons and are included in the analysis.

\subsubsection{Light jet \texorpdfstring{$\pt$}{transverse momentum} spectrum}
\label{sec:lightJetFitting}

To model the light parton jet \pt spectrum of SM processes, we assume that the light-parton jets are produced predominantly by ISR or FSR from \ttbar events and therefore have a steeply falling \pt spectrum. Signal events, on the other hand, are more likely to contain light-parton jets with relatively high $\pt$.

Letting $\pt^{(1)}$ and $\pt^{(2)}$ denote the transverse momenta of, respectively, the highest and second-highest $\pt$ light-parton jets in the event, we use simulated SM events to help choose the form of a two-dimensional probability density function $\rho^\text{SM}_\text{2D}( \pt^{(1)},\pt^{(2)})$ with sufficient flexibility to fit the data well. This distribution is constructed as a sum of three 2D densities with signal region index $i$, $\rho^\text{2D}_i ( \pt^{(1)},\pt^{(2)} )$:
\begin{equation} \label{eq:2dFunctionOrdered}
\rho^\text{SM}_\text{2D}( \pt^{(1)},\pt^{(2)}) = 2\sum_{i=1}^3f_i\rho^\text{2D}_i ( \pt^{(1)},\pt^{(2)} ) \theta( \pt^{(1)}- \pt^{(2)}),
\end{equation}
where $\sum_{i=1}^3f_i = 1$. The 2D distributions are multiplied by a Heaviside step function to enforce the ordering and a factor of 2 to normalize the function. For each component we find from simulation that the 2D density can be expressed as a product of 1D densities:
\begin{equation}
\rho^\text{2D}_i ( \pt^{(1)},\pt^{(2)}
)=\rho^\text{jet}_i( \pt^{(1)}) \rho^\text{jet}_i( \pt^{(2)}).
\end{equation}
The 1D densities all have the same form:
\begin{equation}
\label{eq:2dFunction}
\rho^\text{jet}_i ( \pt ) = \lambda_i\alpha \exp{(- \lambda_i  \pt ^\alpha)}  \pt ^{\alpha-1},
\end{equation}
where $\alpha$ is a parameter between 0 and 1 common to all components, while the $\lambda_i$ are parameters differing in each component. This function has the steeply falling behavior that we expect from ISR or FSR, and potentially a longer tail than a pure exponential distribution, becoming identical to an exponential distribution for $\alpha = 1$. The assumption that the density can be factorized is tested by comparing fit results from an SM MC simulation and found to work well.

\begin{figure}[hbtp]
\centering
\includegraphics[width=0.45\textwidth]{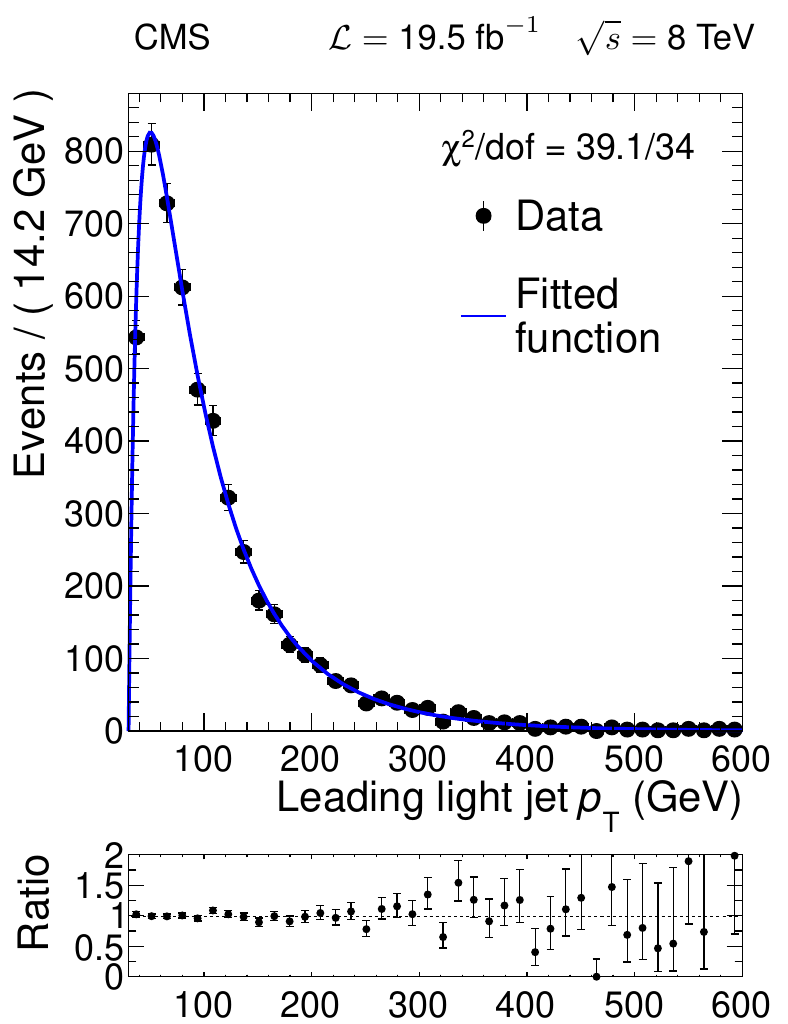}
\includegraphics[width=0.45\textwidth]{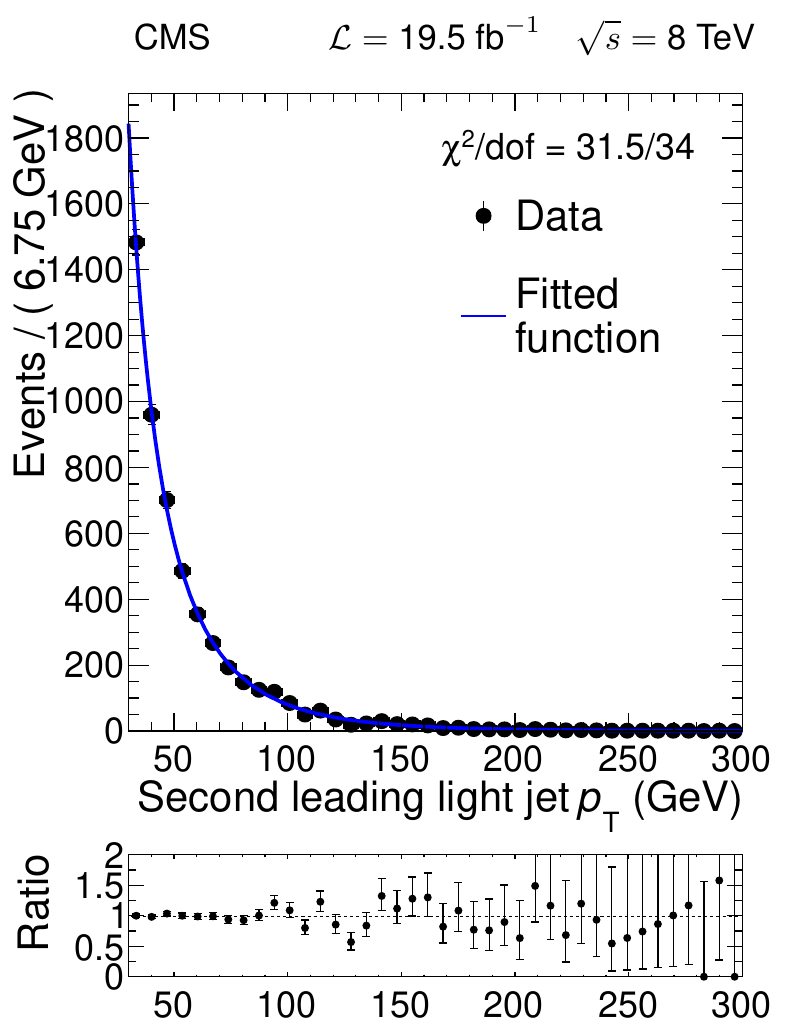}
\caption{Background-only likelihood fits for the light-parton jet $ \pt $ distributions, with signal cross section
  set to zero, for the (\cmsLeft) leading and the (\cmsRight) second-leading light-quark jet. The line represents the fitted function and the points represent the data.  The ratio of the data to the fitted function is also shown.
}
\label{fig:lightJetFit}
\end{figure}

The parameters of these densities are determined by fitting to data, maximizing the likelihood function defined in Section~\ref{sec:likelihood}.  This function is sufficiently flexible to accommodate observed light-parton jet \pt behavior while using a limited number of free parameters. Figure~\ref{fig:lightJetFit} shows the background-only hypothesis fits to the leading and subleading \pt spectra for jets in data, illustrating good agreement.

To parameterize the signal distribution, we model jets as being sampled from the sum of two two-dimensional log-normal distributions and ordered by $\pt$.  The parameters of these two distributions and their relative fractions are determined by fitting to the signal simulation. We call this distribution $\rho^\text{signal}_\text{2D}( \pt^{(1)},\pt^{(2)})$.

\begin{figure}[hbtp]
\centering
\includegraphics[width=0.45\textwidth]{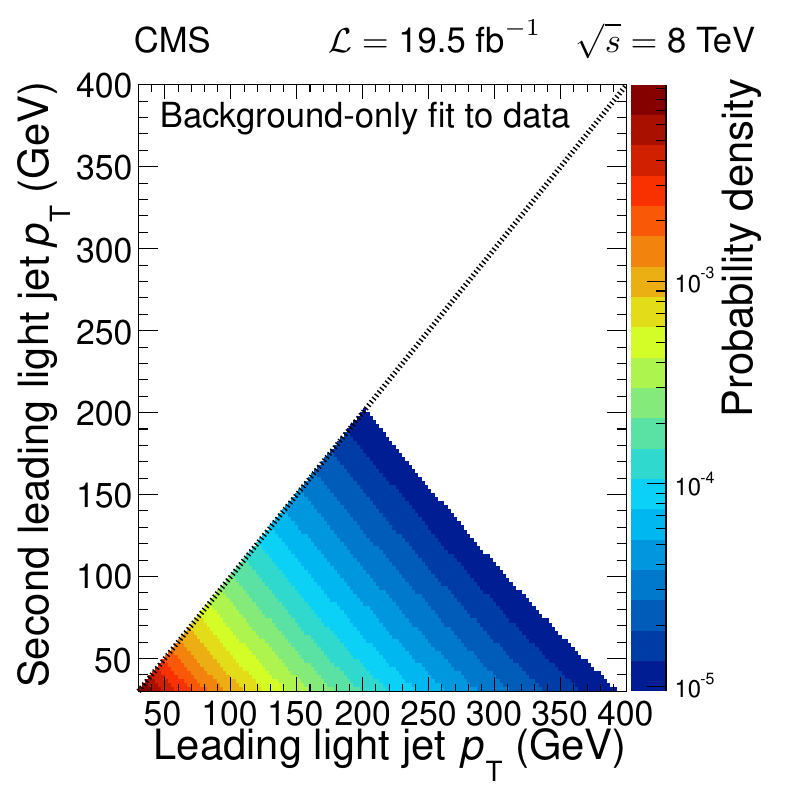}
\includegraphics[width=0.45\textwidth]{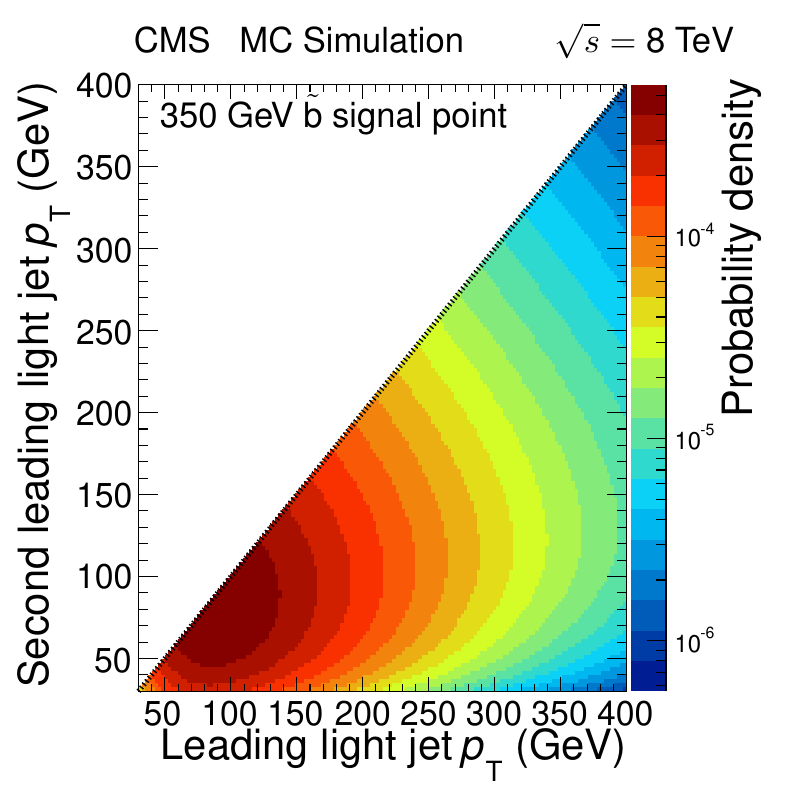}
\caption{Two-dimensional light-parton jet \pt distributions for (\cmsLeft) the background-only hypothesis fit to data and (\cmsRight) the signal with $m_{\PSQb}=350$\GeV.  The scales are logarithmic and a line has been drawn along the diagonal to illustrate the ordering of the jets by \pt.}
\label{fig:2DlightJetShapes}
\end{figure}

Figure~\ref{fig:2DlightJetShapes} illustrates the difference between light-parton jet shapes. It shows the two-dimensional distribution of the background-only hypothesis fitted to the measured \pt distribution and the signal distribution for an example mass point ($m_{\PSQb}=350$\GeV).

\subsubsection{Resonance reconstruction}\label{sec:invMassSpectrum}

To extract the mass of the potential resonance, we reconstruct the bottom squark decay chain. First we reconstruct the top quark pair candidates using the lepton momenta, $\PQb$-tagged jet momenta, and \MET.  Then these candidates are combined with the light-parton jets to reconstruct candidate bottom squark pairs.

In \ttbar decays resulting in two charged leptons, the momenta of the neutrinos can be determined by solving a quartic equation~\cite{Sonnenschein:2006ud,Cheng:2007xv}, given the masses of the top quark and $\PW$ boson.
We assume that all $\MET$ in the event comes from neutrinos associated with leptonic $\PW$ decays and fix all resonance masses to their on-shell values. We solve this quartic equation for both pairings of leptons and $\PQb$-tagged jets yielding eight (possibly unphysical) solutions for the $\ttbar$ candidates in the event.
If none of the eight solutions is physical, we vary the measured jet momenta within their resolutions using the procedure described in Ref.~\cite{Chatrchyan:2012ea}.
We sample 1000 times per event and choose the set of parameters with at least one physical solution for $\ttbar$ candidates and the smallest $\chi^2$ with respect to the measured jets.  The resampling procedure improves the event selection efficiency by a factor of approximately~$1.4$ for both background and simulated signal events.  The efficiency for finding real solutions in signal events that pass all other selection criteria is between $55\%$ and $85\%$, depending on the particular signal model and mass point. The background efficiency is approximately $81\%$.  The additional data events from the resampling procedure improve the signal sensitivity by $15\%$.

The physical solutions for $\ttbar$ candidates are combined with the light-parton jets to form candidate $\PSQb$ resonances, which can include up to 16 solutions.
We select the pair with the smallest absolute value of the logarithm of the ratio of the two candidate resonance masses. For the remainder of the analysis, we substitute the average of these two masses, denoted by $m_{\PQt j}$, for the mass of each candidate within the chosen pair.  It is bounded from below by the top quark mass.

The resulting $m_{\PQt j}$ distributions for background and signal are parameterized by different functions. The background distribution displays a turn-on behavior due to $\pt$ requirements on the leptons and jets, and a falling tail.  It is modeled by the sum of a gamma distribution and a log-normal distribution, constrained to peak at the same value.
There are four parameters in the background model that are left free in the fit: two widths, one peak position, and the relative normalization of the two components.
The signal is parameterized as the sum of two gamma distributions with parameters determined by a fit to the signal simulation.
One of the gamma functions models the correctly paired events and the other models the incorrectly paired events. Events that are correctly paired have a narrower mass width than those that are incorrectly paired, which also have a large tail extending to high mass.
The success rate for correctly matching the reconstructed and true objects increases from 30\% to 50\% with increasing bottom squark mass.

In order to increase the sensitivity of the analysis, events are categorized into different regions according to $\pt^{(2)}$, as illustrated in Table~\ref{tab:fitRegions}. We label these as three signal-enhanced regions (denoted SR1--3) and one signal-depleted control region (denoted CR). The mass spectrum for each region is fit separately.
We write the background distribution as $\rho^\text{SM}_\text{mass}( m_{\PQt j} | \pt^{(2)}) $ and the signal distribution as $\rho^\text{signal}_\text{mass}(m_{\PQt j} | \pt^{(2)}) $.

\begin{table}[htpb]
\centering
\topcaption{\label{tab:fitRegions}Definition of signal and control regions in the dilepton analysis.}
\begin{scotch}{cc}
Second-leading light-parton jet \pt & Region \\
\hline
$30<\pt^{(2)}<50\GeV$  & Control region (CR) \\
$50<\pt^{(2)}<80\GeV$  & Signal region 1 (SR1) \\
$80<\pt^{(2)}<110\GeV$ & Signal region 2 (SR2) \\
$\pt^{(2)}>110\GeV$    & Signal region 3 (SR3) \\
\end{scotch}
\end{table}

\begin{figure*}[hbtp]
\centering
\includegraphics[width=0.4\textwidth]{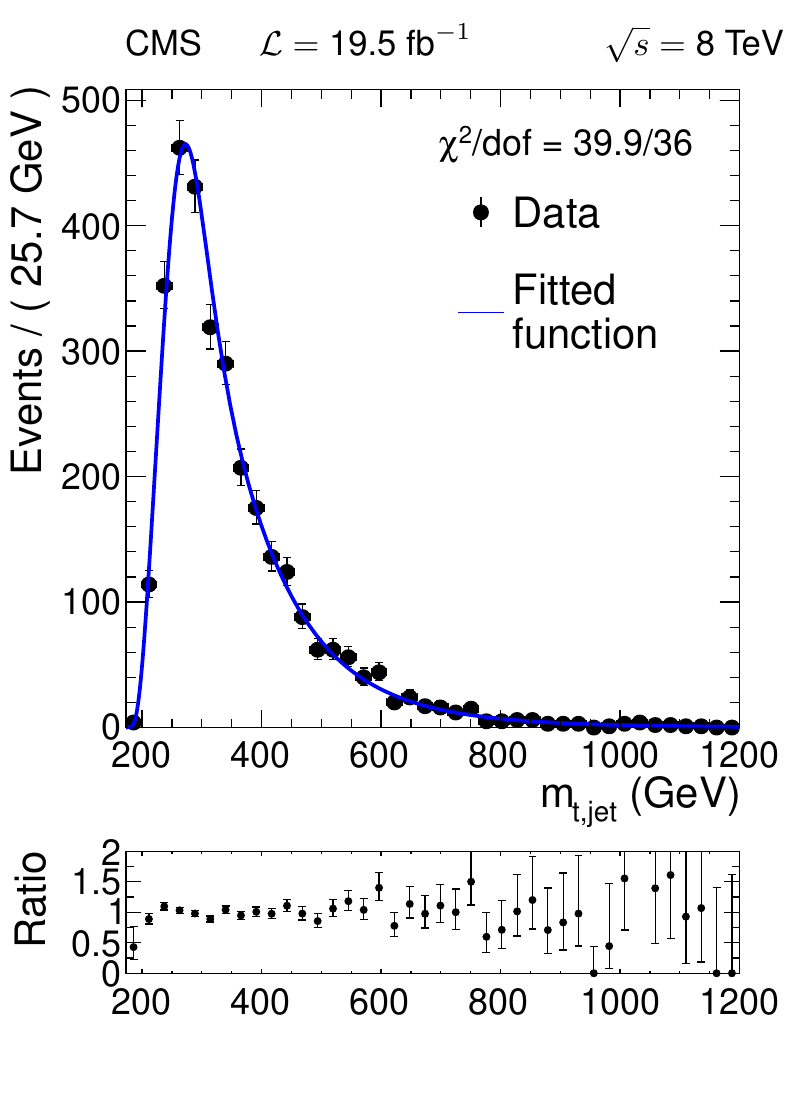}
\includegraphics[width=0.4\textwidth]{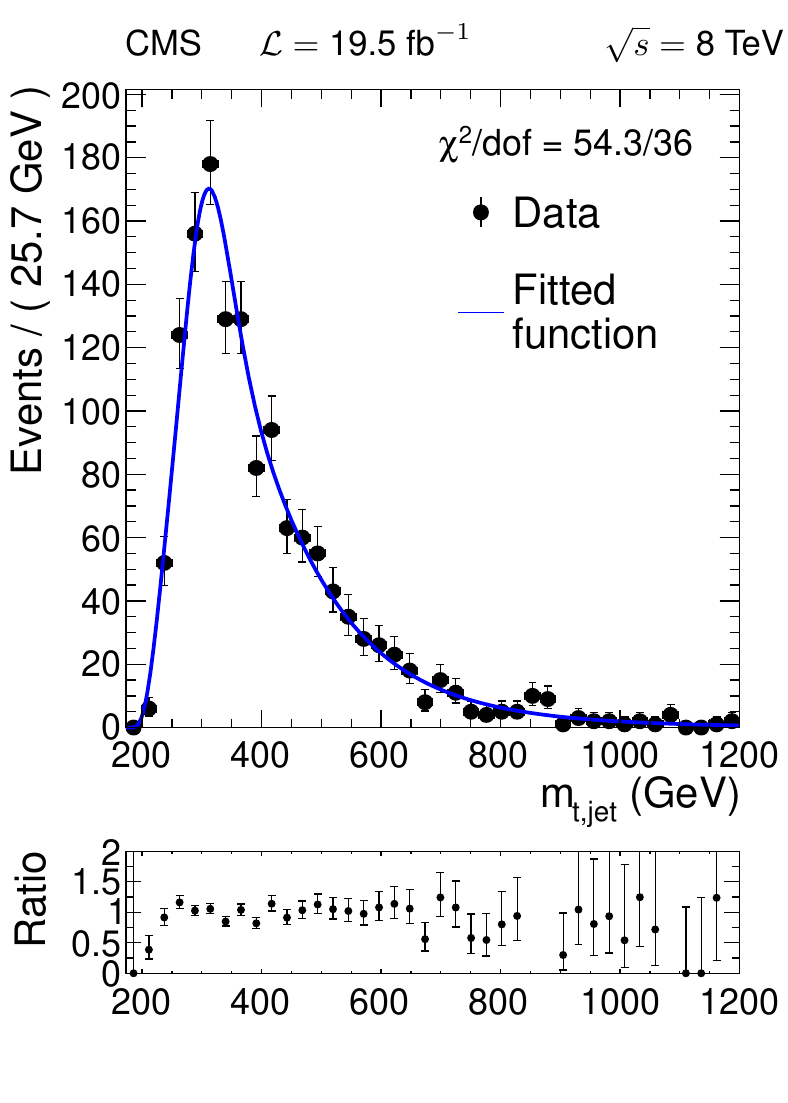}
\includegraphics[width=0.4\textwidth]{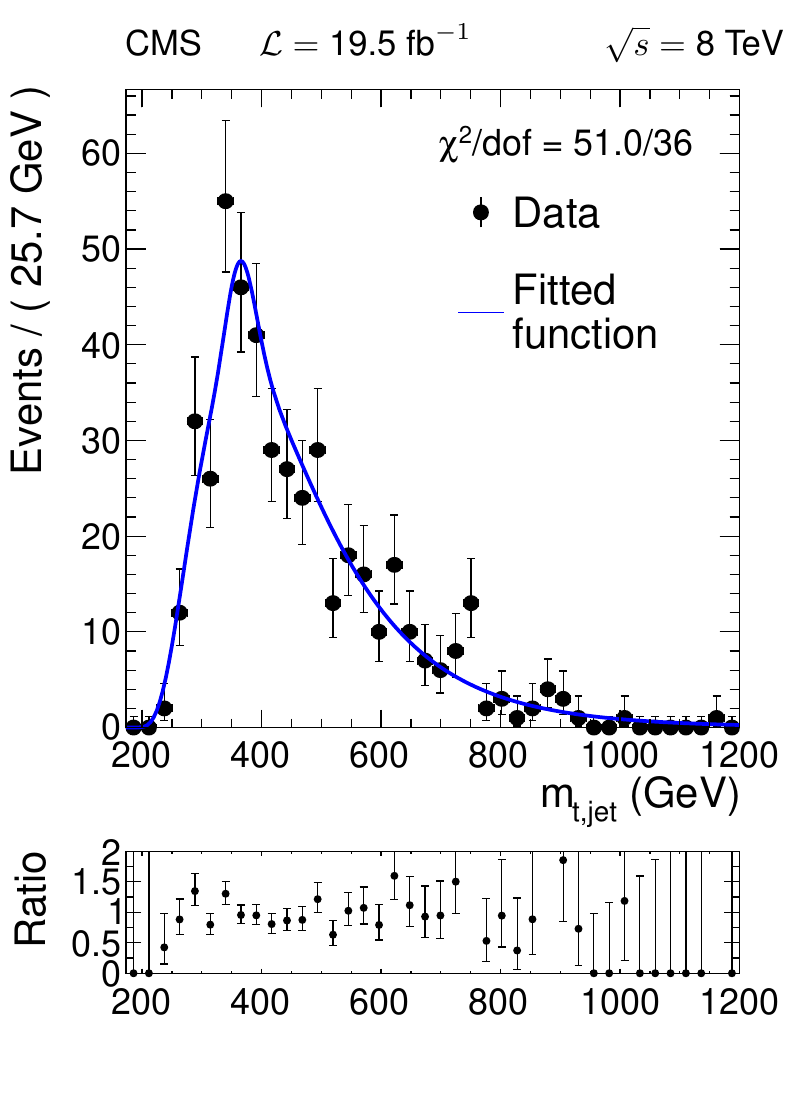}
\includegraphics[width=0.4\textwidth]{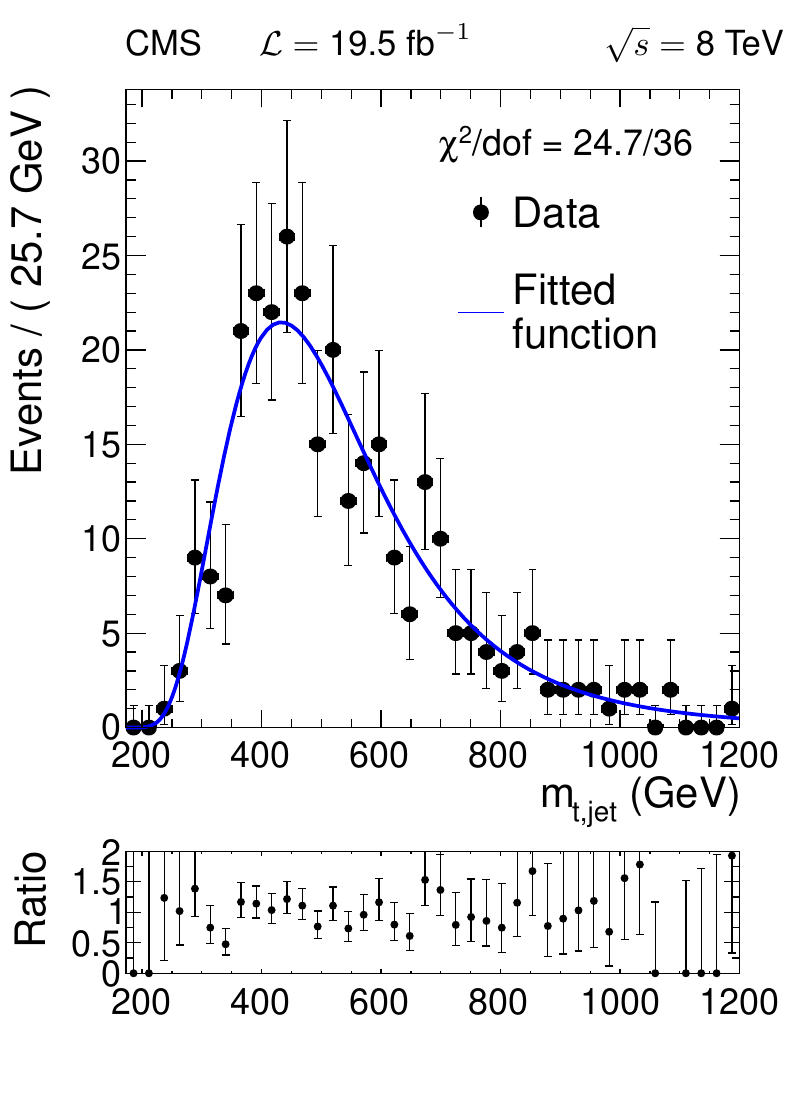}
\caption{Reconstructed invariant mass distributions for data together with the result of the likelihood function maximization with signal cross section set to zero for the four light-parton jet regions defined in Table~\ref{tab:fitRegions}: CR (upper left), SR1 (upper right), SR2 (lower left) and SR3 (lower right).}
\label{fig:invMassFit}
\end{figure*}

Figure~\ref{fig:invMassFit} shows the background-only hypothesis fits to the measured invariant mass spectra in the four light-parton jet regions (SR1--3 and CR).

\subsection{Systematic uncertainties}
\label{sec:systematics}

The fitted parameters of the background model are determined exclusively from data. Potential systematic uncertainties in the background model could arise from the choice of functional forms in Section~\ref{sec:lightJetFitting}.  We perform studies to ensure that the model is sufficiently general to approximate the true shape of the distributions, with any differences negligible compared to the statistical uncertainties.

The signal is modeled by a MC simulation and we consider several systematic uncertainties that can affect the light-parton jet $\pt$ spectrum and the mass reconstruction procedure. For each source of systematic uncertainty, we determine the change in the signal distribution parameters and encode this as a covariance matrix of all parameters. A multivariate normal distribution is constructed from the sum of all such covariance matrices and used to constrain the parameters of the signal distribution.  Allowing the invariant mass parameters to vary within this constraint has a negligible effect on the signal sensitivity. The final constraint function sets these parameters to their maximum likelihood values in signal simulation and does not allow them to vary.  The light-parton jet parameters and selection efficiencies are allowed to vary within their uncertainties.

The total selection efficiencies include the 10.9\% branching fraction~\cite{Khachatryan:2016mqs} of each of the top quarks to a leptonic final state and the production of tau leptons that decay to electrons or muons.
We estimate that in bottom squark pair production with a dileptonic final state, about 60\% of events produce leptons, $\PQb$ quarks, and light partons within the fiducial volume of the detector. About 85\% of these events have the physics objects reconstructed and about 65\% of those events have the jets from $\PQb$ quarks correctly tagged. The trigger efficiency is close to 95\%. Approximately 90\% of the remaining events pass all selection requirements.
The total selection efficiencies are approximately $2.0\times 10^{-3}$, $4.0\times 10^{-3}$, and $1.5\times 10^{-3}$ for the $\Pe\Pe$, $\Pe\mu$, and $\mu\mu$ channels, respectively, and have a slight dependence on the \PSQb mass.

Potential discrepancies between lepton efficiencies in data and simulation are taken into account varying lepton energies using scale factors, which are within 2\% of unity when the \pt is above 10\GeV~\cite{Khachatryan:2015hwa,Chatrchyan:2012xi}. For electrons the energy is varied by 1.5\% in the range $1.5<\abs{\eta}<2.5$ of the detector and 0.6\% in the range $\abs{\eta}<1.5$; for muons this value is 0.2\%.

To match the efficiency for tagging $\PQb$~jets and the mistag rate for charm and light-flavor quarks between data and simulation, a scale factor is applied to the simulation. There is a corresponding uncertainty introduced by using these scale factors. We apply an additional systematic uncertainty to account for the change in efficiency and mistag rates of the b tagging algorithm when applying our specific selection criteria.

Table~\ref{tab:systBreakdown} reports the uncertainty in the signal selection efficiency due to the different sources of systematic uncertainty for an example mass point. In combination these systematic uncertainties change the calculated upper limit on the cross section between 1\% and 10\%, depending on the mass point, compared to the upper limit calculated only with statistical uncertainties. The dominant systematic effect comes from variations of the jet energy scale.

\begin{table}[htbp]
\centering
\topcaption{\label{tab:systBreakdown}Relative systematic uncertainty in the signal selection efficiency
broken down by source of signal systematic uncertainty for a characteristic mass point.}
\begin{scotch}{lc}

Signal simulation uncertainty & $m_{\PSQb}=350\GeV$ \\

\hline

Heavy-flavor SF for \PQb~tagging &    4.9\%  \\
Light-flavor SF for \PQb~tagging &   4.7\%  \\
Jet energy scale                        &    4.6\%  \\
Signal MC statistics                    & 2.1\%  \\
Jet energy resolution                   &   1.8\%  \\
Pileup                                   &  1.5\%  \\
PDFs                   &  1.0\%  \\
MC \PQb~tagging efficiency for \PQb~jets       & 0.4\%  \\
MC \PQb~tagging efficiency for \cPqc~jets      & 0.3\%  \\
MC \PQb~tagging efficiency for light-parton jets  & 0.5\%  \\
Electron energy scale                    & 0.2\%  \\
Muon energy scale                        & $<$0.1\%  \\
Integrated luminosity & 2.6\% \\
\hline
Total & 9.2\% \\
\end{scotch}
\end{table}

\subsection{The likelihood function}
\label{sec:likelihood}

For both signal and background, we define a three-dimensional probability density function constructed using the two-dimensional light-parton jet distributions defined in Section~\ref{sec:lightJetFitting} and the invariant mass distributions defined in Section~\ref{sec:invMassSpectrum}.
These three-dimensional distributions can be written as:
\begin{subequations}
\begin{align}
\rho^\text{SM}_\text{3D}(m_{\PQt j},\pt^{(1)},\pt^{(2)}) &= \rho^\text{SM}_\text{mass}(m_{\PQt j} | \pt^{(2)}) \rho^\text{SM}_\text{2D}( \pt^{(1)},\pt^{(2)})\\
\rho^\text{signal}_\text{3D}(m_{\PQt j},\pt^{(1)},\pt^{(2)}) &= \rho^\text{signal}_\text{mass}(m_{\PQt j} | \pt^{(2)}) \rho^\text{signal}_\text{2D}( \pt^{(1)},\pt^{(2)}),
\end{align}
\end{subequations}
and the complete distribution is:
\begin{equation}
\rho^\text{total}_\text{3D} = (\mu^\text{SM} \rho^\text{SM}_\text{3D} +
\epsilon\mathcal{L} \sigma_\text{signal} \rho^\text{signal}_\text{3D})/(\mu^\text{SM}+\epsilon\mathcal{L} \sigma_\text{signal} ).
\end{equation}
Here $\mu^\text{SM}$ is the SM yield, $\epsilon$ is the signal efficiency, $\mathcal{L}$ is the total integrated luminosity, and $\sigma_\text{signal}$ is the signal cross section.

Constraints on the signal shape parameters are derived as described in Section~\ref{sec:systematics}. We write the constraint distribution as $\rho_\text{syst}$.  There are no constraints on the parameters describing the SM distribution.  We construct an extended unbinned likelihood function from our data and these distributions as
\ifthenelse{\boolean{cms@external}}{
\begin{multline}
L(\sigma_\text{signal},\theta) =\\ \rho_\text{syst} \frac{(\mu^\text{SM}+\epsilon\mathcal{L} \sigma_\text{signal})^N\exp{(-\mu^\text{SM}-\epsilon\mathcal{L} \sigma_\text{signal})}}{N!}\\
\times\prod_{i =0}^N \rho^\text{total}_\mathrm{3D}(m_{\PQt j,i},{\pt^{(1)}}_i,{\pt^{(2)}}_i),
\end{multline}
}{
\begin{equation}
L(\sigma_\text{signal},\theta) = \rho_\text{syst} \frac{(\mu^\text{SM}+\epsilon\mathcal{L} \sigma_\text{signal})^N\exp{(-\mu^\text{SM}-\epsilon\mathcal{L} \sigma_\text{signal})}}{N!}\prod_{i =0}^N \rho^\text{total}_\mathrm{3D}(m_{\PQt j,i},{\pt^{(1)}}_i,{\pt^{(2)}}_i),
\end{equation}
}
where $N$ is the number of events in our sample; $m_{\PQt j,i}$, ${\pt^{(1)}}_i$, and ${\pt^{(2)}}_i$ are, respectively, the mass, $\pt^{(1)}$, and $\pt^{(2)}$ of the $i$th event; $\sigma$ is the cross section for production of the bottom squark resonance pair; and $\theta$ is the set of all nuisance parameters included in $\rho_\text{syst}$.

\subsection{Results for the dilepton final state}

We observe consistency with the SM expectation and set limits on the cross section for each of the signal models. We construct unified intervals~\cite{1998PhRvD..57.3873F} on the signal cross section and observe only intervals with lower edges of zero. The upper edge of these intervals is considered as the upper limit. For each signal mass, pseudoexperiments generated using the background-only model are used to determine the distribution of upper limits in the absence of signal.  The median of each distribution, along with intervals containing the central 68\% and 95\% of each distribution, is found.

\begin{figure}[htbp]
\centering
\includegraphics[width=0.48\textwidth]{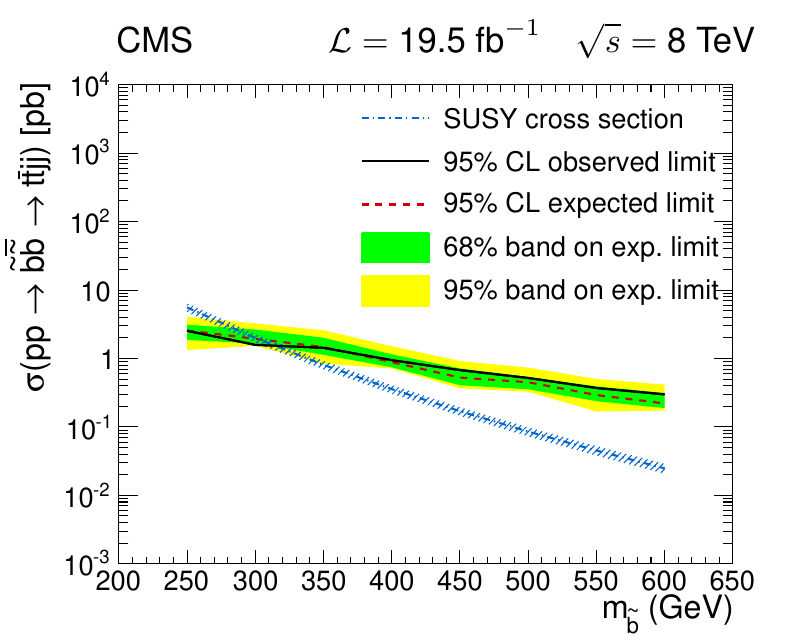}
\caption{Observed and expected 95\% \CL upper limits on the production cross section of $\PSQb$ pairs, where the $\PSQb$ decays to a top and down (strange) quark via the RPV coupling $\lambda''_{331}$ ($\lambda''_{332}$), as a function of $\PSQb$ mass derived using \CLs intervals.  The difference between observed and expected limits is correlated between neighboring signal points. The hatched region represents the theoretical uncertainty on the signal cross section.}
\label{fig:sbottomLimits}
\end{figure}

Figure~\ref{fig:sbottomLimits} shows the observed and expected $95\%$ \CL upper limits on the production cross section of $\PSQb$ pairs using \CLs intervals, assuming $\lambda''_{332}$ or $\lambda''_{331}$ is nonzero.
The $\lambda''_{332}$ and $\lambda''_{331}$ couplings are from the RPV SUSY Lagrangian defined in Eq.~(\ref{eqn:wrpv}) and control the branching fraction of the decay of a bottom squark to a top quark and a light down-type quark, \PQs~and \cPqd, respectively.
We find that the median expected lower limit on the $\PSQb$ mass is 295\GeV with the central 68\% of the limit distribution falling in the range 282--304\GeV. The measured 95\% \CL lower limit on the $\PSQb$ mass is 307\GeV.

\begin{figure*}[htbp]
\centering
\includegraphics[width=0.31\textwidth]{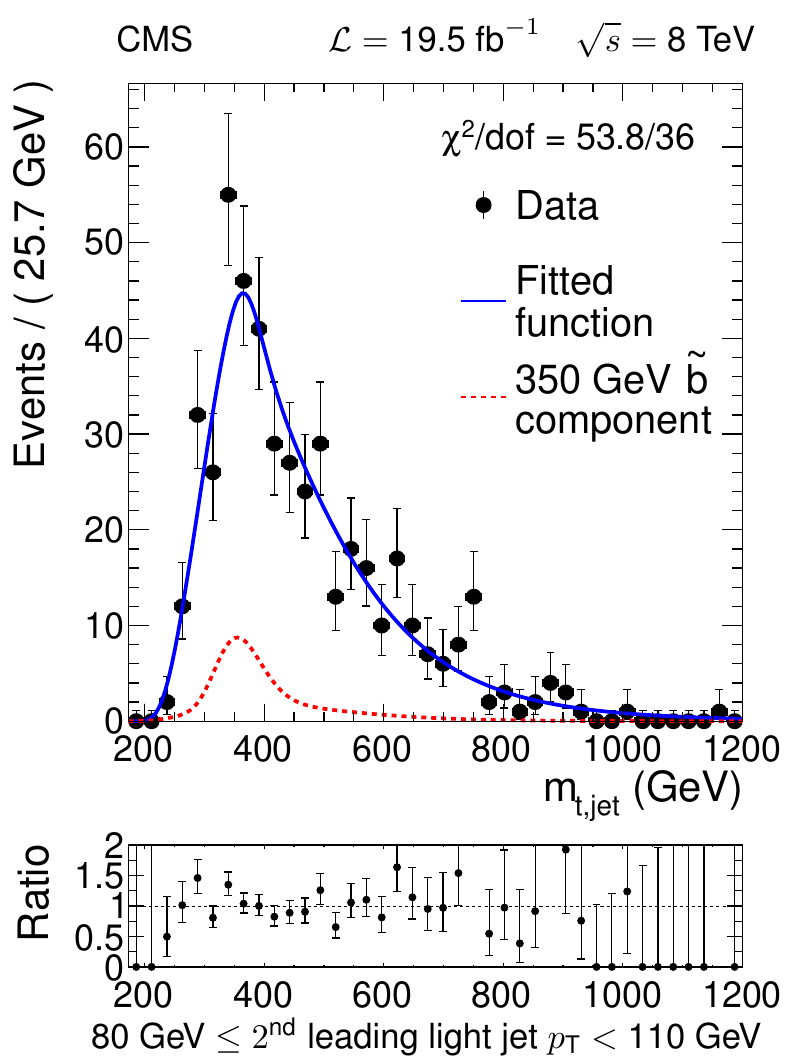}
\includegraphics[width=0.31\textwidth]{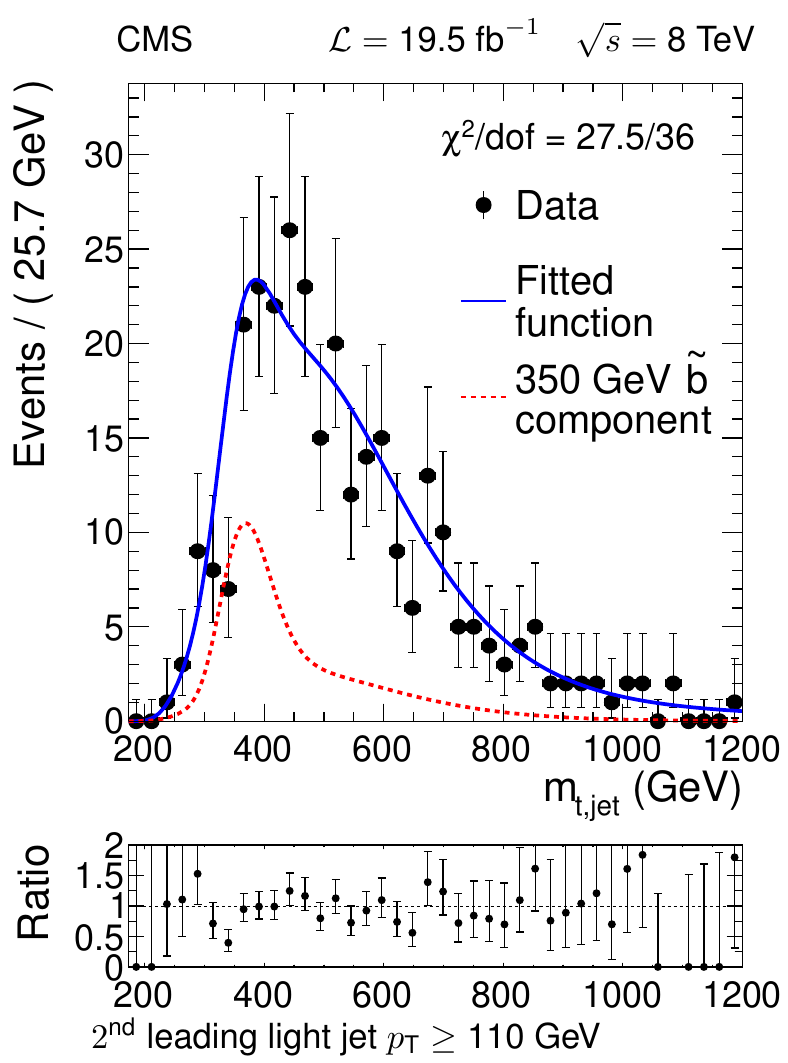}
\includegraphics[width=0.31\textwidth]{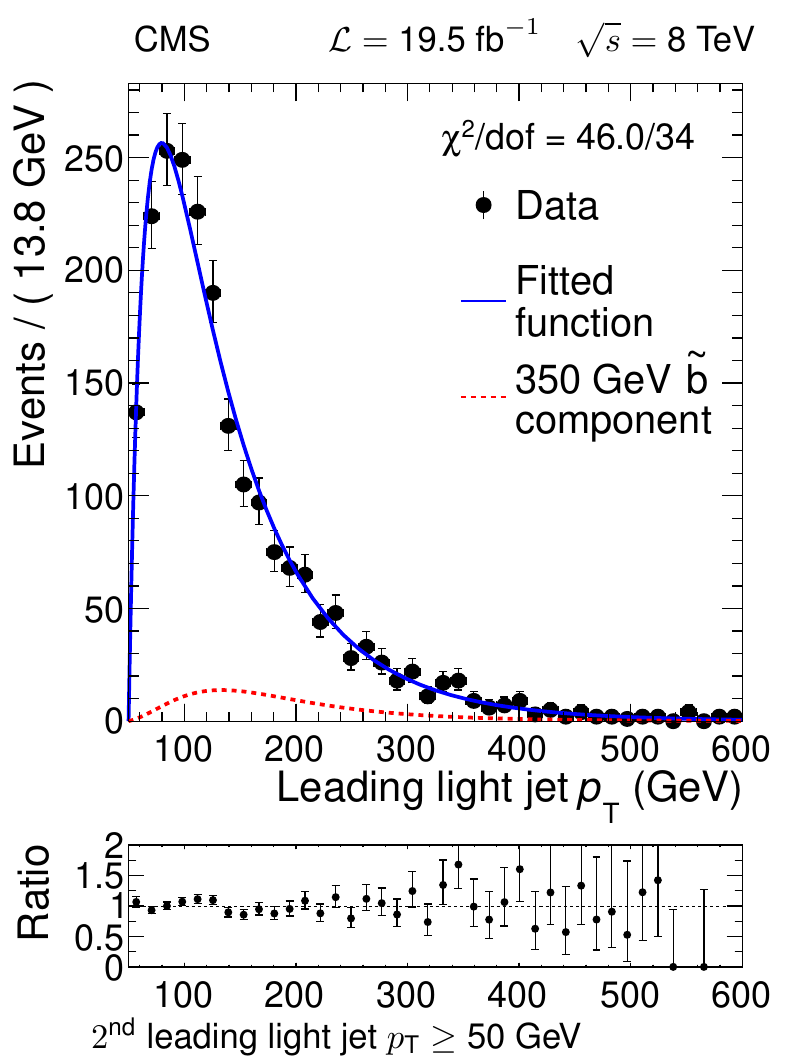}
\caption{Results of likelihood maximization with the signal cross section set to the calculated 95\% \CL upper limit
  of the 350\GeV $\PSQb$ point.  The solid line shows the fitted
function, the dashed line shows the signal component, and the points
show the data. The ratio of the data to the fitted function is also shown.
  From left to right, these are the invariant mass distribution in SR2, the invariant mass distribution in SR3, and
  the leading light jet $\pt$ distribution for events with $\pt^{(2)}>50\GeV$. }
\label{fig:sbottomPair350UpperLimit}
\end{figure*}

Figure~\ref{fig:sbottomPair350UpperLimit} shows the 350\GeV $\PSQb$ point where the invariant mass distributions peak in similar locations to the background and the search sensitivity is mainly attributable to the two-dimensional light-parton jet distribution.

The limits at large masses ($m_{\tilde \PQb} > 400$\GeV) are correlated as the signal distributions are all concentrated in the SR3 region and the invariant mass shapes have a large overlap.

\section{Four-lepton final state via strong production}
\label{sec:multilepton}
Multilepton final states, having clean identification criteria and well-understood background sources, present an important sector for searching for BSM physics. The SUSY models that contain leptonic or semileptonic RPV couplings produce multilepton events with very little \MET.

\begin{figure}[!htp]
  \centering
    \includegraphics[width=0.48\textwidth]{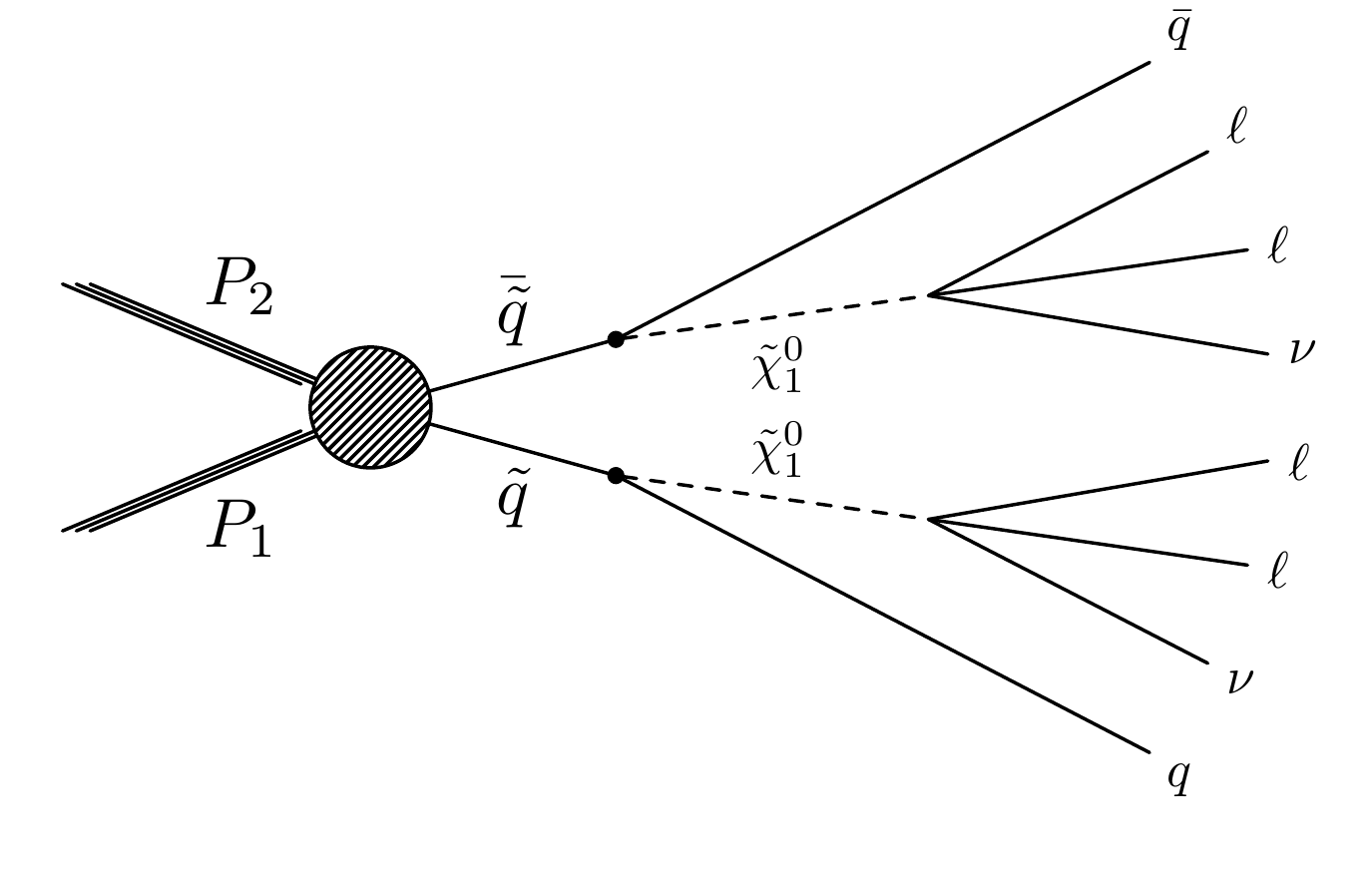}
    \includegraphics[width=0.48\textwidth]{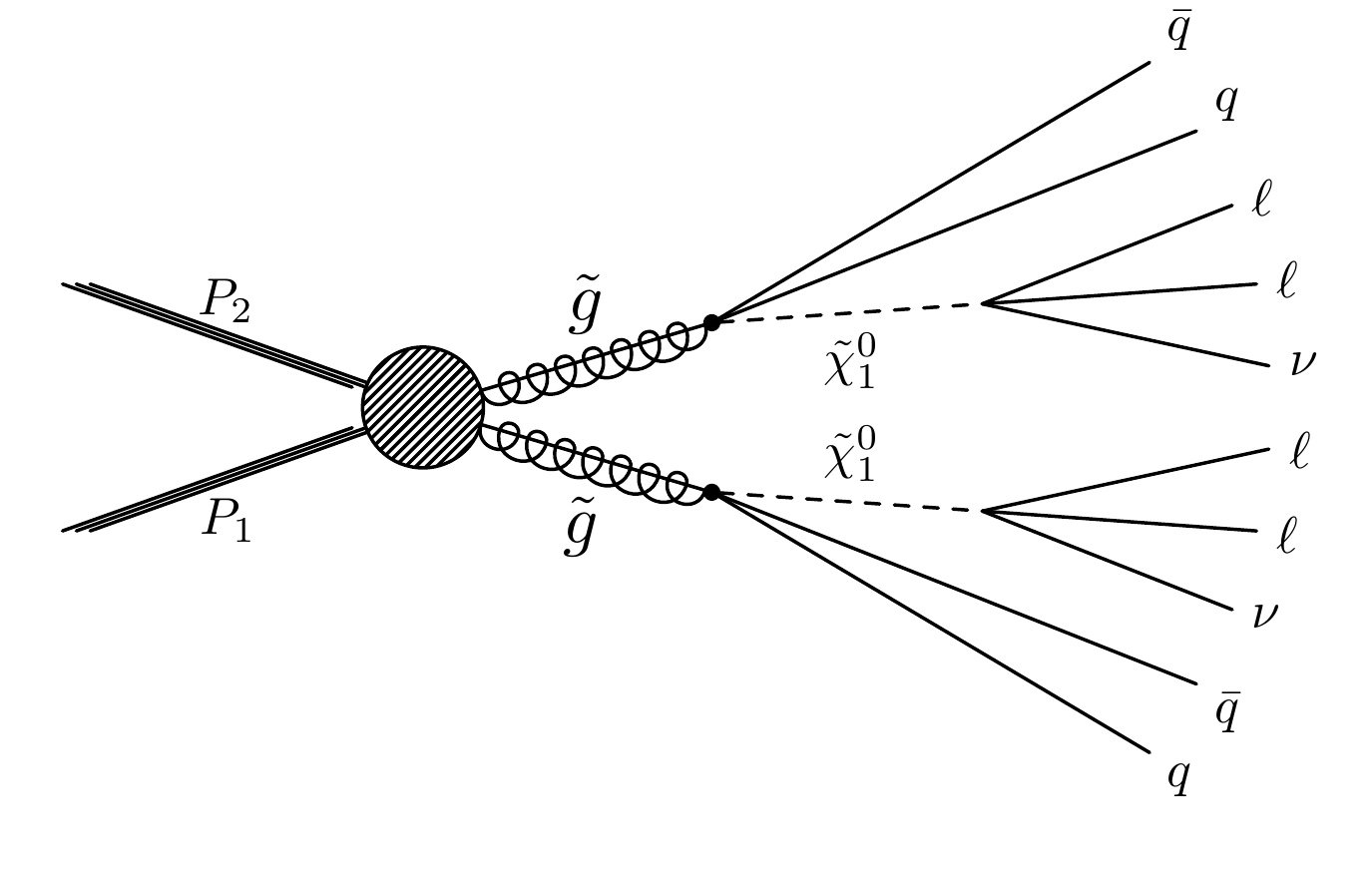}
    \caption{
      Diagrams of two simplified models with RPV~\cite{Alves:2011wf}.
      \cmsLLeft: A simplified model featuring squark pair production,
      with $\PSQ\to\cPq \PSGczDo$, and $m(\PSg) \gg m(\PSQ)$.
      \cmsRRight: A simplified model featuring gluino pair production,
      with $\PSg\to\cPq \cPaq \PSGczDo$, and $m(\PSQ) \gg m(\PSg)$.
      In both models the neutralinos decay to two charged leptons and a neutrino via a RPV term.
    }
    \label{fig:T2LRPV}
\end{figure}

In this section we describe a search designed explicitly to look for evidence of nonzero leptonic RPV couplings $\lambda_{122}$ and $\lambda_{121}$ in models with a winolike neutralino and different types of strong SUSY production: gluino pairs, top-squark pairs and squark pairs composed of an equal mixture of up, down, strange, and charm. Examples of this production are shown in the diagrams in Fig.~\ref{fig:T2LRPV}. In these events, the final state includes four light charged leptons (electrons and muons), and two neutrinos from the RPV decay of the neutralinos, as well as two or more jets from cascade decays of the strongly produced superpartners. The four leptons have a total charge of zero.

The main SM contribution to the multilepton final state comes from processes with $\Z$ bosons decaying into lepton pairs. In contrast, the four leptons produced in RPV neutralino decays are not expected to accumulate at the $\Z$ boson mass in dilepton spectra. The leptons from both the SM and RPV decays are produced at the proton-proton interaction vertex; we refer to these leptons as prompt.

We select events from a dilepton-triggered data set with exactly four isolated light leptons containing at least one opposite-sign, same-flavor (OSSF) lepton pair. For events with four isolated light leptons, the dilepton triggers are close to 100\% efficient.  
We define $M_1$ as the invariant mass of the OSSF lepton pair.  If there are two OSSF lepton pairs in the event, $M_1$ is the mass of the pair that is closest to the $\cPZ$ boson mass of 91\GeV.
The invariant mass of the remaining lepton pair in the event is defined as $M_2$.

Then we classify each mass according
to:
\begin {itemize}
\item ``below \Z'': $M<75$\GeV;
\item ``at \Z'': $75<M<105$\GeV;
\item ``above \Z'': $M>105$\GeV.
\end{itemize}

\begin{figure*}[tbh]
  \centering
    \includegraphics[width=0.4\textwidth]{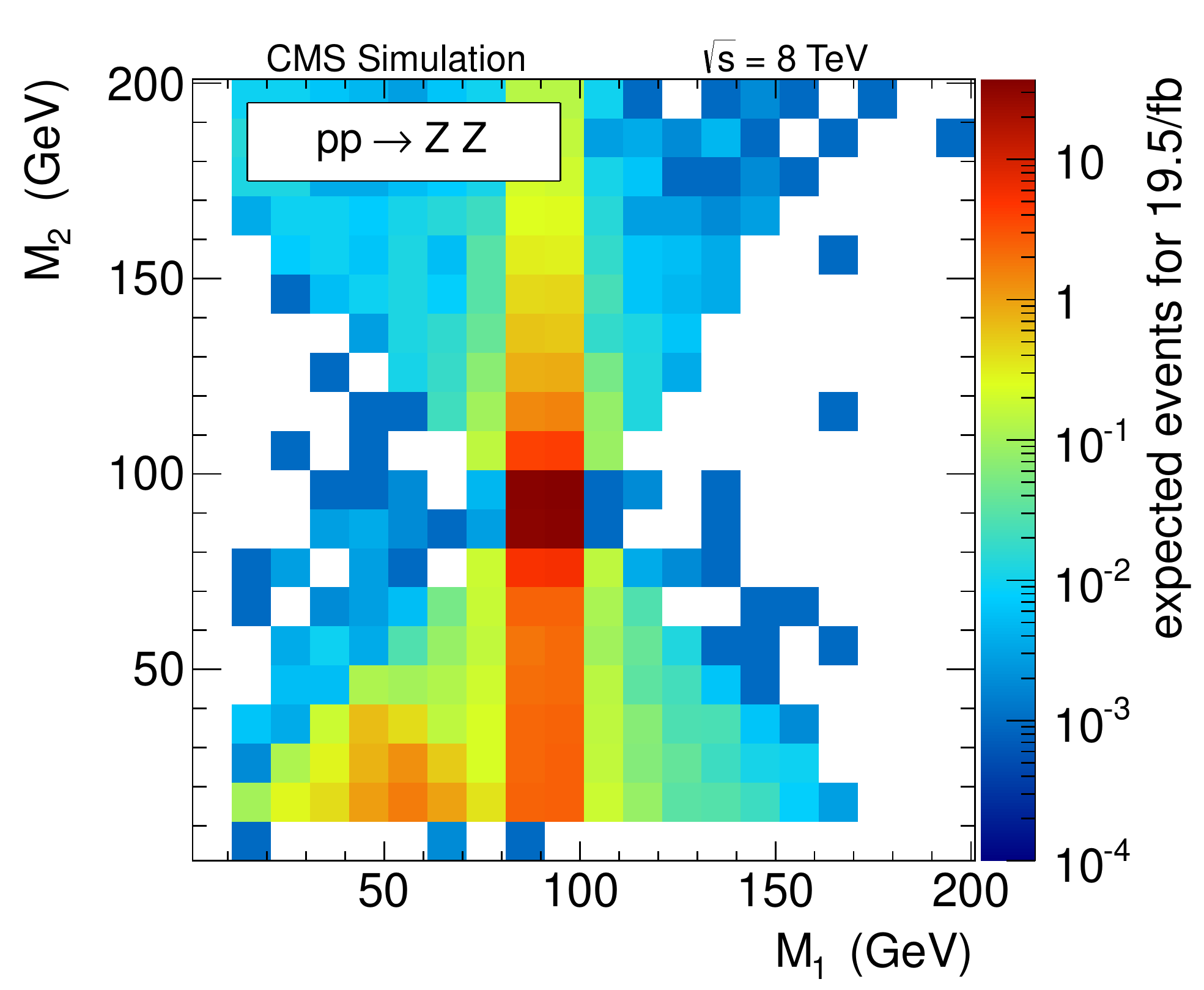}
    \includegraphics[width=0.4\textwidth]{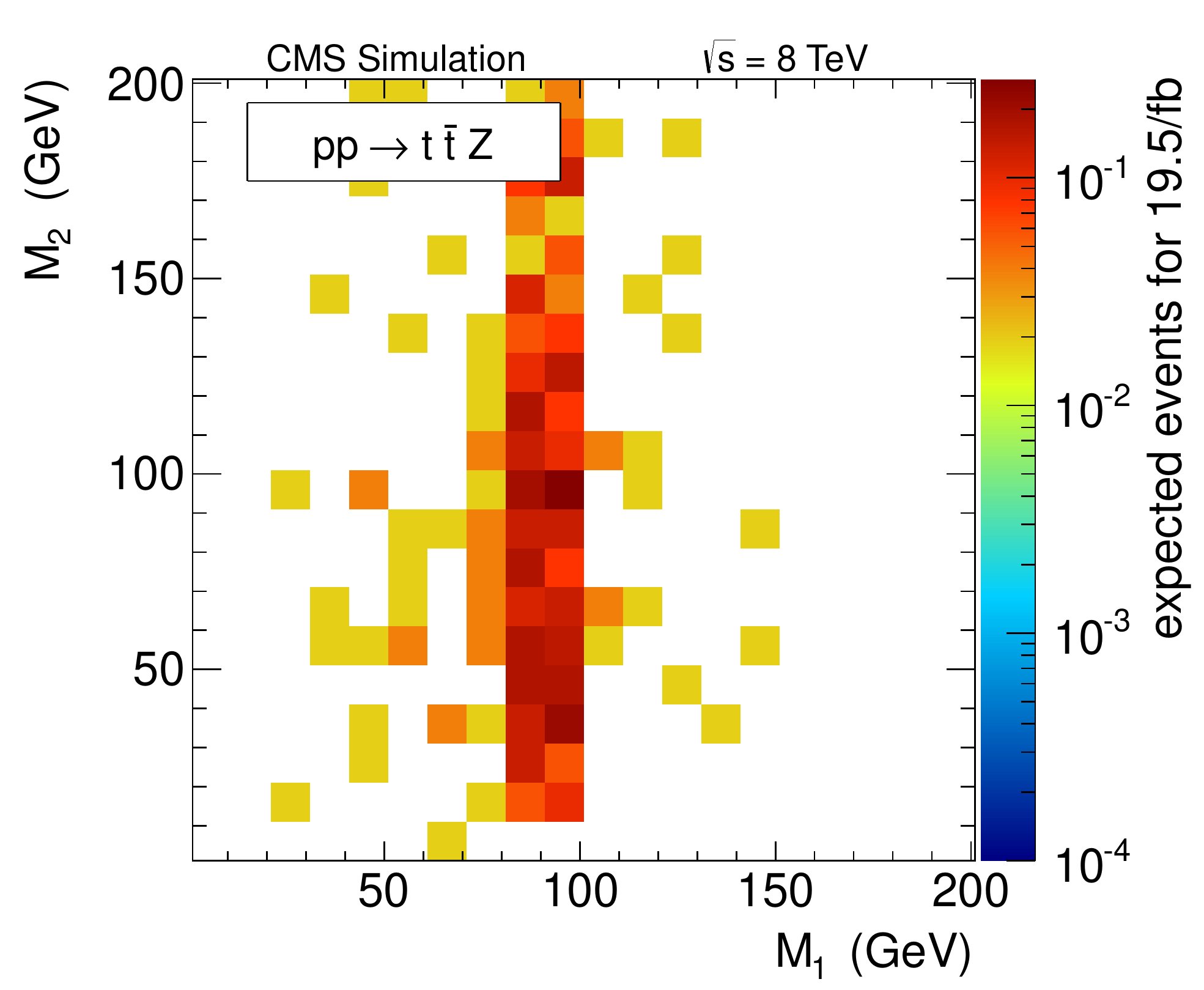}
    \includegraphics[width=0.4\textwidth]{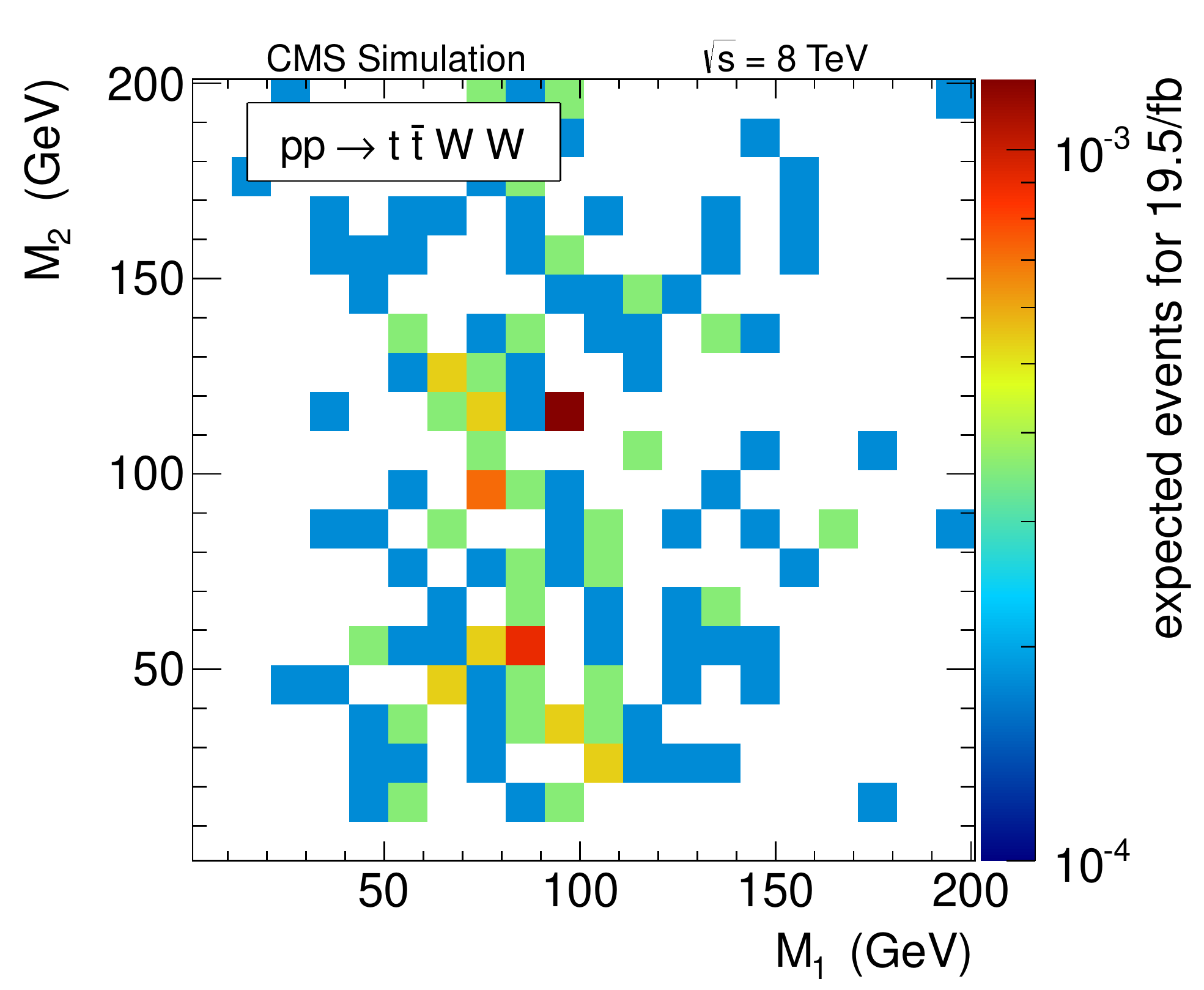}
    \includegraphics[width=0.4\textwidth]{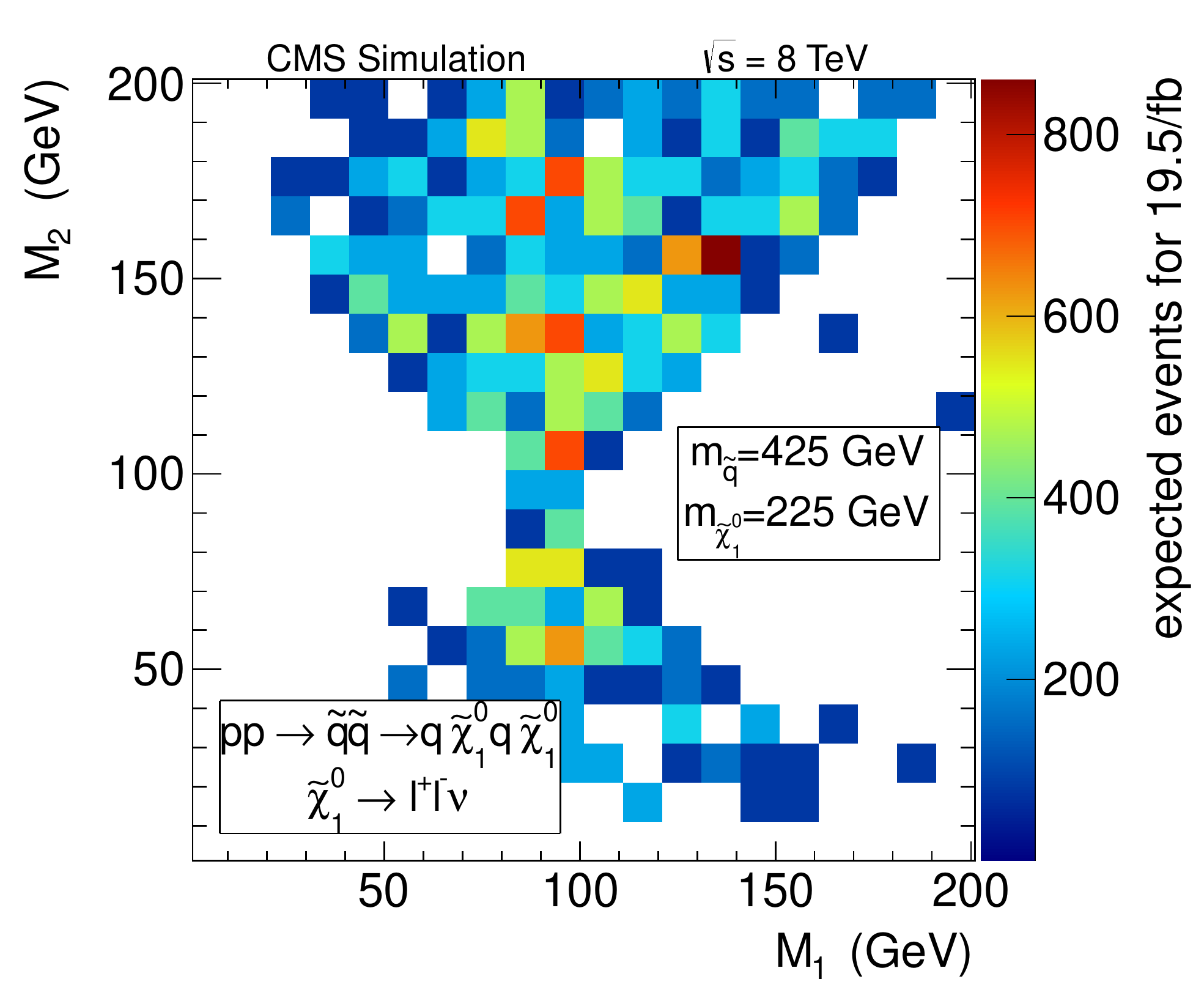}
    \caption{Expected shapes of the $(M_1, M_2)$ distribution for the $\Z\Z$ background (upper left), $\ttbar\Z$ background (upper right), $\ttbar\PW\PW$ background (lower left), as well as for the  model of Fig.~\ref{fig:T2LRPV} (\cmsLeft) with $m_{\PSQ}=425$\GeV and $m_{\PSGczDo}=225$\GeV (lower right).}
    \label{fig:ZZM1M2}
\end{figure*}

This defines nine regions reflecting different kinds of resonant and nonresonant four-lepton production. Figure~\ref{fig:ZZM1M2} presents the distribution of expected yields in the $(M_1, M_2)$ plane for different SM processes and for an example RPV scenario. Based on the occupancy of different regions for typical $\Z\Z$ production events, we define our signal region as ``$M_1$ above \Z'' or  ``$M_1$ below \Z and $M_2$ above \Z.''

\subsection {Backgrounds}
\label{sec:backgrounds}
The presence of four prompt leptons in the final state is a strong discriminant.
The SM processes contributing to this signature are:
\begin {itemize}
\item processes producing exactly four prompt leptons: $\Z\Z^{(*)}$ and $\ttbar\Z$;
\item more rare processes producing four or more prompt leptons: $\ttbar\PW\PW$, $\PW\PW\Z$, $\PW\Z\Z$, and $\Z\Z\Z$;
\item processes such as $\PW\Z$+jets, $\ttbar\PW$+jets, and $\ttbar\Z$ that can produce three genuine prompt leptons and one candidate (nonprompt lepton) that arises primarily from the decay of a hadron of heavy flavor or from particle misidentification;
\item Drell--Yan production with two extra nonprompt leptons.
\end {itemize}

We use MC simulated samples to estimate the backgrounds with four or more genuine prompt leptons. The primary background contribution is $\Z\Z$ production. We normalize this sample to the $\Z\Z$ production cross section measured by CMS \cite{Chatrchyan:2013oev}, so our observation in the region with both $M_1$ and $M_2$ at \Z  is correlated with that measurement. Based on the consistency of our observations with the predictions in control regions, we assign a systematic uncertainty of $25\%$ to the MC predictions of $\Z\Z$ production contributions.
The yields of the remaining background sources with four or more genuine leptons are estimated from MC simulation normalized to the theoretical cross sections, which are known to 5--10\%.
This uncertainty, combined with the $25\%$ systematic uncertainty in the mismodeling observed in the $\Z\Z$ sample, motivates the assignment of a conservative systematic uncertainty of $50\%$ to background contributions of other rare processes.

\begin{figure}[tbh]
  \centering
    \includegraphics[width=0.48\textwidth]{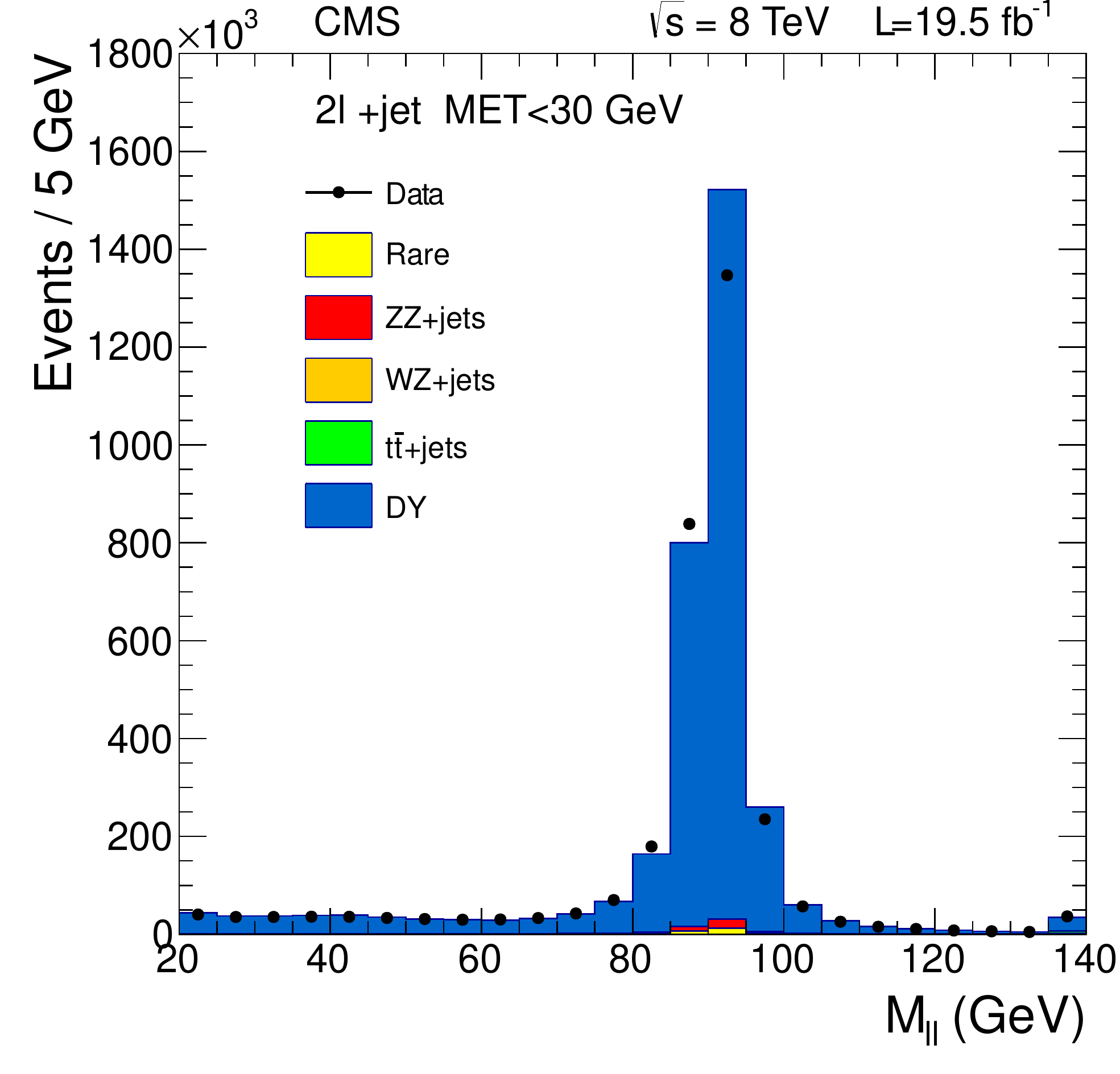}
    \includegraphics[width=0.48\textwidth]{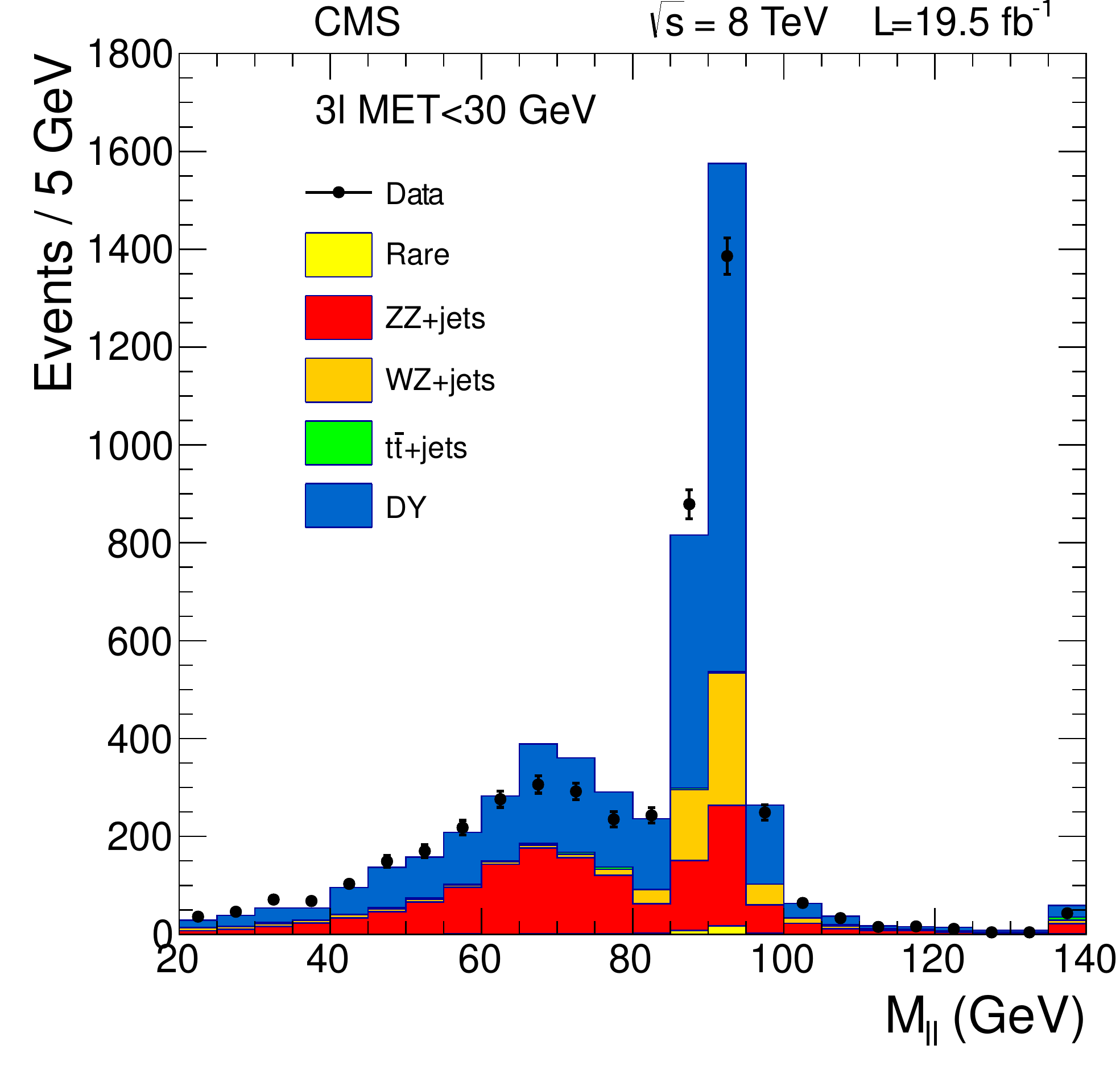}
    \caption {
      Invariant mass of OSSF lepton pairs in events with $\MET < 30$\GeV. Data are overlaid with contributions from different sources, predicted by simulation. \cmsLLeft: Events with two isolated leptons and exactly one jet with $\pt > 30$\GeV. \cmsRRight: Invariant mass of the two OSSF isolated leptons closest to the $\Z$ boson mass in events with three isolated leptons.
}
    \label{fig:DY_fakes}
\end{figure}

We estimate the contribution of nonprompt leptons with a sample of events containing an OSSF lepton pair with the requirement $\MET < 30$\GeV, to increase the purity of the Drell-Yan control sample. The left plot in Fig.~\ref{fig:DY_fakes} presents the dilepton mass distribution of these events in which a single jet with $\pt > 30$\GeV is present. The right plot presents the OSSF dilepton mass distribution for events in which a third isolated lepton is present. By comparing the yield of events in the \Z peak in the left plot with that of events in the \Z peak less the expected prompt lepton contribution in the right plot, we obtain a jet-to-lepton nonprompt rate of $(0.103 \pm 0.003\stat)\%$.

\begin{figure}[tbh]
  \centering
    \includegraphics[width=0.45\textwidth]{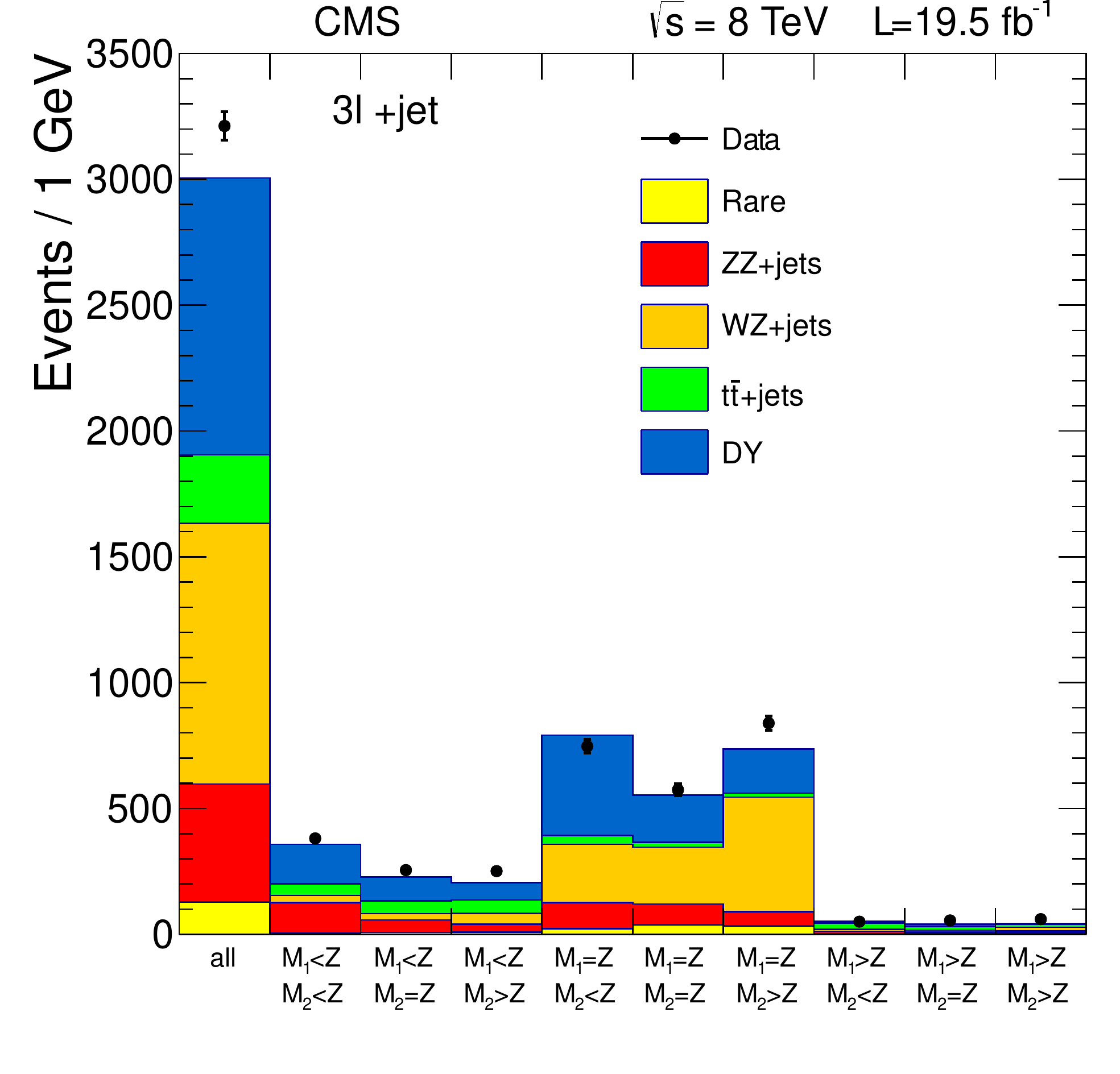}
    \caption {Number of events with three isolated leptons and one jet to different $(M_1, M_2)$ regions. Data are overlaid with predicted contributions from different sources. }
    \label{fig:4L_fakes}
\end{figure}

Figure~\ref{fig:4L_fakes} shows the contribution of events with three isolated leptons and one jet to the different regions of this search.
These numbers are used as a reference for determining the contribution of nonprompt leptons to the four-lepton selection using the nonprompt rate evaluation of 0.103\%.
The jet momentum is used to calculate $M_2$ and classify the events in Fig.~\ref{fig:4L_fakes}.
As the actual nonprompt lepton takes only a fraction of the heavy-flavor jet momentum, this procedure overestimates $M_2$, which brings significant systematic uncertainty into this background estimation.
We account for this by assigning a 100\% systematic uncertainty to this contribution.
Despite this large uncertainty, we tested that varying the estimation of this background down to zero keeps the exclusion result stable within 1\%.

\subsection{Observations}
\label{sec:observations}

Table \ref{table:backgroundsAll} shows the observed number of events in different regions together with the expectations from SM processes. The observations are consistent with the SM background expectations. Based on the observations in the search region (``$M_1$ above \Z'', or ``$M_1$ below \Z and $M_2$ above \Z''), the 95\% \CL upper limit on cross section times integrated luminosity times efficiency ($\sigma\mathcal{L}\varepsilon$) for any physics process beyond the SM contributing to this search region is 3.4 events. The expected upper limit for this observation is 4.7 events.

\begin{table*}[!htp]
  \centering
  \topcaption{\label{table:backgroundsAll}Expected background contributions from different SM sources and experimentally observed events in all analysis regions. The $\Z\Z$ prediction in the region with both $M_1$ and $M_2$ falling in the ``at \Z'' region is based on simulation normalized to the CMS $\Z\Z$ production cross section measurement and is therefore correlated with the observation in this analysis. The uncertainties take into account statistical and systematic contributions, combined quadratically. Mass ranges are given in \GeVns.}
    \begin{scotch}{clccc}
      \multicolumn{2}{c}{} & $M_1<75$ & $75<M_1<105$ & $M_1>105$ \\ \hline
                      & $\Z\Z$   & 10$\pm$2    & 32$\pm$8         & 1.0 $\pm$0.2 \\
                      & Rare & 0.3$\pm$0.1& 3$\pm$1      & 0.01$\pm$0.01 \\
      $M_2<75$        & Nonprompt& 0.3$\pm$0.3  & 0.8$\pm$0.8      & 0.06$\pm$0.06 \\
                      & All backgrounds & 10$\pm$2   & 35$\pm$8  & 1.0$\pm$0.2 \\
                      & Observed        & 14  & 30 & 1 \\ \hline
                      & $\Z\Z$   & 0.10$\pm$0.03 & 150       & 0.05$\pm$0.01 \\
                      & Rare & 0.12$\pm$0.05& 3$\pm$1      & 0.06$\pm$0.03 \\
   $75<M_2<105$       & Nonprompt& 0.3$\pm$0.3  & 1$\pm$1      & 0.05$\pm$0.05 \\
                      & All backgrounds & 0.5$\pm$0.3& 153 & 0.16$\pm$0.06 \\
                      & Observed        & 0           & 160        & 0 \\ \hline
                      & $\Z\Z$   & 0.8$\pm$0.2& 15$\pm$4       & 0.30$\pm$0.07 \\
                      & Rare & 0.3$\pm$0.1& 3$\pm$1      & 0.12$\pm$0.05 \\
      $M_2>105$       & Nonprompt& 0.4$\pm$0.4  & 0.7$\pm$0.7      & 0.05$\pm$0.05 \\
                      & All backgrounds & 1.4$\pm$0.5 & 18$\pm$4 & 0.5$\pm$0.1 \\
                      & Observed        & 0           & 20         & 0  \\
    \end{scotch}
\end{table*}

\subsection{Generalized efficiency}
\label{sec:acceptance}
To understand how our results can be applied to a generic SUSY model with RPV, we study how the signal efficiency varies across about 7300 sets of RPC phenomenological MSSM (pMSSM) \cite{Djouadi:1998di}  model parameters, each one containing 10000 events, selected to fulfill different pre-CMS observations~\cite{Khachatryan:2016nvf}.

The RPV leptonic decay of the pair of neutralinos leads to four prompt leptons. The kinematics of these leptons are generally driven by the momentum distribution of the decaying neutralinos and by their masses. In most of the SUSY scenarios the lepton momentum is well above the required threshold, which results in high efficiency before the isolation requirement. The isolation efficiency depends on the hadronic activity in the event.

We find that nearly all models have a four-lepton isolation efficiency in the range 50--100\%. Therefore we consider an efficiency band spanning this range, using a 30\% uncertainty when combining with other uncertainties.

\subsection{Interpretations of the four-lepton results}
\label{sec:interpretations}
Once an upper limit on $\sigma\mathcal{L}\varepsilon$ is extracted from the observations,
and the efficiency is evaluated, the corresponding limit on the cross section, $\sigma^\mathrm{SUSY}_\text{total}$ can be calculated.

\begin{figure}[tbh]
  \centering
    \includegraphics[width=0.45\textwidth]{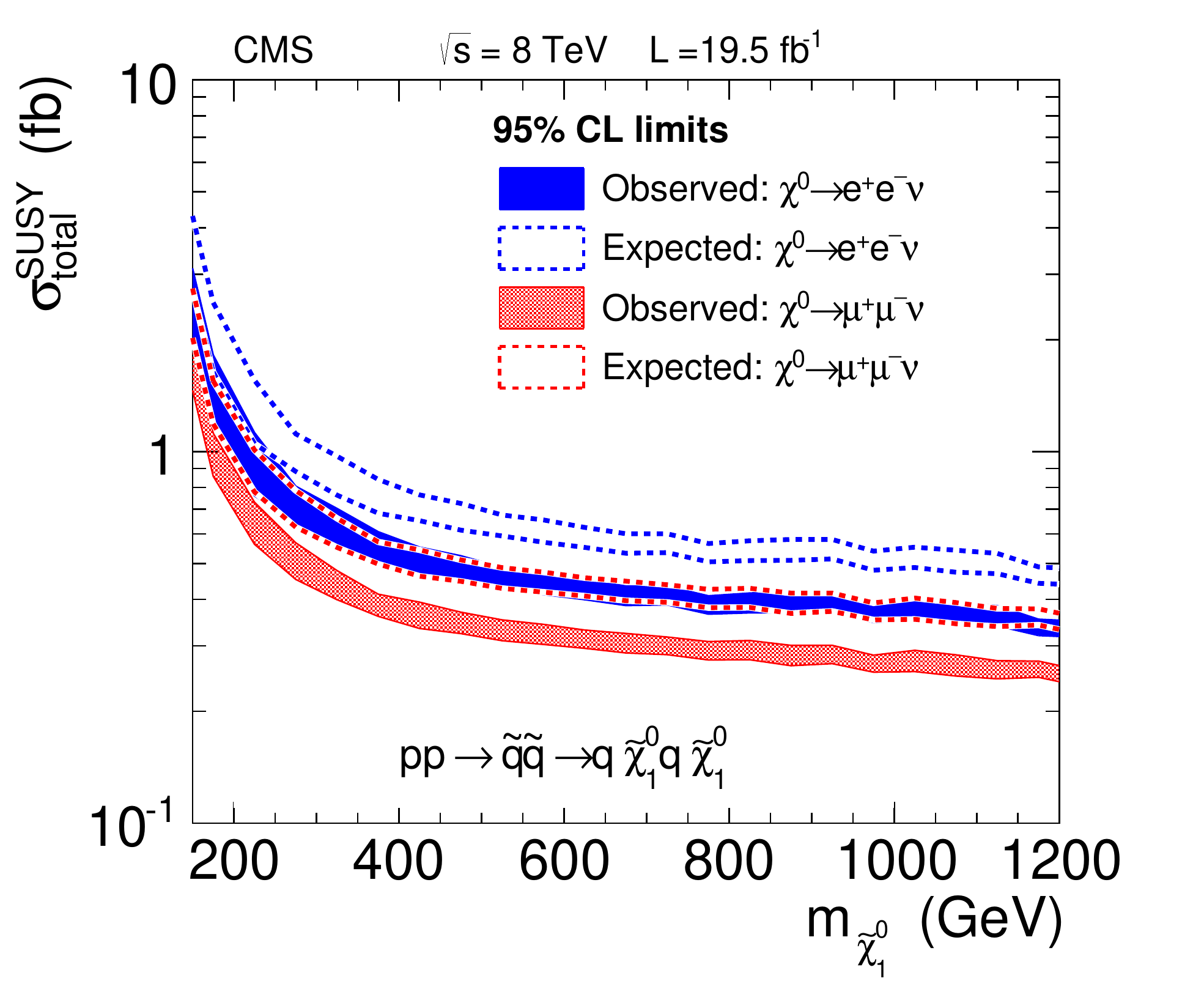}
    \includegraphics[width=0.45\textwidth]{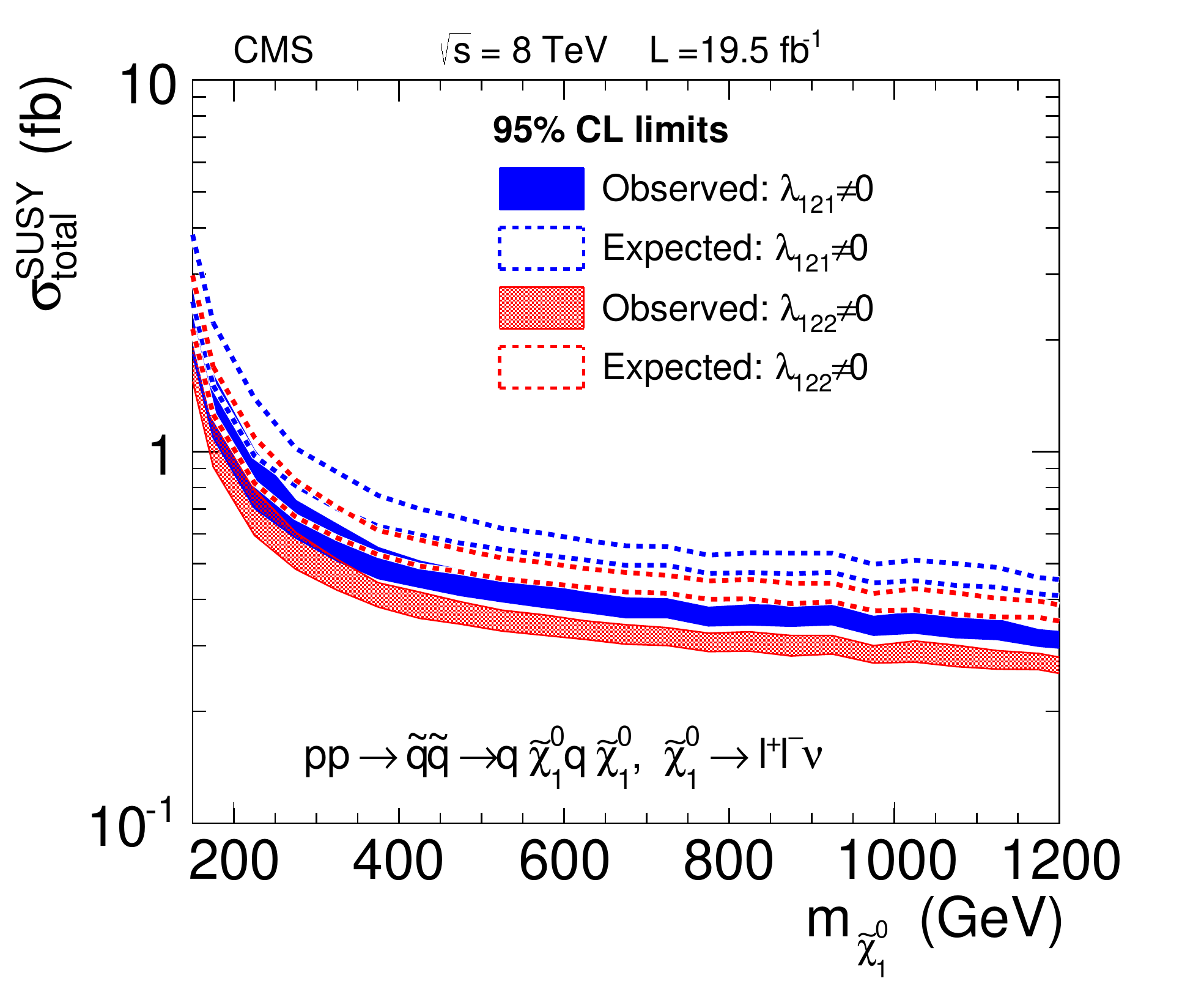}
    \caption{The 95\% \CL upper limit on the total cross section of the process sketched in Fig.~\ref{fig:T2LRPV} (\cmsLeft), as a function of the mass of the RPV decaying neutralino.
   The band corresponds to the component of uncertainty from the lepton efficiency.
   Results are shown for neutralinos decaying exclusively to electrons or muons (\cmsLeft) and decaying with the appropriate lepton flavor mixture corresponding to $\lambda_{121} \neq 0$ and
   $\lambda_{122} \neq 0$ RPV scenarios (\cmsRight).
    }
    \label{fig:T2sigmaBand}
\end{figure}

The cross section exclusion limits for squark pair production with leptonic RPV SUSY (Fig.~\ref{fig:T2LRPV}, \cmsLeft) are presented in Fig.~\ref{fig:T2sigmaBand}. They are obtained using the exclusion limit on the number of events in the two signal regions. Here the bands show only the component of the uncertainty attributable to the lepton ID efficiencies. The edges of the bands correspond to the two efficiency profiles, that is, the width reflects the uncertainty introduced by changing the amount of hadronic activity in the signal model. The left plot of Fig.~\ref{fig:T2sigmaBand} presents
results for neutralinos decaying exclusively to electrons or muons. The right plot takes the
appropriate mixture of electrons and muons in neutralino decays for $\lambda_{121}\neq0$ and $\lambda_{122}\neq0$ cases into account.

\begin{figure}[tbh]
  \centering
    \includegraphics[width=0.45\textwidth]{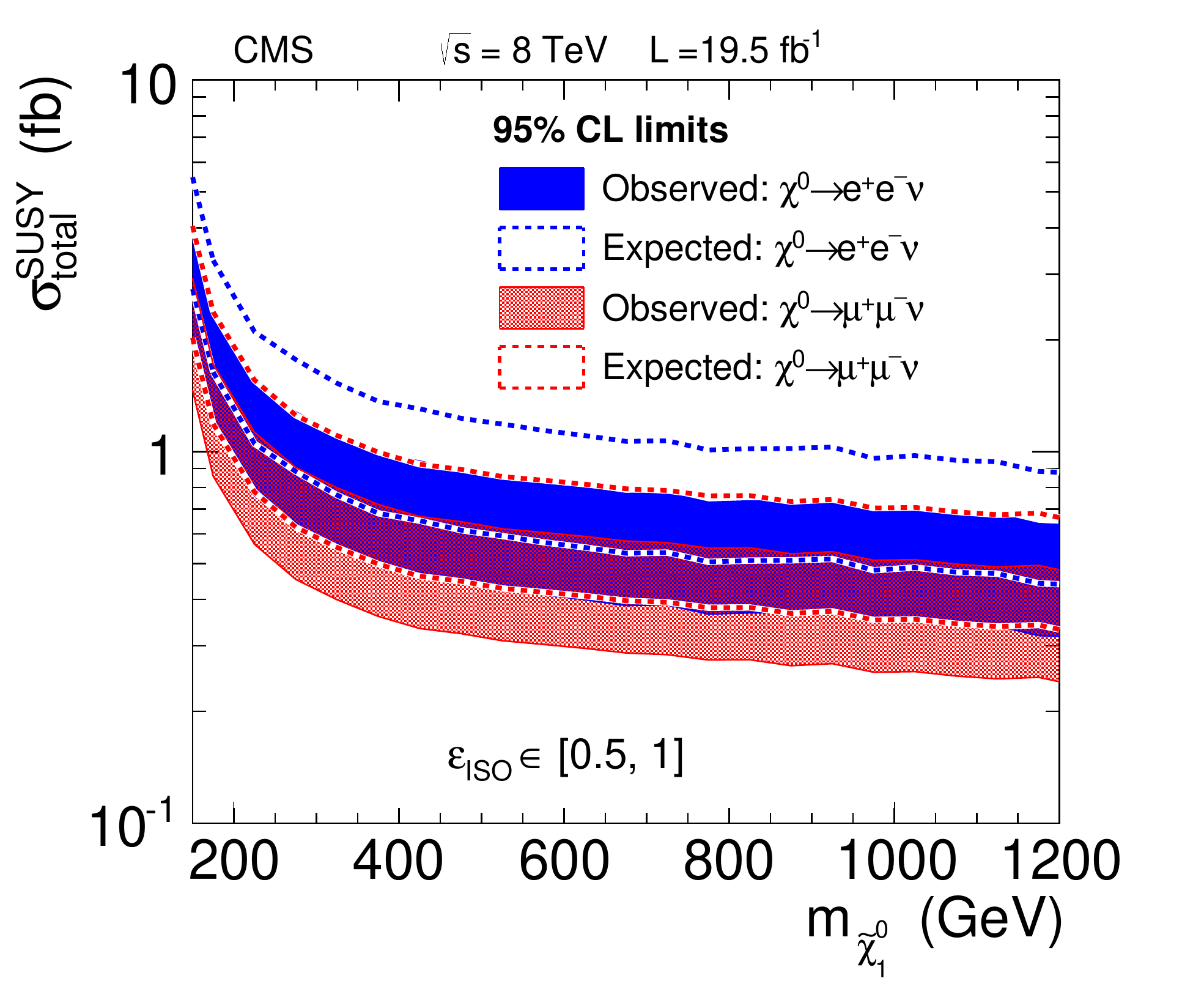}
    \includegraphics[width=0.45\textwidth]{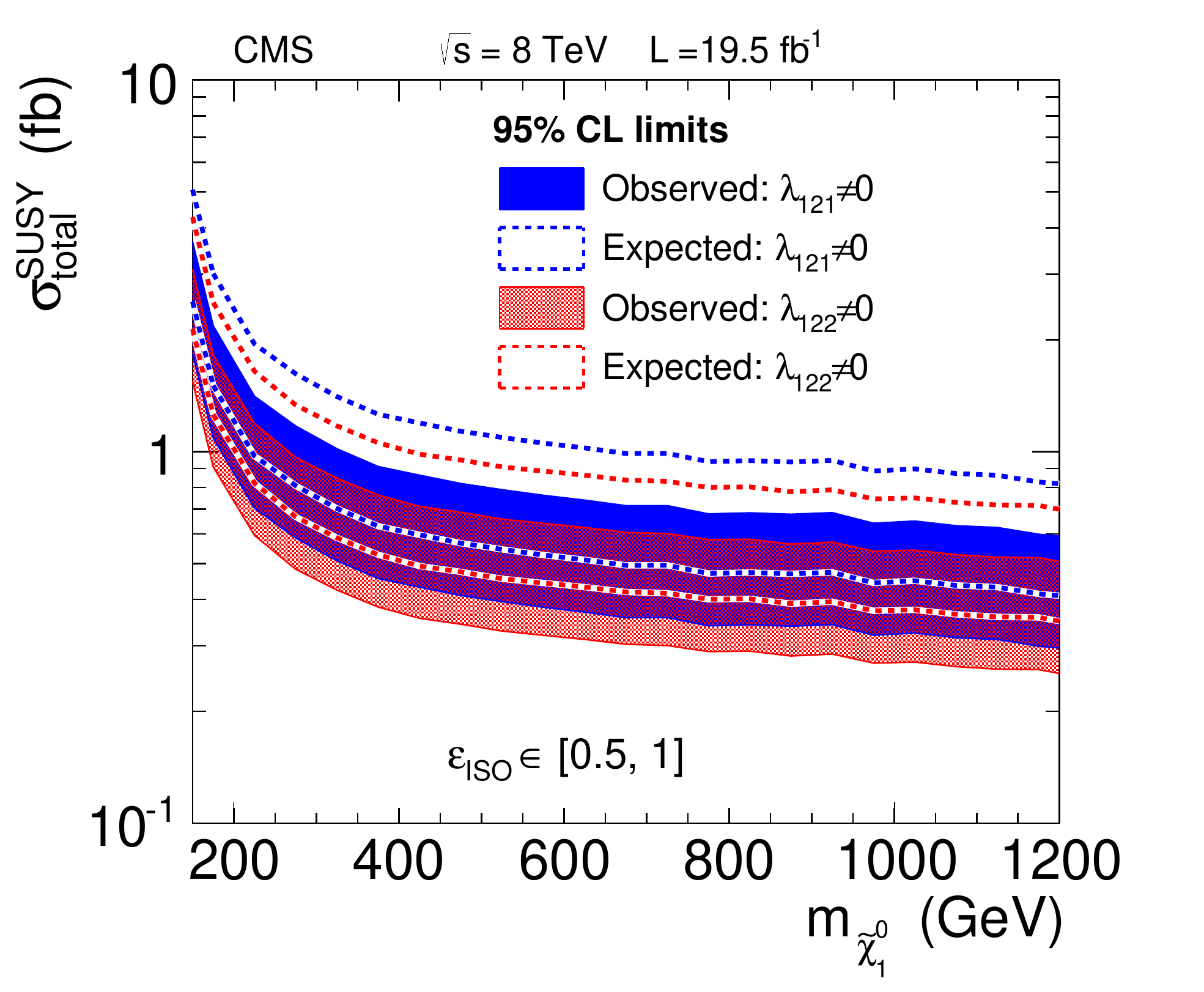}
    \caption{The 95\% \CL upper limit on total cross sections for generic SUSY models.
   The band corresponds to the efficiency uncertainty from the various pMSSM models.
   Results are shown for neutralinos decaying exclusively to electrons or muons (\cmsLeft) and decaying with the appropriate lepton flavor mixture corresponding to $\lambda_{121} \neq 0$ and
   $\lambda_{122} \neq 0$ RPV scenarios (\cmsRight).
    }
    \label{fig:sigmaBand}
\end{figure}

The experimental observations described in Section~\ref{sec:observations}, together with the pMSSM-based~\cite{Djouadi:1998di} efficiency estimation, drive the exclusion for the cross section of total leptonic RPV SUSY production, which is presented in Fig.~\ref{fig:sigmaBand}. The bands in Fig.~\ref{fig:sigmaBand} correspond to varying the four-lepton isolation efficiency from 50\% to 100\%. Note that this band covers  RPV models with a wide range of underlying RPC SUSY models.
We demonstrate that the kinematic efficiency is controlled by the neutralino mass and only weakly dependent on the neutralino momentum.

\begin{figure}[tbh]
  \centering
    \includegraphics[width=0.45\textwidth]{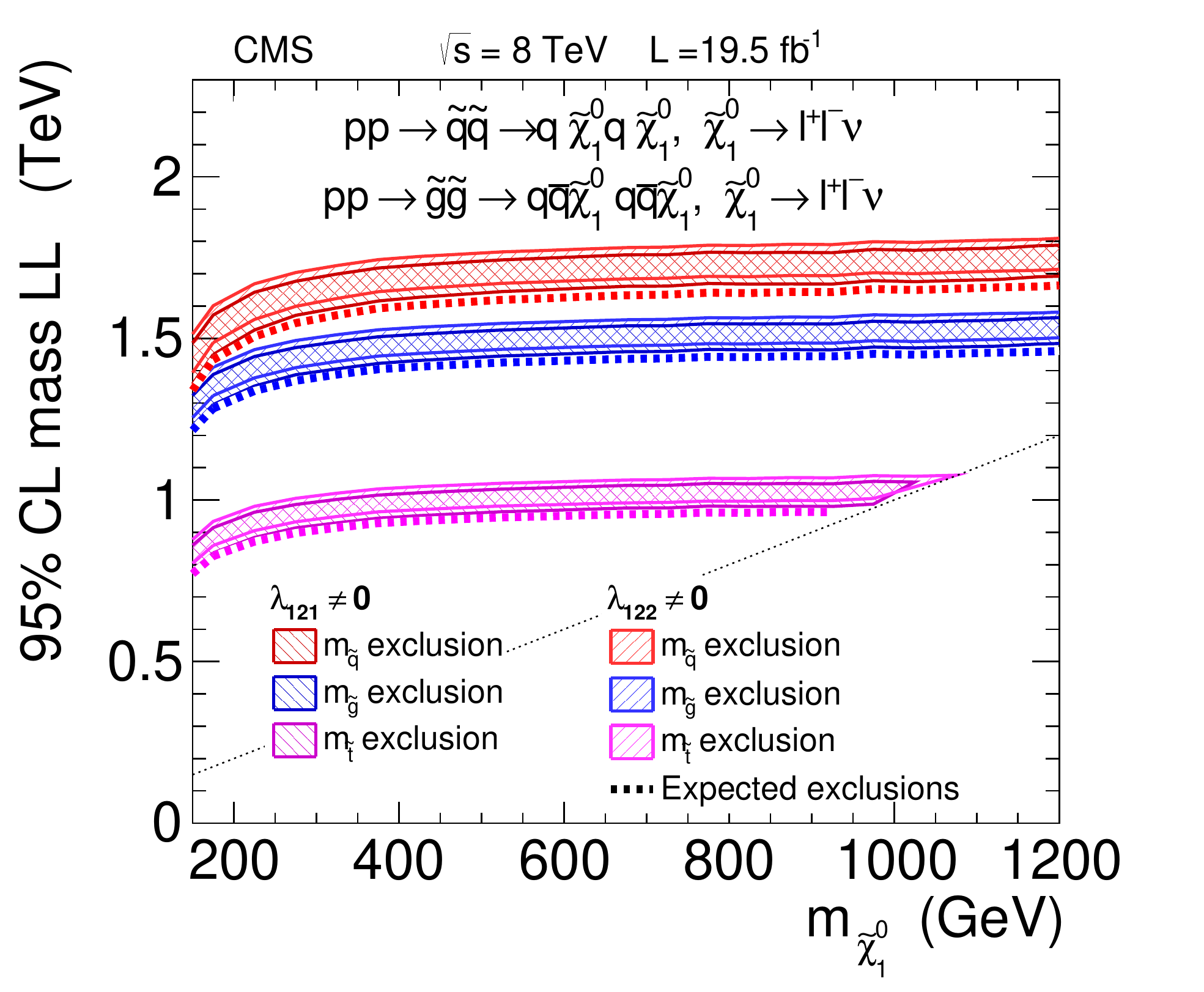}
    \includegraphics[width=0.45\textwidth]{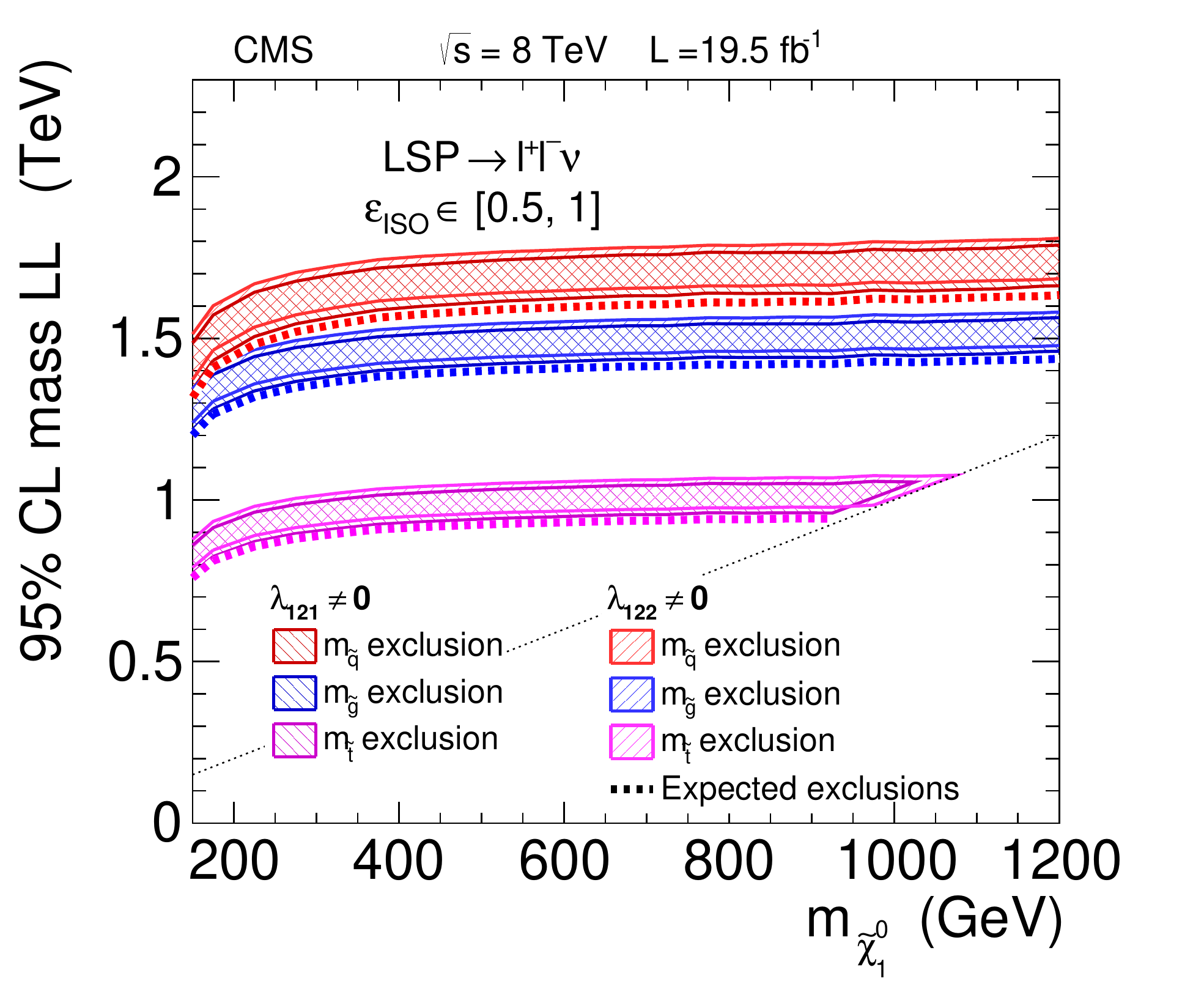}
    \caption{
    Lower limits on the masses of intermediate-state particles for various SUSY production mechanisms as a function of the neutralino mass when the neutralino is the LSP. The limits shown in the \cmsLeft plot on the gluino mass, top-squark mass, and squark mass are for gluino pair production, top-squark pair production, and first and second generation squark pair production, respectively. The first and second generation squark mass exclusion is obtained from the cross section limit of Fig.~\ref{fig:T2sigmaBand}~(\cmsRight), combined with the corresponding cross section for this model, displayed in Fig.~\ref{fig:T2LRPV}~(\cmsLeft). The other exclusions are obtained in a similar fashion. The limits shown in the \cmsRight plot are obtained from the generic total RPV SUSY cross section limit from Fig.~\ref{fig:sigmaBand}~(\cmsRight). The bands include the uncertainty in SUSY production cross sections, along with the efficiency uncertainties. The models below the diagonal line were not investigated as the neutralino is no longer the LSP.
   }
    \label{fig:massBands}
\end{figure}

To further convert the cross section limit into a mass exclusion, we consider several SUSY production mechanisms: gluino pair production, top-squark pair production, and first and second generation squark pair production.
Using these total cross sections as a function of the mass of the corresponding SUSY particle, we convert the cross section limit bands in Figs.~\ref{fig:T2sigmaBand} and~\ref{fig:sigmaBand} into exclusion bands for the masses of the parent SUSY particles, as a function of the LSP mass. These results are presented in Fig.~\ref{fig:massBands}. A 30\% theoretical uncertainty in NLO+NLL calculations of SUSY production cross sections is included in the uncertainty band, along with the efficiency uncertainties.

The weaker limits at low neutralino mass are a consequence of the low efficiency in this region. For the cases $\lambda_{121} \ne 0$ or $\lambda_{122}\ne 0$, and $\lambda$ sufficiently large to lead to prompt neutralino decays, this model is excluded at a 95\% \CL for gluino masses below about 1.4\TeV for a neutralino mass higher than 400\GeV. With only top-squark production, a top-squark mass below about 950\GeV is generally excluded. The squark mass exclusion depends significantly on the gluino mass in the model, as the squark pair production cross section increases as gluino mass decreases. For the 2.4\TeV gluino benchmark point, squarks with a mass below about 1.6\TeV are excluded.

\section{Multilepton final state via electroweak production}
\label{sec:multilepton_broad}

When leptonic or semileptonic RPV couplings are allowed, LSP pair production can result in a final state with three or more leptons.
In this section we expand the results from a previous search~\cite{Chatrchyan:2013xsw} to this class of RPV models.
Our focus here is on models with a Higgsino- or winolike LSP, and we explore the leptonic RPV couplings $\lambda_{122}$,  $\lambda_{123}$, and $\lambda_{233}$ and the semileptonic couplings $\lambda^\prime_{131}$, $\lambda^\prime_{233}$, $\lambda^\prime_{331}$, and $\lambda^\prime_{333}$.

\subsection{Event selection}

We select events with three or more leptons (including up to one hadronically decaying tau lepton $\tauh$) that are accepted by a trigger requiring two light leptons, which may be electrons or muons. Any OSSF pair of electrons or muons must have an invariant mass $m_{\ell \ell} > 12$\GeV, removing low-mass bound states and most $\gamma^{*}\to\ell^{+} \ell^{-}$ events.

We define 32 exclusive signal regions depending on the total number of leptons, the number of $\tauh$ candidates in the event, the number of $\PQb$-tagged jets in the event, the number of OSSF pairs, and in the cases where an OSSF pair is present, different bins in the dilepton invariant mass, which is calculated only for light leptons (electrons and muons). When there is more than one OSSF pair, if any of these dilepton masses are within 15\GeV of the $\Z$ mass, the event is called on-\Z. The binning is shown in the first four columns of Table~\ref{tab:fullResults}. Each of these 32 signal regions is divided into five different $\ST$ bins, where $\ST$ is defined as the sum of the missing transverse momentum and the scalar sum of jet and charged lepton \pt: $\ST=$[0--300], [300--600], [600--1000], [1000--1500], and [$>$1500]\GeV.

\subsection{Backgrounds}
Multilepton searches have two main sources of background, the first arising from processes that produce genuine multilepton events. The most abundant examples are $\PW\Z$ and $\Z\Z$ production, but rare processes such as $\ttbar\PW$ and $\ttbar\Z$\  also contribute.
Samples simulating $\PW\Z$ and $\Z\Z$ production have been validated in control regions in data. For the rarer background processes, we rely solely on simulation.

The second source of background originates from objects that are misclassified as prompt, isolated leptons, but are actually hadrons, leptons from a hadron decay, etc. Misidentified leptons are classified in three categories: misidentified light leptons, misidentified $\tauh$ leptons, and light leptons originating from asymmetric internal conversions. The methods used to estimate these backgrounds are described in more detail in Ref.~\cite{Chatrchyan:2012mea}.

We estimate the contribution of misidentified light leptons by measuring the number of isolated tracks and applying a scale factor between isolated leptons and isolated tracks. These scale factors are measured in control regions that contain leptonically decaying \Z~bosons and a third, isolated track, as well as in control regions with opposite-sign, different-flavor leptons, an isolated track, and a $\PQb$-tagged jet, which are dominated by \ttbar production. The scale factor is then the probability for the third track to pass the lepton identification criteria. We find the scale factors to be ($0.9 \pm 0.2$)\% for electrons and ($0.7 \pm 0.2$)\% for muons. The scale factors are applied to the number of events in the sideband region with two light leptons and an isolated track. The scale factors depend on the heavy-flavor content in the different signal regions. We parameterize this dependence as a function of the impact parameter distribution of nonisolated tracks. The $\ttbar$ contribution is taken from simulation.

The $\tauh$ misidentification rate is measured in a jet-dominated data control sample by comparing the number of $\tauh$ candidates in the isolated region defined by $\ETcone < 2$\GeV to the number of nonisolated $\tauh$ candidates, which have $6 < \ETcone < 15$\GeV. We measure the average misidentification rate as 15\% with a systematic uncertainty of 30\% based on the variation in different control samples. We apply this scale factor to the number of events in the sideband region with two light leptons and one nonisolated $\tauh$ candidate.

Another source of background leptons is internal conversions of a virtual photon to a dilepton pair.
We measure the conversion factors of photons to light leptons in a control region with low \MET and low hadronic activity. The ratio of the number of $\ell^+\ell^-\ell^{\pm}$ candidates to the number of $\ell^+\ell^-\gamma$ candidates in \Z boson decays defines the conversion factor, which is  $ 2.1\%\pm1.0\%$ ($0.5\%\pm0.3\%$) for electrons (muons).

\subsection{Systematic uncertainties}

We scale the $\PW\Z$ and $\Z\Z$ simulation samples to match the data in control regions. The overall systematic uncertainty in $\PW\Z$ and $\Z\Z$ contributions to the signal regions varies between 15\% and 30\%, depending on the kinematics, and is the combination of the normalization uncertainties with resolution uncertainties~\cite{Chatrchyan:2013xsw}.
 An uncertainty of 50\% for the $\ttbar$ background contribution is due to the low event counts in the isolation distributions in high-$\ST$ bins, which are used to validate the misidentification rate. We apply a 50\% uncertainty to the normalization of all rare processes to cover cross section and PDF uncertainties.

\subsection{Results and interpretations for the multilepton final state}

The observed and expected yields in the regions described above are shown in Table \ref{tab:fullResults}. We find good agreement between the SM predictions and observed data.

\begin{table*}[htbp]
\centering
\topcaption{Expected and observed yields for three- and four-lepton events. The channels are split by the number of leptons (N$_\ell$), the number of $\tauh$ candidates (N$_{\tau}$), whether the event contains $\PQb$-tagged jets (N$_{\PQb}$), the number of OSSF pairs (N$_\text{OSSF}$), binning in the dilepton invariant mass ($m_{\ell\ell}$ of light leptons only), and the $\ST$. Events are considered on-\Z if $75 < m_{\ell\ell} < 105\GeV$. Expected yields are the sum of simulation and estimates of backgrounds from data in each channel.  The channels are mutually exclusive. The uncertainties include statistical and systematic uncertainties. The $\ST$ and $m_{\ell\ell}$ values are given in\GeV. Reproduced from Ref.~\cite{Chatrchyan:2013xsw}.
\label{tab:fullResults}}
\cmsTableResize{
\begin{scotch}{@{\,}c@{\;}c@{\;}c@{\;}c@{\;}c@{\,\,}c@{\,\,}c@{\,\,}c@{\;}c@{\;}c@{\;}c@{\;}c@{\;}c@{\;}c@{\;}c@{\,}}
N$_\ell$ & N$_{\tau}$ & N$_{\PQb}$ & N$_\text{OSSF}$ & $m_{\ell\ell}$& \multicolumn{2}{c}{$0 < \ST < 300$} &\multicolumn{2}{c}{$300 < \ST < 600$}& \multicolumn{2}{c}{$600 < \ST < 1000$}& \multicolumn{2}{c}{$1000 < \ST < 1500$} & \multicolumn{2}{c}{$\ST > 1500$} \\
\hline
& & & & & obs & exp & obs & exp & obs & exp & obs & exp & obs & exp\\
\hline
4 & 0 & 0 & 0 & \NA &0 & 0.06 $\pm$ 0.06 & 0 & 0.09 $\pm$ 0.07 & 0 & 0.00 $\pm$ 0.03 & 0 & 0.00 $\pm$ 0.03 & 0 & 0.00 $\pm$ 0.03 \\
4 & 0 & 1 & 0 & \NA &0 & 0.00 $\pm$ 0.03 & 0 & 0.00 $\pm$ 0.03 & 0 & 0.06 $\pm$ 0.05 & 0 & 0.00 $\pm$ 0.03 & 0 & 0.00 $\pm$ 0.03 \\
4 & 0 & 0 & 1 & on-\Z & 2 & 3.1 $\pm$ 0.90 & 5 & 1.9 $\pm$ 0.48 & 0 & 0.44 $\pm$ 0.16 & 1 & 0.06 $\pm$ 0.06 & 0 & 0.00 $\pm$ 0.03 \\
4 & 0 & 1 & 1 & on-\Z & 2 & 0.07 $\pm$ 0.05 & 2 & 1.1 $\pm$ 0.53 & 0 & 0.57 $\pm$ 0.30 & 0 & 0.12 $\pm$ 0.09 & 0 & 0.02 $\pm$ 0.03 \\
4 & 0 & 0 & 1 & off-\Z & 2 & 0.48 $\pm$ 0.18 & 0 & 0.27 $\pm$ 0.11 & 0 & 0.07 $\pm$ 0.05 & 0 & 0.00 $\pm$ 0.02 & 0 & 0.00 $\pm$ 0.03 \\
4 & 0 & 1 & 1 & off-\Z & 0 & 0.04 $\pm$ 0.04 & 0 & 0.34 $\pm$ 0.17 & 0 & 0.06 $\pm$ 0.08 & 0 & 0.04 $\pm$ 0.04 & 0 & 0.00 $\pm$ 0.03 \\
4 & 0 & 0 & 2 & on-\Z & 135 & 120 $\pm$ 29 & 26 & 43 $\pm$ 10 & 4 & 6.0 $\pm$ 2.0 & 1 & 0.63 $\pm$ 0.26 & 0 & 0.06 $\pm$ 0.04 \\
4 & 0 & 1 & 2 & on-\Z & 1 & 1.0 $\pm$ 0.27 & 4 & 3.2 $\pm$ 1.1 & 1 & 1.1 $\pm$ 0.39 & 0 & 0.11 $\pm$ 0.06 & 0 & 0.04 $\pm$ 0.04 \\
4 & 0 & 0 & 2 & off-\Z & 7 & 8.3 $\pm$ 2.3 & 3 & 1.1 $\pm$ 0.30 & 0 & 0.11 $\pm$ 0.05 & 0 & 0.01 $\pm$ 0.02 & 0 & 0.00 $\pm$ 0.02 \\
4 & 0 & 1 & 2 & off-\Z & 0 & 0.18 $\pm$ 0.07 & 1 & 0.22 $\pm$ 0.11 & 0 & 0.15 $\pm$ 0.08 & 0 & 0.00 $\pm$ 0.03 & 0 & 0.00 $\pm$ 0.03 \\
4 & 1 & 0 & 0 & \NA &2 & 1.1 $\pm$ 0.46 & 1 & 0.54 $\pm$ 0.20 & 0 & 0.12 $\pm$ 0.12 & 0 & 0.00 $\pm$ 0.03 & 0 & 0.00 $\pm$ 0.03 \\
4 & 1 & 1 & 0 & \NA &0 & 0.26 $\pm$ 0.16 & 0 & 0.29 $\pm$ 0.13 & 0 & 0.13 $\pm$ 0.11 & 0 & 0.01 $\pm$ 0.02 & 0 & 0.00 $\pm$ 0.03 \\
4 & 1 & 0 & 1 & on-\Z & 43 & 42 $\pm$ 11 & 10 & 12 $\pm$ 3.1 & 0 & 1.8 $\pm$ 0.63 & 0 & 0.11 $\pm$ 0.07 & 0 & 0.02 $\pm$ 0.03 \\
4 & 1 & 1 & 1 & on-\Z & 2 & 1.0 $\pm$ 0.40 & 2 & 1.7 $\pm$ 0.5 & 0 & 0.78 $\pm$ 0.33 & 0 & 0.04 $\pm$ 0.04 & 0 & 0.01 $\pm$ 0.03 \\
4 & 1 & 0 & 1 & off-\Z & 18 & 8.4 $\pm$ 2.2 & 4 & 2.1 $\pm$ 0.52 & 2 & 0.48 $\pm$ 0.18 & 0 & 0.13 $\pm$ 0.08 & 0 & 0.01 $\pm$ 0.03 \\
4 & 1 & 1 & 1 & off-\Z & 1 & 0.64 $\pm$ 0.31 & 0 & 1.2 $\pm$ 0.44 & 0 & 0.30 $\pm$ 0.13 & 0 & 0.02 $\pm$ 0.03 & 0 & 0.00 $\pm$ 0.03 \\
3 & 0 & 0 & 0 & \NA &72 & 80 $\pm$ 23 & 32 & 27 $\pm$ 11 & 3 & 3.1 $\pm$ 1.00 & 0 & 0.22 $\pm$ 0.18 & 0 & 0.07 $\pm$ 0.06 \\
3 & 0 & 1 & 0 & \NA &37 & 33 $\pm$ 16 & 42 & 39 $\pm$ 19 & 2 & 5.0 $\pm$ 2.0 & 0 & 0.36 $\pm$ 0.14 & 0 & 0.06 $\pm$ 0.07 \\
3 & 0 & 0 & 1 & on-\Z & 4255 & 4400 $\pm$ 690 & 669 & 740 $\pm$ 170 & 106 & 110 $\pm$ 41 & 11 & 15 $\pm$ 6.9 & 3 & 1.3 $\pm$ 0.76 \\
3 & 0 & 1 & 1 & on-\Z & 140 & 150 $\pm$ 25 & 122 & 110 $\pm$ 25 & 16 & 25 $\pm$ 7.0 & 2 & 3.3 $\pm$ 1.2 & 1 & 0.32 $\pm$ 0.22 \\
3 & 0 & 0 & 1 & $m_{\ell\ell} < 75$ & 617 & 640 $\pm$ 100 & 84 & 86 $\pm$ 21 & 14 & 11 $\pm$ 3.6 & 0 & 1.2 $\pm$ 0.39 & 1 & 0.12 $\pm$ 0.09 \\
3 & 0 & 1 & 1 & $m_{\ell\ell} < 75$  & 62 & 74 $\pm$ 28 & 52 & 57 $\pm$ 23 & 4 & 8.3 $\pm$ 2.7 & 1 & 0.69 $\pm$ 0.28 & 0 & 0.08 $\pm$ 0.06 \\
3 & 0 & 0 & 1 & $m_{\ell\ell} > 105$ & 180 & 200 $\pm$ 34 & 63 & 66 $\pm$ 12 & 13 & 10 $\pm$ 2.5 & 2 & 1.1 $\pm$ 0.40 & 0 & 0.16 $\pm$ 0.09 \\
3 & 0 & 1 & 1 & $m_{\ell\ell} > 105$ & 17 & 17 $\pm$ 6.5 & 36 & 35 $\pm$ 14 & 7 & 7.4 $\pm$ 2.5 & 0 & 0.54 $\pm$ 0.23 & 0 & 0.08 $\pm$ 0.05 \\
3 & 1 & 0 & 0 & \NA &1194 & 1300 $\pm$ 330 & 289 & 290 $\pm$ 130 & 26 & 28 $\pm$ 12 & 2 & 2.6 $\pm$ 1.3 & 0 & 0.23 $\pm$ 0.20 \\
3 & 1 & 1 & 0 & \NA &316 & 330 $\pm$ 160 & 410 & 480 $\pm$ 240 & 46 & 58 $\pm$ 28 & 2 & 3.9 $\pm$ 2.0 & 0 & 0.46 $\pm$ 0.32 \\
3 & 1 & 0 & 1 & on-\Z & 49916 & 49000 $\pm$ 15000 & 2099 & 2700 $\pm$ 770 & 108 & 70 $\pm$ 17 & 9 & 6.0 $\pm$ 1.6 & 0 & 0.33 $\pm$ 0.18 \\
3 & 1 & 1 & 1 & on-\Z & 795 & 830 $\pm$ 230 & 325 & 280 $\pm$ 74 & 17 & 17 $\pm$ 4.8 & 1 & 1.8 $\pm$ 0.64 & 0 & 0.30 $\pm$ 0.14 \\
3 & 1 & 0 & 1 & $m_{\ell\ell} < 75$ & 10173 & 9200 $\pm$ 2700 & 290 & 280 $\pm$ 72 & 21 & 11 $\pm$ 3.5 & 1 & 0.97 $\pm$ 0.44 & 0 & 0.04 $\pm$ 0.06 \\
3 & 1 & 1 & 1 & $m_{\ell\ell} < 75$ & 297 & 290 $\pm$ 97 & 167 & 170 $\pm$ 87 & 14 & 12 $\pm$ 6.0 & 0 & 1.1 $\pm$ 0.74 & 0 & 0.06 $\pm$ 0.08 \\
3 & 1 & 0 & 1 & $m_{\ell\ell} > 105$ & 1620 & 1700 $\pm$ 480 & 285 & 370 $\pm$ 96 & 21 & 23 $\pm$ 7.2 & 1 & 1.4 $\pm$ 0.61 & 0 & 0.22 $\pm$ 0.23 \\
3 & 1 & 1 & 1 & $m_{\ell\ell} > 105$ & 97 & 79 $\pm$ 36 & 169 & 190 $\pm$ 94 & 23 & 28 $\pm$ 14 & 1 & 2.2 $\pm$ 1.3 & 0 & 0.20 $\pm$ 0.18\\
\end{scotch}}
\end{table*}

We explore SUSY models featuring production of electroweak boson superpartners, where these sparticles can either be wino- or Higgsino-like. We investigate cases where one leptonic or semileptonic coupling is nonzero. The selected models with leptonic RPV contain sleptons with masses of 1.5\TeV to mediate the decay. In the semileptonic models, these mediators are bottom and top squarks that also have masses of 1.5\TeV. In all cases, we assume that the decays of the neutralinos are prompt, and the difference between the neutralino and chargino masses is roughly 1\GeV.

\begin{figure*}[htbp]
\centering
 \includegraphics[width=0.40\textwidth]{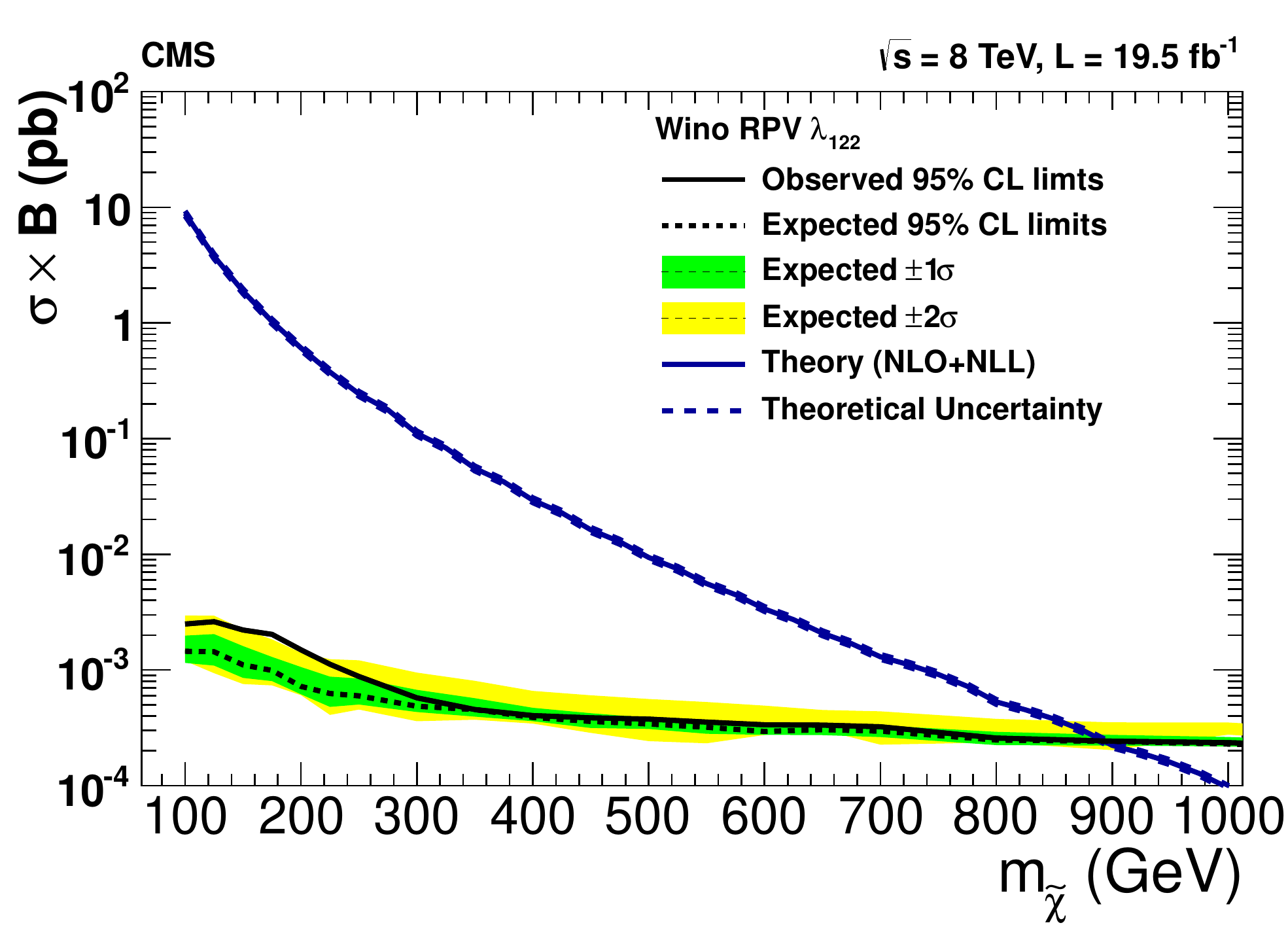}
  \includegraphics[width=0.40\textwidth]{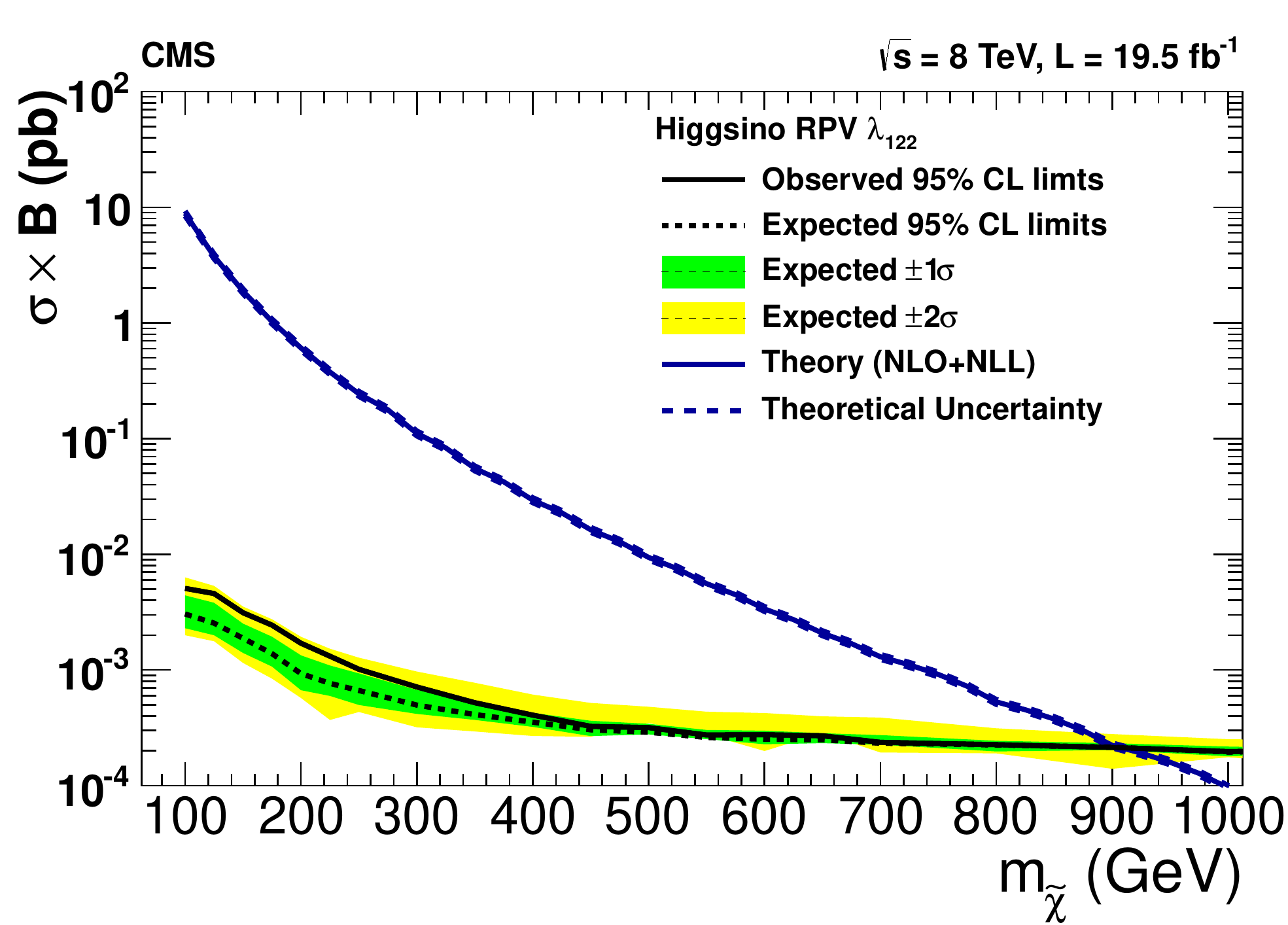}
 \includegraphics[width=0.40\textwidth]{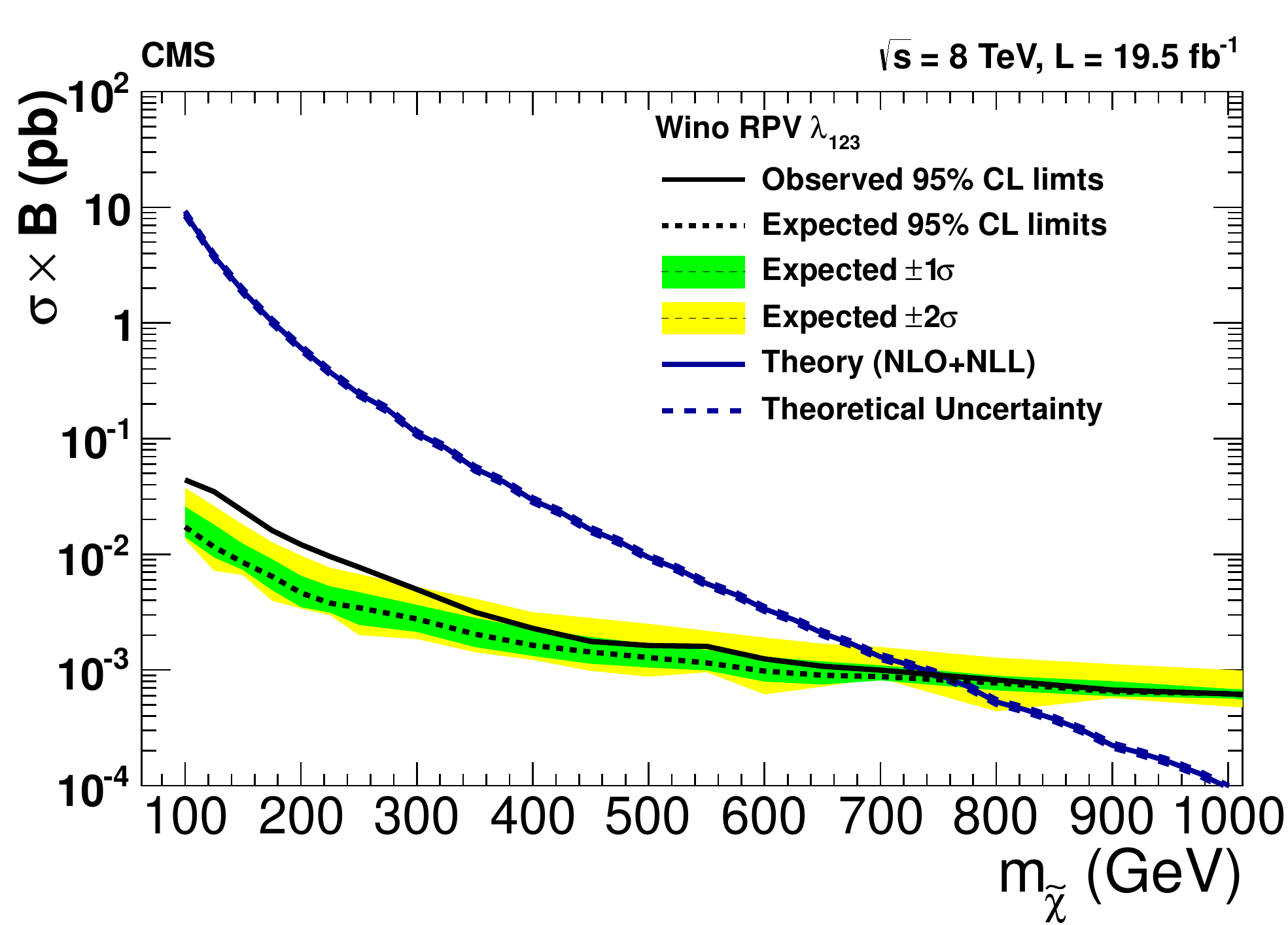}
  \includegraphics[width=0.40\textwidth]{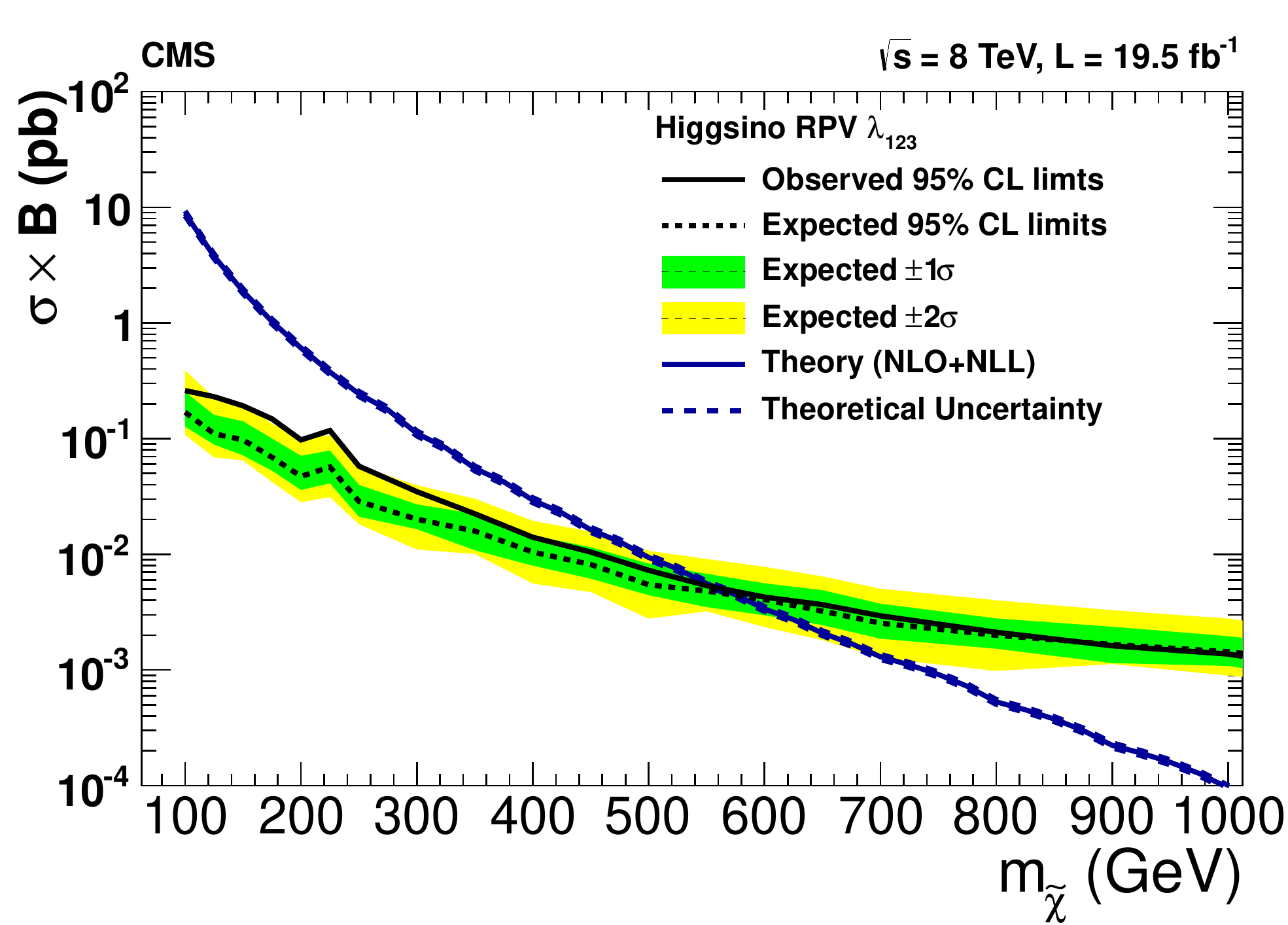}
 \includegraphics[width=0.40\textwidth]{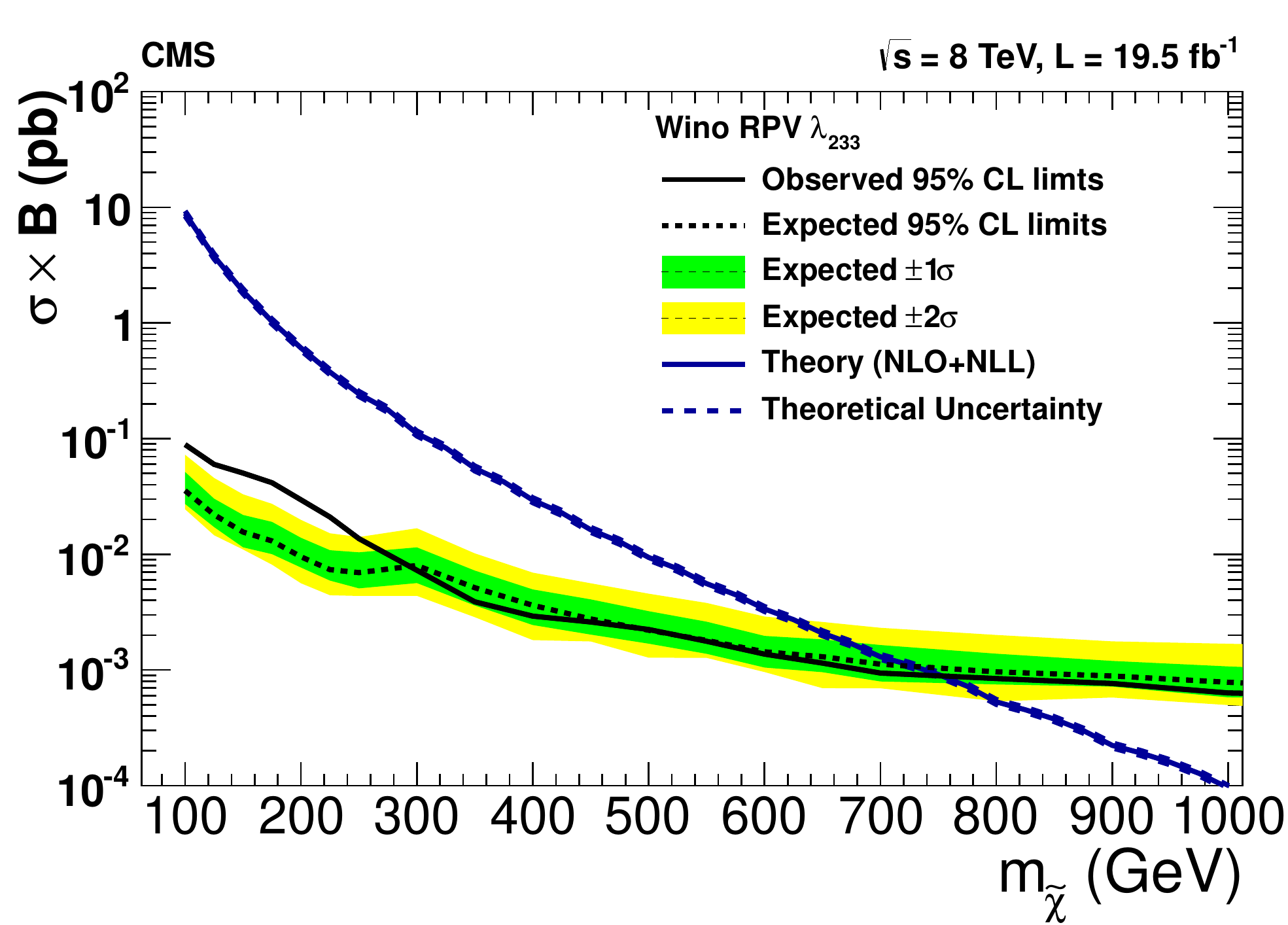}
  \includegraphics[width=0.40\textwidth]{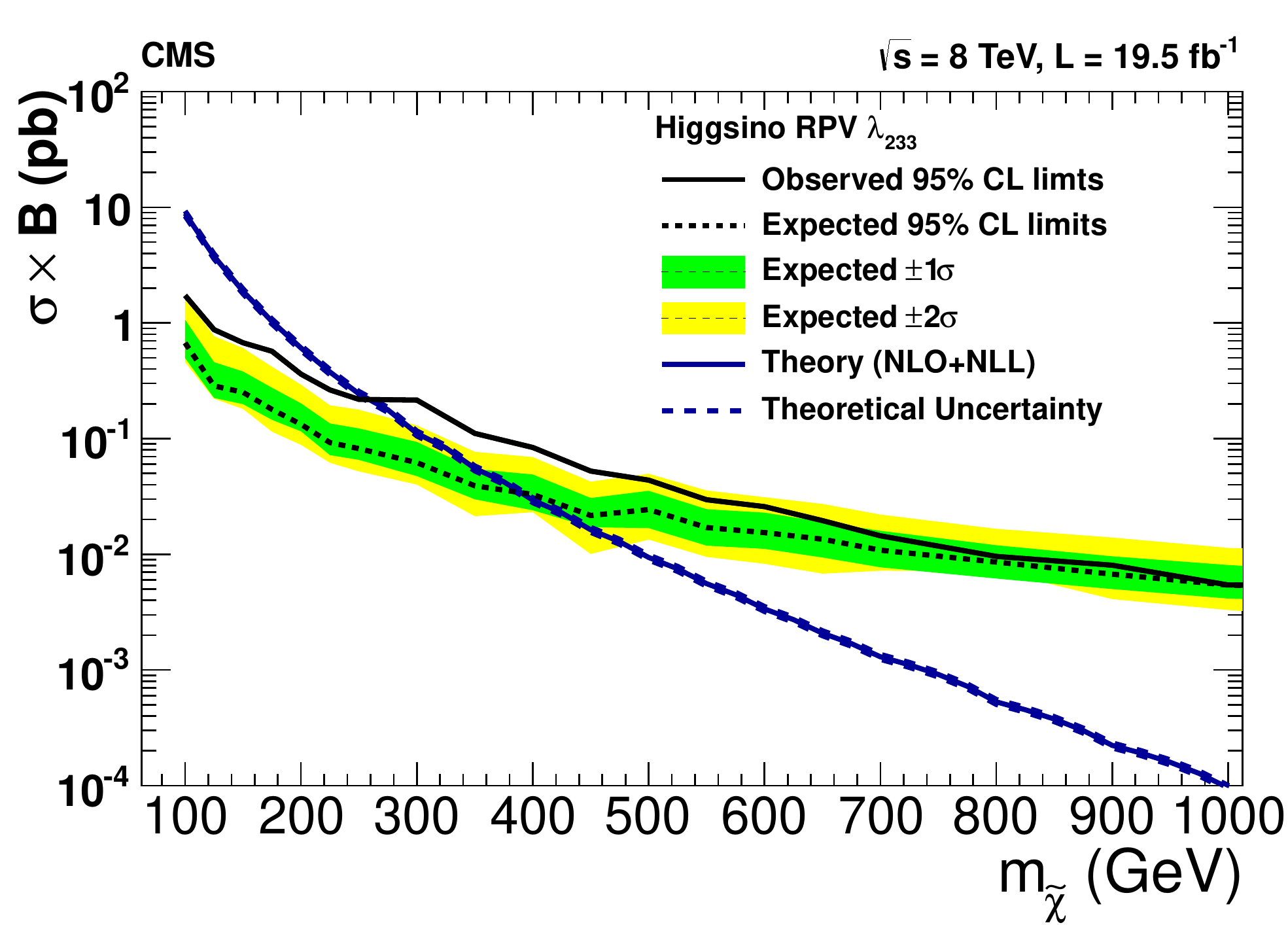}
\caption{\label{fig:LLE} Upper 95\% \CL cross section times branching ratio limits as a function of the neutralino mass in models with wino production (left) and Higgsino production (right), assuming nonzero $\lambda_{ijk}$ couplings: $\lambda_{122}$ (top), $\lambda_{123}$ (middle), and $\lambda_{233}$ (bottom). The decays proceed promptly through slepton mediators.}
\end{figure*}

Figure~\ref{fig:LLE} contains plots that show where we exclude wino- and Higgsino-like neutralinos when there are nonzero RPV couplings, which from top to bottom are $\lambda_{122}$,  $\lambda_{123}$, and $\lambda_{233}$. In the models with winolike neutralinos, the lower limit on the neutralino mass ranges from approximately 700 to 875\GeV. In the case of Higgsino-like neutralinos, the range is much larger, from approximately 300 to 900\GeV.

\begin{figure*}[htbp]
\centering
 \includegraphics[width=0.40\textwidth]{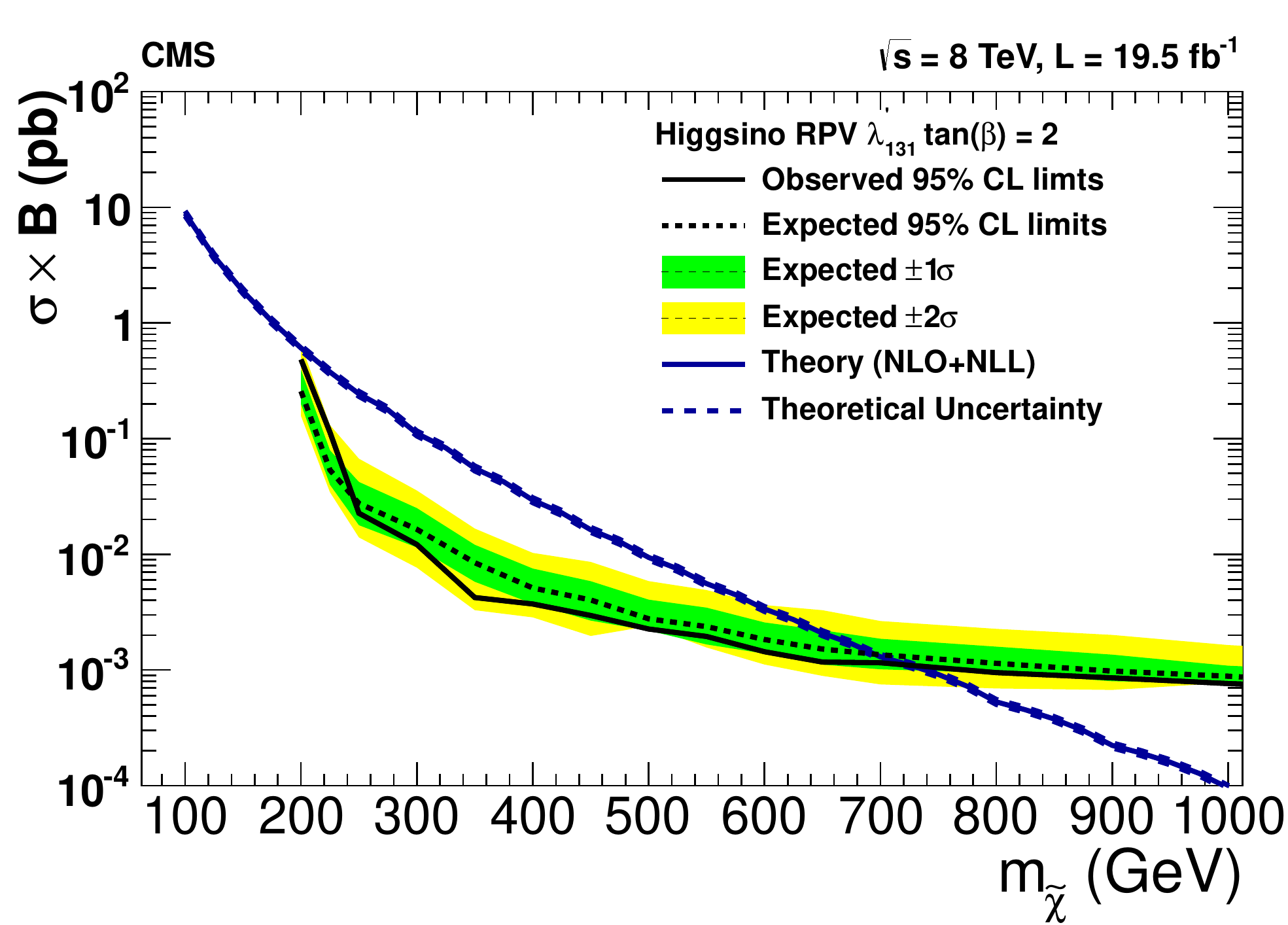}
 \includegraphics[width=0.40\textwidth]{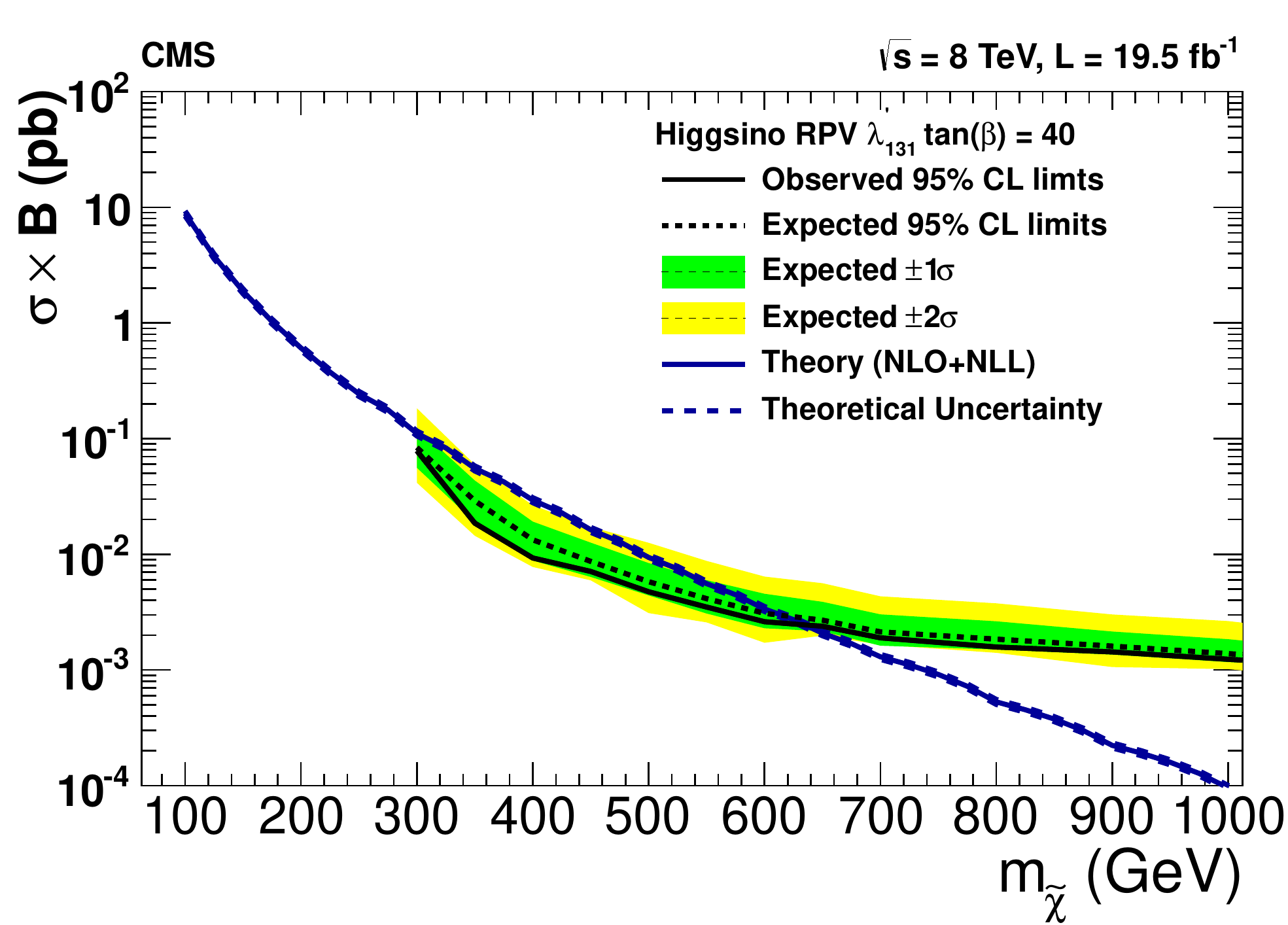}
  \includegraphics[width=0.40\textwidth]{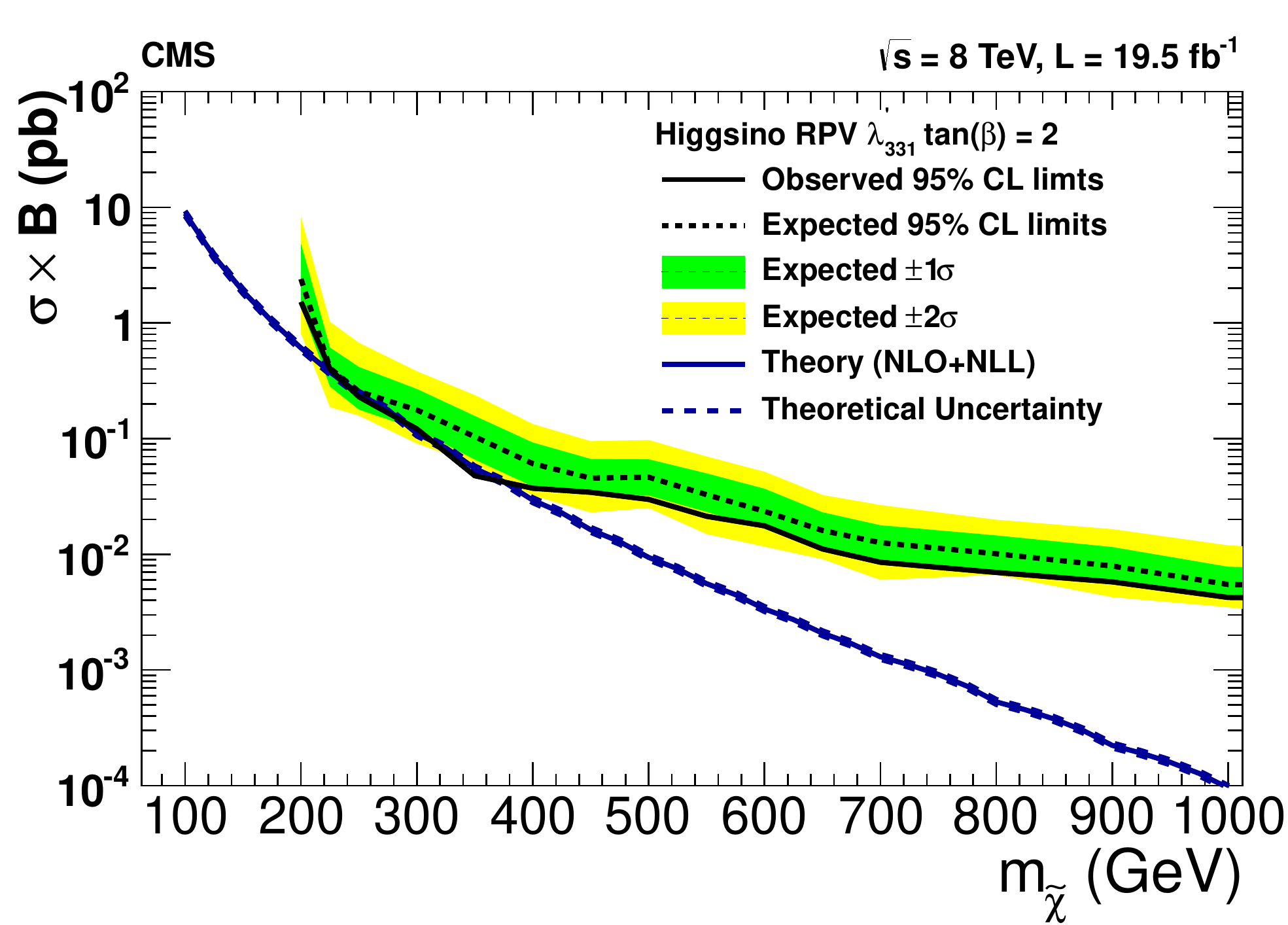}
  \includegraphics[width=0.40\textwidth]{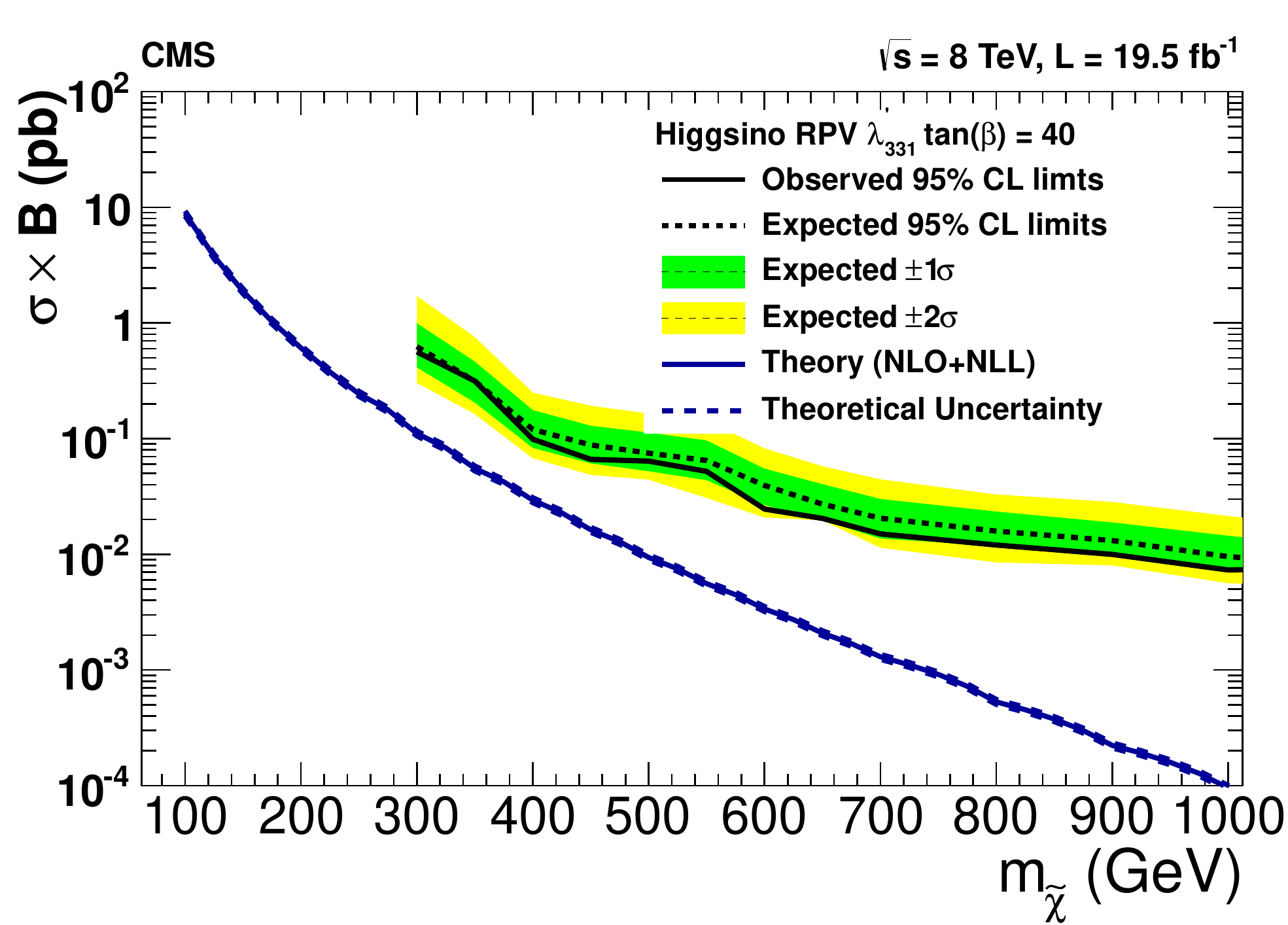}
 \includegraphics[width=0.40\textwidth]{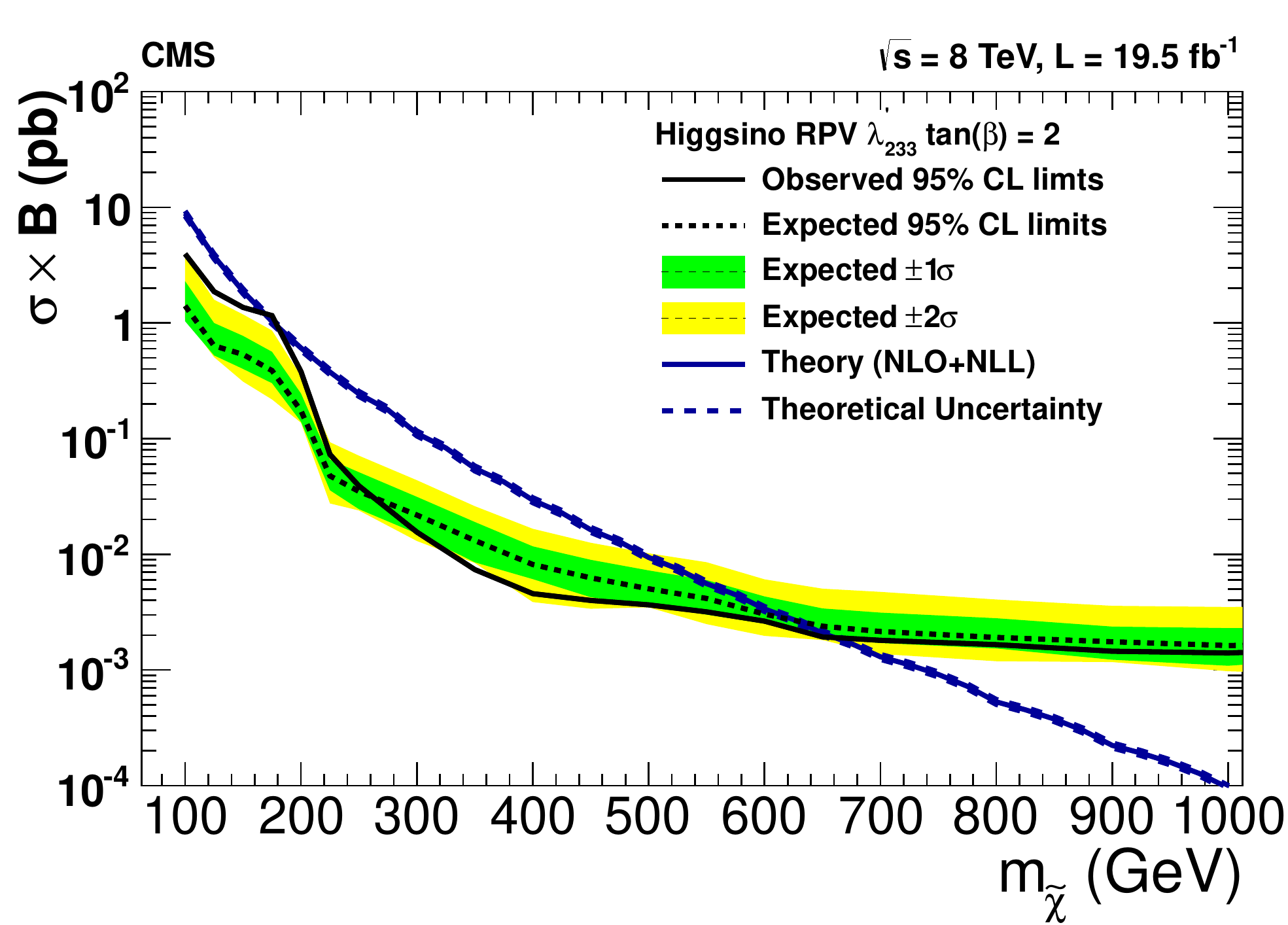}
 \includegraphics[width=0.40\textwidth]{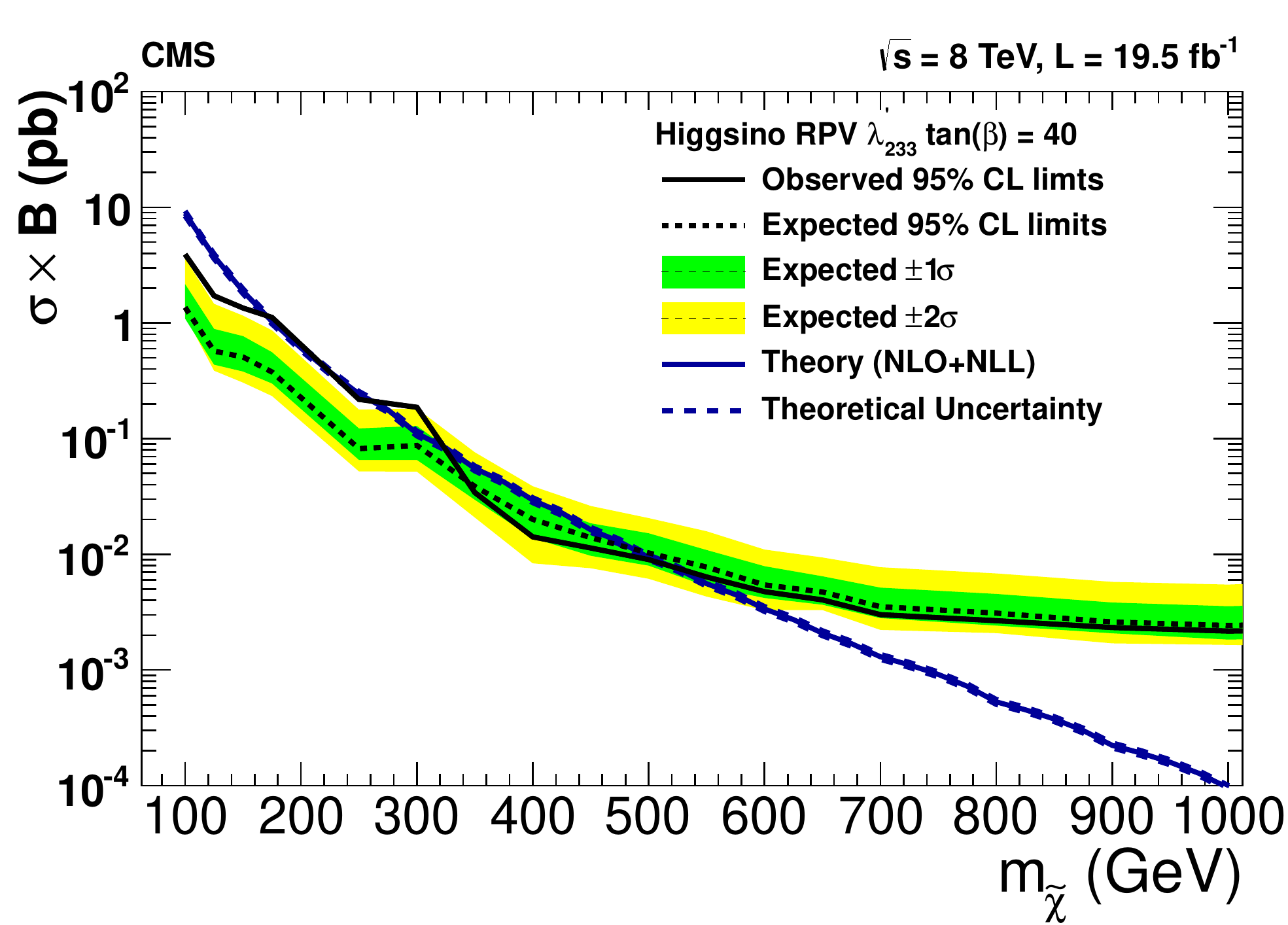}
  \includegraphics[width=0.40\textwidth]{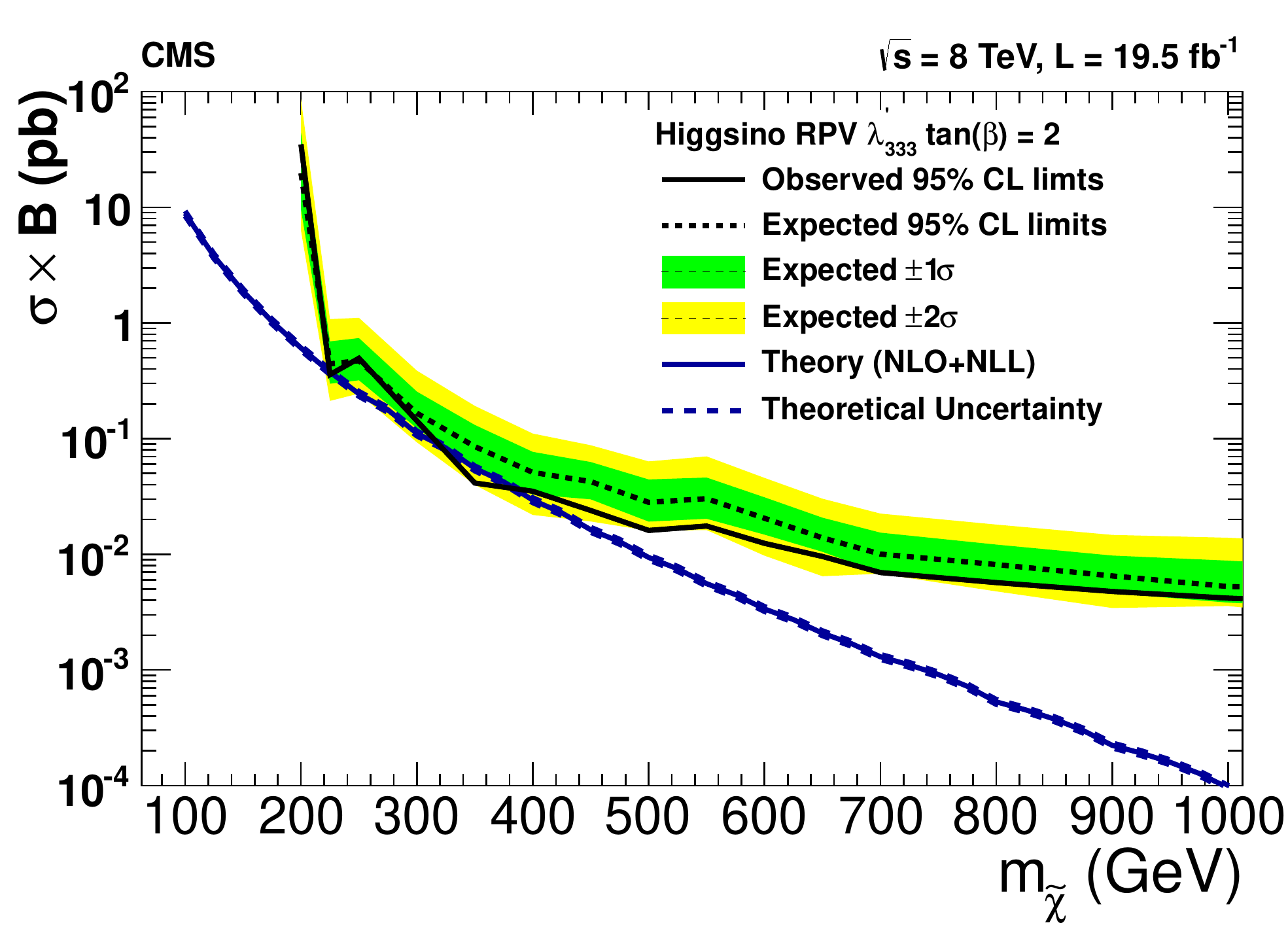}
  \includegraphics[width=0.40\textwidth]{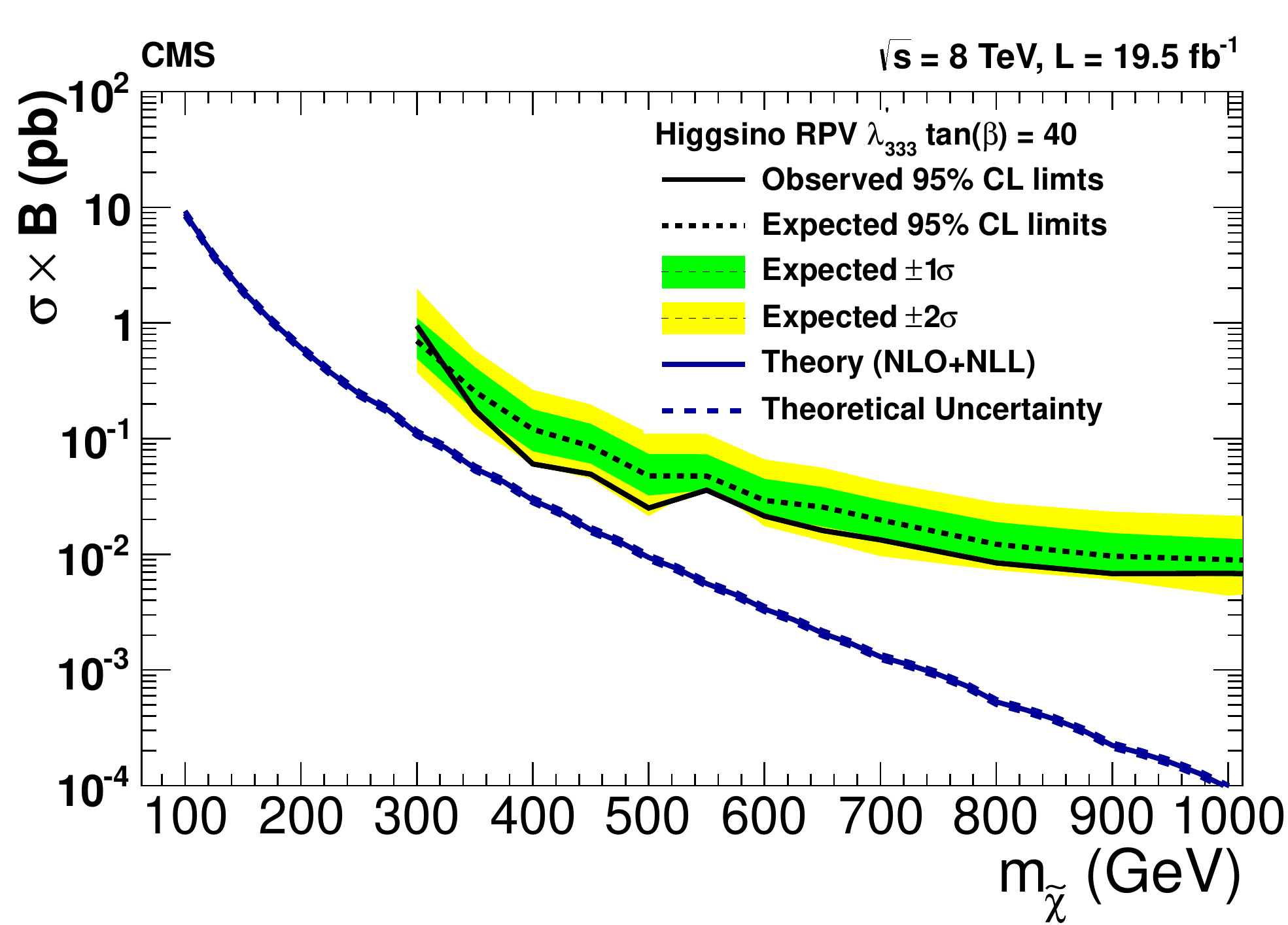}
\caption{\label{fig:HiggsinoLQD} Upper 95\% \CL cross section times branching fraction limits as a function of the neutralino mass in models with Higgsino production and nonzero $\lambda^\prime_{ijk}$ couplings with (left) $\tan \beta = 2$ and (right) $\tan \beta = 40$: $\lambda^\prime_{131}$ (top), $\lambda^\prime_{331}$ (second row), $\lambda^\prime_{233}$ (third row), and $\lambda^\prime_{333}$ (bottom). The decays proceed promptly through bottom and top squark mediators. The couplings of the Higgsino to the mediator particles varies with $\tan \beta$, affecting the branching fraction to multileptons and the acceptance.}
\end{figure*}

We show limits in models that have Higgsino-like neutralinos with semileptonic RPV couplings $\lambda^\prime_{131}$, $\lambda^\prime_{233}$, $\lambda^\prime_{331}$, and $\lambda^\prime_{333}$ in Fig.~\ref{fig:HiggsinoLQD}. We also feature two different values of $\tan \beta$ (2 and 40), which changes the relative coupling between the Higgsino and the mediator particles, which in turn changes the branching fraction to multileptons. The sensitivity of this analysis is best for the couplings that produce a light lepton in the final state (the first and third rows of Fig.~\ref{fig:HiggsinoLQD}).

\section{Summary}
\label{sec:conclusions}
This paper explores a variety of final states where $R$-parity-violating (RPV) supersymmetry could appear. Using data samples corresponding to 19.5\fbinv~of proton-proton collision data collected with the CMS detector at $\sqrt{s} = 8\TeV$, no discrepancies from the expectations of the standard model are found. Limits on the masses of supersymmetric particles are set at 95\% confidence level in several models that exhibit different RPV couplings and contain different lightest supersymmetric particles (LSP).

The consequences of minimal flavor violation are explored in analyses that consider pair production of either gluinos or bottom squarks. The $\PQb$-tagged and total jet multiplicity distributions are used to set limits on the mass of a gluino that decays to a top, a bottom, and a strange quark via the superpotential coupling $\lambda''_{332}$. Gluinos with masses less than 0.98\TeV are excluded. Using a search region characterized by one lepton and high multiplicity of jets and $\PQb$-tagged jets, gluinos with masses less than 1.03\TeV are excluded in the same model. Another model assumes pair production of a bottom squark LSP that decays via $\lambda''_{332}$ or  $\lambda''_{331}$. Using the reconstructed resonance mass distribution in the dilepton final state, bottom squark production is excluded for masses less than 307\GeV.

Multilepton final states are sensitive to models with a variety of different leptonic or semileptonic RPV couplings.
Limits on the mediator masses are established in a search in the four-lepton channel for strong production of neutralinos and their decay via the leptonic couplings $\lambda_{121}$ and $\lambda_{122}$.
In a  model of squark pair production with a neutralino LSP, lower limits on the squark mass are set at about 1.6\TeV.
In models that feature electroweak production via leptonic couplings with wino- or Higgsino-like LSPs, limits are set on the LSP mass that range from 300 to 900\GeV. In those with semileptonic couplings, there are a variety of scenarios. In the regions of highest sensitivity, lower limits on the Higgsino masses reach approximately 700\GeV.

\begin{acknowledgments}
\hyphenation{Bundes-ministerium Forschungs-gemeinschaft Forschungs-zentren} We congratulate our colleagues in the CERN accelerator departments for the excellent performance of the LHC and thank the technical and administrative staffs at CERN and at other CMS institutes for their contributions to the success of the CMS effort. In addition, we gratefully acknowledge the computing centers and personnel of the Worldwide LHC Computing Grid for delivering so effectively the computing infrastructure essential to our analyses. Finally, we acknowledge the enduring support for the construction and operation of the LHC and the CMS detector provided by the following funding agencies: the Austrian Federal Ministry of Science, Research and Economy and the Austrian Science Fund; the Belgian Fonds de la Recherche Scientifique, and Fonds voor Wetenschappelijk Onderzoek; the Brazilian Funding Agencies (CNPq, CAPES, FAPERJ, and FAPESP); the Bulgarian Ministry of Education and Science; CERN; the Chinese Academy of Sciences, Ministry of Science and Technology, and National Natural Science Foundation of China; the Colombian Funding Agency (COLCIENCIAS); the Croatian Ministry of Science, Education and Sport, and the Croatian Science Foundation; the Research Promotion Foundation, Cyprus; the Secretariat for Higher Education, Science, Technology and Innovation, Ecuador; the Ministry of Education and Research, Estonian Research Council via IUT23-4 and IUT23-6 and European Regional Development Fund, Estonia; the Academy of Finland, Finnish Ministry of Education and Culture, and Helsinki Institute of Physics; the Institut National de Physique Nucl\'eaire et de Physique des Particules~/~CNRS, and Commissariat \`a l'\'Energie Atomique et aux \'Energies Alternatives~/~CEA, France; the Bundesministerium f\"ur Bildung und Forschung, Deutsche Forschungsgemeinschaft, and Helmholtz-Gemeinschaft Deutscher Forschungszentren, Germany; the General Secretariat for Research and Technology, Greece; the National Scientific Research Foundation, and National Innovation Office, Hungary; the Department of Atomic Energy and the Department of Science and Technology, India; the Institute for Studies in Theoretical Physics and Mathematics, Iran; the Science Foundation, Ireland; the Istituto Nazionale di Fisica Nucleare, Italy; the Ministry of Science, ICT and Future Planning, and National Research Foundation (NRF), Republic of Korea; the Lithuanian Academy of Sciences; the Ministry of Education, and University of Malaya (Malaysia); the Mexican Funding Agencies (BUAP, CINVESTAV, CONACYT, LNS, SEP, and UASLP-FAI); the Ministry of Business, Innovation and Employment, New Zealand; the Pakistan Atomic Energy Commission; the Ministry of Science and Higher Education and the National Science Centre, Poland; the Funda\c{c}\~ao para a Ci\^encia e a Tecnologia, Portugal; JINR, Dubna; the Ministry of Education and Science of the Russian Federation, the Federal Agency of Atomic Energy of the Russian Federation, Russian Academy of Sciences, and the Russian Foundation for Basic Research; the Ministry of Education, Science and Technological Development of Serbia; the Secretar\'{\i}a de Estado de Investigaci\'on, Desarrollo e Innovaci\'on and Programa Consolider-Ingenio 2010, Spain; the Swiss Funding Agencies (ETH Board, ETH Zurich, PSI, SNF, UniZH, Canton Zurich, and SER); the Ministry of Science and Technology, Taipei; the Thailand Center of Excellence in Physics, the Institute for the Promotion of Teaching Science and Technology of Thailand, Special Task Force for Activating Research and the National Science and Technology Development Agency of Thailand; the Scientific and Technical Research Council of Turkey, and Turkish Atomic Energy Authority; the National Academy of Sciences of Ukraine, and State Fund for Fundamental Researches, Ukraine; the Science and Technology Facilities Council, United Kingdom; the U.S. Department of Energy, and the U.S. National Science Foundation.

Individuals have received support from the Marie-Curie program and the European Research Council and EPLANET (European Union); the Leventis Foundation; the A. P. Sloan Foundation; the Alexander von Humboldt Foundation; the Belgian Federal Science Policy Office; the Fonds pour la Formation \`a la Recherche dans l'Industrie et dans l'Agriculture (FRIA-Belgium); the Agentschap voor Innovatie door Wetenschap en Technologie (IWT-Belgium); the Ministry of Education, Youth and Sports (MEYS) of the Czech Republic; the Council of Science and Industrial Research, India; the HOMING PLUS program of the Foundation for Polish Science, cofinanced from European Union, Regional Development Fund, the Mobility Plus program of the Ministry of Science and Higher Education, the OPUS program Contract No. 2014/13/B/ST2/02543 and Contract No. Sonata-bis DEC-2012/07/E/ST2/01406 of the National Science Center (Poland); the Thalis and Aristeia programs cofinanced by EU-ESF and the Greek NSRF; the National Priorities Research Program by Qatar National Research Fund; the Programa Clar\'in-COFUND del Principado de Asturias; the Rachadapisek Sompot Fund for Postdoctoral Fellowship, Chulalongkorn University and the Chulalongkorn Academic into Its 2nd Century Project Advancement Project (Thailand); and the Welch Foundation, Contract No. C-1845.
\end{acknowledgments}
\bibliography{auto_generated}

\cleardoublepage \appendix\section{The CMS Collaboration \label{app:collab}}\begin{sloppypar}\hyphenpenalty=5000\widowpenalty=500\clubpenalty=5000\textbf{Yerevan Physics Institute,  Yerevan,  Armenia}\\*[0pt]
V.~Khachatryan, A.M.~Sirunyan, A.~Tumasyan
\vskip\cmsinstskip
\textbf{Institut f\"{u}r Hochenergiephysik der OeAW,  Wien,  Austria}\\*[0pt]
W.~Adam, E.~Asilar, T.~Bergauer, J.~Brandstetter, E.~Brondolin, M.~Dragicevic, J.~Er\"{o}, M.~Flechl, M.~Friedl, R.~Fr\"{u}hwirth\cmsAuthorMark{1}, V.M.~Ghete, C.~Hartl, N.~H\"{o}rmann, J.~Hrubec, M.~Jeitler\cmsAuthorMark{1}, A.~K\"{o}nig, M.~Krammer\cmsAuthorMark{1}, I.~Kr\"{a}tschmer, D.~Liko, T.~Matsushita, I.~Mikulec, D.~Rabady, N.~Rad, B.~Rahbaran, H.~Rohringer, J.~Schieck\cmsAuthorMark{1}, J.~Strauss, W.~Treberer-Treberspurg, W.~Waltenberger, C.-E.~Wulz\cmsAuthorMark{1}
% ...
\vskip\cmsinstskip
% ...
1:~~Also at Vienna University of Technology, Vienna, Austria\\
% ...

\end{sloppypar}
\end{document}